\documentclass[12pt]{article}

\usepackage{amsmath,amssymb,amsfonts}
\usepackage{epsfig}

\numberwithin{equation}{section}
\newcommand{\be}{\begin{equation}} \newcommand{\ee}{\end{equation}}
\newcommand{\bea}{\begin{eqnarray}} \newcommand{\eea}{\end{eqnarray}}
\title{
\begin{center}
 { \bf  The Topological} ${\mathbf {G_2}}$  {\bf  String}
\end{center}
}

\begin{document}

\thispagestyle{empty} \setcounter{page}{0}



\rightline{hep-th/0506211} \rightline{ITFA-2005-23}

\vskip 1.5 cm

\renewcommand{\thefootnote}{\fnsymbol{footnote}}
\centerline{\Large \bf The Topological $G_2$ String} \vskip 1.5 cm
 \centerline{{\bf Jan de Boer${}^1$\footnote{\tt
jdeboer@science.uva.nl}, Asad Naqvi${}^1$\footnote{\tt
anaqvi@science.uva.nl} and Assaf Shomer${}^2$\footnote{\tt
shomer@scipp.ucsc.edu}}}

\vskip .5 cm \centerline{\it ${}^1$ Instituut voor Theoretische Fysica}
\centerline{\it Valckenierstraat 65, 1018XE Amsterdam, The
Netherlands} \vskip .5 cm

 \centerline{\it ${}^2$
Santa Cruz Institute for Particle Physics}
\centerline {\it 1156 High Street, Santa Cruz, 95064 CA, USA }

\begin{abstract}

We construct new topological theories related to sigma models
whose target space is a seven dimensional manifold of $G_2$
holonomy. We define a new type of topological twist and identify
the BRST operator and the physical states. Unlike the more
familiar six dimensional case, our topological model is defined in
terms of conformal blocks and not in terms of local operators of
the original theory. We also present evidence that one can extend
this definition to all genera and construct a seven-dimensional
topological string theory. We compute genus zero correlation
functions and relate these to Hitchin's functional for three-forms
in seven dimensions. Along the way we develop the analogue of
special geometry for $G_2$ manifolds. When the seven dimensional
topological twist is applied to the product of a Calabi-Yau
manifold and a circle, the result is an interesting combination of
the six dimensional A- and B-models.

\end{abstract}
\newpage

\setcounter{footnote}{0}
\renewcommand{\thefootnote}{\arabic{footnote}}

\section{Introduction}

Topological strings on Calabi-Yau manifolds describe certain
solvable sectors of superstrings. In particular, various BPS
quantities in string theory can be exactly computed using their
topological twisted version. Also, topological strings provide
simplified toy examples of string theories which are still rich
enough to exhibit interesting stringy phenomena  in a more
controlled setting. There are two inequivalent ways to twist the
Calabi-Yau $\sigma$ model which leads to the celebrated A and B
model \cite{Witten}. The metric is not a fundamental degree of
freedom in these models. Instead, the A-model apparently only
involves the K\"ahler moduli and the B-model only the complex
structure moduli. However, the roles interchange once branes are
included, and it has even been conjectured that there is a version
of S-duality which maps the A-model to the B-model on the same
Calabi-Yau manifold \cite{sdual}. This is quite distinct from
mirror symmetry which relates the A-model on $X$ to the B-model on
the mirror of $X$. Subsequently, several authors found evidence
for the existence of seven and/or eight dimensional theories that
unify and extend the A and B-models\
\cite{gerasimov,vafa,nekrasov,grassi,sinkovics}. This was one of
our motivations to take a closer look at string theory on
seven-dimensional manifolds of $G_2$ holonomy and see whether it
allows for a topological twist. We were also motivated by other
issues, such as applications to M-theory compactifications on
$G_2$-manifolds, and the possibility of improving our
understanding of the relation between supersymmetric gauge
theories in three and four dimensions.

In this paper, we study the construction of a topological string
theory on a seven dimensional manifold of $G_2$ holonomy. Our
approach is to define a topological twist of the $\sigma$ model on
$G_2$ manifolds. On such manifolds, the $(1,1)$ world-sheet
supersymmetry algebra gets extended to a non-linear algebra, which
has a $c={7 \over 10}$ minimal model subalgebra \cite{sv}. We use
this fact to define the topological twist of the $\sigma$ model.
This is a particular realization of a more generic result: On an
orientable $d$ dimensional manifold which has holonomy group $H$
which is a subgroup of $SO(d)$, the coset CFT $SO(d)_1 /H_1$ with
its chiral algebra appears as a building block of the
corresponding sigma model, at least at large volume. It is natural
to conjecture that this building block persists at finite volume
(i.e. to all orders in $\alpha'$). It therefore gives rise to
extra structure in the world sheet theory which corresponds to
geometrical constructions in the target space. For example, for
Calabi-Yau three folds, this extra structure is given by the
$U(1)$ R-symmetry current, which can be used to Hodge decompose
forms of total degree $p+q$ into $(p,q)$ forms. The exterior
derivative has a corresponding decomposition as $d=
\partial + \bar{\partial}$, and physical states in
the world sheet theory correspond to suitable Dolbeault cohomology
groups $H_{\bar{\partial}}^*(X,V)$. A $G_2$ manifold has an
analogous refinement of the de Rham cohomology \cite{dolbeault}.
Differential forms can be decomposed into irreducible
representations of $G_2$. The exterior derivative can be written
as the sum of two nilpotent operators $d= \check{d} + \hat{d}$,
where $\check{d}$ and $\hat{d}$ are obtained from $d$ by
restricting its action on differential forms to two disjoint
subsets of $G_2$ representations.
This leads to a natural question:
is there a topologically twisted theory such that the BRST
operator in the left (or right) sector maps to $\check{d}$. We
will see that the answer to this question is yes, and in this
paper, we give the explicit construction of such a theory.

The outline of the paper is as follows. In section 2, we start by
reviewing $\sigma$ models on  target spaces of $G_2$ holonomy. We
discuss the relation between covariantly constant p-forms on
target spaces and holomorphic currents in the world sheet theory:
every covariantly constant p-form leads to the existence of a
chiral current supermultiplet \cite{Howe} (at least classically).
A $G_2$ manifold has a covariantly constant 3 and 4 form leading
to extra currents in the chiral algebra extending it from a
$(1,1)$ super-conformal algebra to a non-linear algebra generated
by 6 currents. As expected, this algebra contains the chiral
algebra of the coset $SO(7)_1/(G_2)_1$, which by itself is another
${\cal N}=1$ superconformal algebra with central charge $c={7
\over 10}$. This is a minimal model, called the tri-critical Ising
model, which plays a crucial role in defining the twisted theory.
In fact, the tri-critical Ising model is what replaces the $U(1)$
R-symmetry of the ${\cal N}=2$ superconformal algebra. The full
$c={21\over 2}$ Virasoro algebra with generators $L_n$ splits into
two commuting Virasoro algebras, $L_n=L^I_n+L^r_n$, with $L^I_n$
the generators of the $c={7\over 10}$ tri-critical Ising model.
This means that we can label highest weight states by their
$L_0^{I}$ and $L_0^r=L_0-L_0^{I}$ eigenvalues. We also review some
facts about the tri-critical Ising model. In the NS sector, there
are primary fields of weights $0,{1 \over 10}$, ${6 \over 10}$ and
${3 \over 2}$ and in the Ramond sector, there are two primary
fields of weights ${7 \over 16}$ and ${3 \over 80}$. We discuss
the fusion rules in this model, which helps us identify the
conformal block structure of various fields. This structure plays
an important role in definition of the twisted theory.


In section 3, we derive a unitarity bound for the algebra which
provides a non-linear inequality (a BPS bound) between the total
weight of the state and its tri-critical Ising model weight. We
define a notion of chiral primary states for $G_2$ sigma model by
requiring that they saturate this bound. We also discuss the
special chiral primary states in the CFT which correspond to the
metric moduli that preserve the $G_2$ holonomy.

In section 4, we define the topological twisting of the $G_2$
$\sigma$-model. We define correlation functions in the twisted
theory by relating them to certain correlation functions in the
untwisted theory with extra insertion of a certain Ramond sector
spin field. The twisting acts on different conformal blocks of the
same local operators in a different way.  We also define the BRST
operator $Q$ as a particular conformal block of the original
${\cal N}=1$ supercharge.
The BRST cohomology consists precisely of the chiral primary
states. We discuss the chiral ring, descent relations and a
suggestive localization argument which shows that the path
integral localizes on constant maps. Finally, we analyze some of
the putative properties of the twisted stress tensor of the
theory.

In section 5, we go on to discuss the geometric interpretation of
the BRST cohomology. To make this connection, we use the fact that
p-forms on the $G_2$ manifold transforming in different $G_2$
representations correspond to operators in the CFT which carry
different tri-critical Ising model weight ($L_0^I$ eigenvalue).
Using this we can identify how the BRST operator acts on p-forms.
We find that the BRST cohomology in the left or the right moving
sector is a Dolbeault type cohomology of the differential complex
$
 0 \rightarrow \Lambda^0_1 {\rightarrow} \Lambda^1_7
{\rightarrow} \Lambda^2_7 {\rightarrow}\Lambda^3_1 {\rightarrow}0
$ where the differential operator is the usual exterior derivative
composed with various projection operators to particular
representations of $G_2$ as indicated by the subscript. When we
combine the left and the right movers, the BRST cohomology is just
as a vector space equal to the total de Rham cohomology $H^*(M)$.
The BRST cohomology includes the metric moduli that preserve the
$G_2$ holonomy. These are in one-to-one correspondence with
elements of $H^3(M)$. We also compute three point functions at
genus 0 and show that these can be written as appropriate triple
derivatives of a suitable generalization of Hitchin's functional.
To show this, we develop an analogue of special geometry for $G_2$
manifolds by defining coordinates on the moduli space of $G_2$
metrics as periods of the $G_2$ invariant three form and the dual
four form. As in the case of Calabi-Yau manifolds, the dual
periods are derivatives of a certain pre-potential, which is
proportional to the Hitchin's functional. We also argue that the
partition function should be viewed as a wave function in a
quantum mechanics corresponding to the phase space $H^2\oplus H^3
\oplus H^4 \oplus H^5 $, where the symplectic form is given by
integrating the wedge product of two forms over the seven
manifold. We also consider the special case of the $G_2$ manifold
being a product of Calabi-Yau and a circle and show that the
twisted $G_2$ theory is an interesting and non-trivial combination
of the A and the B model.

There is extensive literature about string theory and M-theory
compactified on $G_2$ manifolds. The first detailed study of the
world-sheet formulation of strings on $G_2$ manifolds appeared in
\cite{sv}. The world-sheet chiral algebra was studied in some
detail in \cite{sv,blumenhagen,figueroa,noyvert}. For more about
type II strings on $G_2$ manifolds and their mirror symmetry, see
e.g.
\cite{9604133,9707186,0108091,0110302,0111012,0111048,Aganagic,roiban,0204213,0301164,gaberdiel}.
A review of M-theory on $G_2$ manifolds with many references can
be found in \cite{0409191}.


\section{${\mathbf{G_2}}$ sigma models}

A supersymmetric $\sigma$ model on a generic Riemannian manifold
has $(1,1)$ world sheet supersymmetry. However, existence of
covariantly constant p-forms implies the existence of an extended
symmetry algebra \cite{Howe}. This symmetry algebra is a priori
only present in the classical theory. Upon quantization, it could
either be lost or it could be preserved up to quantum
modifications. However, since the extended symmetry is typically
crucial for many properties of the theory such as spacetime
supersymmetry, it is natural to postulate the extended symmetry
survives quantization. To determine the quantum version of the
algebra, one can for example study the most general quantum
algebra with the right set of generators. For the generators
expected in the $G_2$ case this was done in \cite{blumenhagen}
(though not with this motivation). It turns out that there is a
two-parameter family of algebras with the right generators. By
requiring the right value of the total central charge, and by
requiring that it contains the tri-critical Ising model (which is
crucial for space-time supersymmetry), both parameters are fixed
uniquely leading to what we call the $G_2$ algebra.

Alternatively, one could have started with the special case of
$\mathbb R^7$ as a model of a $G_2$ manifold in the infinite
volume limit. This is simply a theory of free fermions and bosons,
and one can easily find a quantum algebra with the right number of
generators using the explicit form of the covariantly closed three
and four form for $G_2$ manifolds written in terms of a local
orthonormal frame. From this large volume point of view it is
natural to expect the coset $SO(7)_1/(G_2)_1$ to appear, since
$SO(7)_1$ is just a theory of free fermions and bosons. In any
case, this leads to the same result for the $G_2$ algebra as the
approach described in the previous paragraph. In the remainder of
this section we will briefly describe the large volume approach.

\subsection{Covariantly Constant p-forms and Extended Chiral Algebras}

We start from a sigma model with $(1,1)$ supersymmetry, writing its action in superspace:
\be
S= \int d^2z~ d^2 \theta ~(G_{\mu \nu} +B_{\mu \nu}) D_\theta \mathbf{X}^\mu D_{\bar{\theta}} \mathbf{X}^\nu
\ee
where
\[ D_\theta = {\partial \over \partial \theta} + {\theta} {\partial \over \partial z} ~~~~~~,~~~~~~
 D_{\bar{\theta}} = {\partial \over \partial \bar{\theta}} + {\bar{\theta}} {\partial \over \partial \bar{z}}
\]
and $\mathbf{X}$ is a superfield, which, on shell can be taken to be chiral:
\[
\mathbf{X}^\mu=\phi^\mu(z) + \theta \psi^\mu(z)
\]
 For now, we  set $B_{\mu \nu}=0$. This model generically has $(1,1)$ superconformal symmetry classically. The super stress-energy tensor is given by
\[
\mathbf{T}(z,\theta)=G(z)+ \theta T(z) = -{1\over 2}  G_{\mu \nu} D_\theta\mathbf{X}^\mu \partial_z \mathbf{X}^\nu
\]
This $\mathcal{N}=(1,1)$ sigma model can be formulated on an
arbitrary target space. However, generically the target space theory
will not be supersymmetric. For the target space theory to be
supersymmetric the target space manifold must be of special
holonomy. This ensures that covariantly constant spinors, used to
construct supercharges, can be defined. The existence of covariantly
constant spinors on the manifold also implies the existence of
covariantly constant \emph{p-forms} given by \be\label{peeform}
\phi_{(p)}=\epsilon^T \Gamma_{i_1 \dots i_p} \epsilon ~dx^{i_1}
\wedge \cdots \wedge dx^{i_p}. \ee This expression may be
identically zero. The details of the holonomy group of the target
space manifold dictate which p-forms are actually present.

The existence of such covariantly constant p-forms on the target
space manifold implies the existence of extra elements in the
chiral algebra \cite{Howe}. For example, given a covariantly
constant $p$ form, $\phi_{(p)}=\phi_{i_1 \cdots  i_p} dx^{i_p}
\wedge \cdots \wedge dx^{i_p} $ satisfying $\nabla \phi_{i_1
\cdots  i_p}=0$, we can construct a holomorphic superfield current
given by
\[
\mathbf{J}_{(p)}(z,\theta)=\phi_{i_1 \cdots  i_p}  D_\theta \mathbf{X}^{i_1} \cdots D_{\theta}\mathbf{X}^{i_p}
\]
which satisfies $D_{\bar{\theta}} \mathbf{J}_{(p)}=0$ on shell. In
components, this implies the existence of a dimension ${p \over
2}$ and a dimension ${p+1 \over 2}$ current. For example, on a
K\"ahler manifold, the existence of a covariantly constant
K\"ahler two form $\omega= g_{i\bar{j}}(d \phi^i \wedge
d\phi^{\bar{j}}-d\phi^{\bar{j}} \wedge d\phi^i) $ implies the
existence of a dimension 1 current $J=g_{i\bar{j}} \psi^i
\psi^{\bar{j}}$ and a dimension ${3 \over 2}$ current
$G'(z)=g_{i\bar{j}}(\psi^i \partial_z
\phi^{\bar{j}}-\psi^{\bar{j}}
\partial_z \phi^i)$, which add to the $(1,1)$ superconformal
currents $G(z)$ and $T(z)$ to give a $(2,2)$ superconformal
algebra. In fact, there is a non-linear extension of the $(2,2)$
algebra even in the case of Calabi-Yau by including generators
corresponding to the (anti)holomorphic three-form. This algebra
was studied in \cite{Odake}.

\subsection{Extended algebra for ${\mathbf{G_2}}$ sigma models}

A generic seven dimensional Riemannian manifold has $SO(7)$
holonomy. A $G_2 $ manifold has holonomy which sits in a $G_2$
subgroup of $SO(7)$. Under this embedding, the eight dimensional
spinor representation $\mathbf{8}$ of $SO(7)$ decomposes into a
$\mathbf{7}$ and a singlet of $G_2$:
\[\mathbf{ 8} \rightarrow \mathbf{7} \oplus \mathbf{1} \]
The singlet corresponds to a covariantly constant spinor $\epsilon$ on the manifold satisfying
\[ \nabla  \epsilon =0. \]

For ${G_2}$ manifolds (\ref{peeform}) is non-zero only when $p=0,
~3,~ 4$ and $7$ since an anti-symmetrized product of $p$
fundamentals ($\mathbf{7}$) of $SO(7)$ has a $G_2$ singlet for
these $p$. The zero and the seven forms just correspond to
constant functions and the volume form. In addition to these,
there is a covariantly constant 3-form $\phi^{(3)}=
\phi_{ijk}^{(3)} dx^i \wedge dx^j \wedge dx^k $ and its Hodge dual
4-form, $\phi^{(4)}=*\phi_{(3)}= \phi^{(4)}_{ijkl} dx^i \wedge
dx^j \wedge dx^k \wedge dx^l$. By the above discussion, the 3-form
implies the existence of a superfield current
$\mathbf{J}_{(3)}(z,\theta)=\phi_{ijk}^{(3)} D_\theta \mathbf{X}^i
D_\theta \mathbf{X}^j D_\theta \mathbf{X}^k \equiv  \Phi +\theta
K$. Explicitly, $\Phi$ is a  dimension ${3 \over 2}$ current \be
\Phi=\phi_{ijk}^{(3)} \psi^i \psi^j \psi^k \ee and $K$ is its
dimension 2 superpartner \be K=\phi_{ijk}^{(3)} \psi^i \psi^j
\partial \phi^k. \ee Similarly, the 4-from implies the existence of
a dimension 2 current \be Y=\phi_{ijkl}^{(4)} \psi^i \psi^j \psi^k
\psi^l\ee
 and its dimension $ {5 \over 2}$ superpartner
\be N=\phi_{ijkl}^{(4)} \psi^i \psi^j \psi^k \partial \phi^l.\ee
However,  as it will become clear later, instead of $Y$ and $N$, it
is more useful to use the following basis of chiral currents \be
X=-Y-{1 \over 2} G_{ij} \psi^i \partial \psi^j \label{xdef} \ee and
its superpartner \be M=-N-{1 \over 2} G_{ij} \partial \phi^i
\partial \psi^j + {1 \over 2} G_{ij} \psi^i \partial^2 \phi^j. \ee
So in summary, the $G_2$ sigma model has a chiral algebra generated
by the following six currents
\begin{center}
\begin{tabular}{cccc}
$\quad ~~~~~~~~~~$  & $\quad ~~~~~~~~~~$  & $\quad ~~~~~~~~~~$ & $\quad ~~~~~~~~~~$ \\
$h=\frac{3}{2}$ &${G(z)}$  & $ {\Phi(z)}$ & \\
$\quad$ & $\quad$  & $\quad $ & $\quad $ \\
  $h=2$ & $ {T(z)}$ & $ {K(z)}$ & $ {X(z)}$ \\
  $\quad$ & $\quad$  & $\quad $ & $\quad $ \\
 $h=\frac{5}{2}$ & $\quad$ & $\quad$ & $ {M(z)}$ \\
 $\quad$ & $\quad$  & $\quad $ & $\quad $ \\
\end{tabular}
\end{center}
These six generators form a closed algebra which appears
explicitly e.g. in \cite{sv,figueroa} (see also \cite{noyvert}).
We have reproduced the algebra in appendix \ref{algebra}. As
explained in the beginning of section~2, the existence of this
algebra can be taken as the definition of string theory on $G_2$
manifolds.

\subsection{The Tri-critical Ising Model}\label{tim}

An important fact, which will be crucial in almost all the remaining
analysis, is that the generators $\Phi$ and $X$ form a closed
sub-algebra:
\begin{eqnarray*}\label{timope}
\Phi(z) \Phi(0)&=&-{7 \over z^3 }+ {6 \over z}X(0)\\
\Phi(z) X(0)&=&-{15 \over 2z^2}\Phi(0) -{5 \over 2z} \partial \Phi(0)\\
X(z) X(0)&=&{35 \over 4z^4} -{10 \over z^2}X(0)-{5 \over
z}\partial X(0).
\end{eqnarray*} Defining the supercurrent $G_{I}={i \over \sqrt{15}} \Phi$ and
stress-energy tensor $T_I=-{1 \over 5}X$ this is recognized to be
the unique ${\cal N}=1$ super-conformal algebra of the minimal
model with central charge $c={7 \over 10}$ known as the
tri-critical Ising Model. This sub-algebra plays a similar role to
the one played by the $U(1)$ R-symmetry in the case of Calabi-Yau
target spaces. The extended chiral algebra contains two ${\cal
N}=1$ superconformal sub-algebras: the original one generated by
$(G,T)$ and the ${\cal N}=1$ superconformal sub-algebra generated
by $(\Phi,X)$.

In fact, with respect to the conformal symmetry, the full Virasoro
algebra decomposes in two commuting Virasoro algebras: $T=T_I+T_r$
with \be\label{teerest} T_I(z) T_r(w)= {\mathrm {regular}}. \ee
This means we can classify conformal primaries by two quantum
numbers, namely its tri-critical Ising model highest weight and
its highest weight with respect to $T_r$: $|{\rm primary} \rangle
= | h_I,h_r \rangle $. The Virasoro modules decompose accordingly
as
\begin{equation}\label{decomp}
{\cal M}_{c=\frac{21}{2}}={\cal M}^{ I}_{c=\frac{7}{10}} \otimes {\cal M}^{rest}_{c=\frac{98}{10}}.
\end{equation}
Notice that this decomposition is with respect to the Virasoro
algebras and not with respect to the $\mathcal{N}=1$ structures,
which in fact do not commute. For e.g., the superpartner of $\Phi$
with respect to the full ${\cal N}=1$ algebra is $K$ whereas its
superpartner with respect to the ${\cal N}=1$ of the tri-critical
Ising model is $X$.


\subsection{Tri-critical Ising and Unitary Minimal Models}

We now review a few facts about the tri-critical Ising that we
will use later in the paper.

Unitary minimal models are labelled by a positive integer
$p=2,3,\dots$ and occur only on the ``discrete series" at central
charges $c=1-\frac{6}{p(p+1)}$. The tri-critical Ising model is
the second member ($p=4$) which has central charge
$c=\frac{7}{10}$. In fact, it is also a minimal model for the
$\mathcal{N}=1$ superconformal algebra.

The conformal primaries of unitary minimal models are labelled by
two integers $1\leq n'\leq p$ and $1\leq n<p$. Primaries with
label $(n',n)$ and $(p+1-n',p-n)$ are identical and should be
identified with each other. Therefore, there are in total
$p(p-1)/2$ primaries in the theory. The weights of the primaries
are conveniently arranged into a Kac table. The conformal weight
of the primary $\Phi_{n'n}$ is
$h_{n'n}=\frac{[pn'-(p+1)n]^2-1}{4p(p+1)}.$ In the tri-critical
Ising model $(p=4)$ there are 6 primaries of weights
$0,\frac{1}{10},\frac{6}{10},\frac{3}{2},\frac{7}{16},\frac{3}{80}$.
Below we write the Kac table for the tri-critical Ising model.
Beside the Identity operator $(h=0)$ and the ${\cal N}=1$
supercurrent $(h=\frac{3}{2})$ the NS sector (first and third
columns) contains a primary of weight $h=\frac{1}{10}$ and its
${\cal N}=1$ superpartner $(h=\frac{6}{10})$. The primaries of
weight $\frac{7}{16},\frac{3}{80}$ are in the Ramond sector
(middle column).

\begin{table}
\begin{center}
 \begin{tabular}{|c|c|c|c|}
   \hline
  &&& \\
 $n' \setminus n $ & $\quad 1\quad$ & $\quad 2\quad$ & $\quad 3\quad$ \\
  &&&\\
   \hline
 &&&\\
   1 & $\mathbf{0}$ & $\mathbf{\dfrac{7}{16}}$\  & $\frac{3}{2}$\\
 &&&\\
 \hline
 &&&\\
   2 & $\mathbf{\dfrac{1}{10}\ }$ & $ \mathbf{\dfrac{3}{80}\ }$ &
 $\frac{6}{10}\ $ \\  &&&\\ \hline &&&\\
   3 & $ \mathbf{\dfrac{6}{10}\ }$ & $\frac{3}{80}\ $ &
 $\frac{1}{10}\ $ \\  &&&\\\hline &&&\\
   4 & $ \mathbf{\dfrac{3}{2}\ }$ & $\frac{7}{16}\ $ & 0 \\  &&&\\
 \hline
   \end{tabular}
\end{center}
\caption{Kac table for the tri-critical Ising model}
\end{table}

The Hilbert space of the theory decomposes in a similar way,
${\cal H}=\oplus_{n,n'} {\cal H}_{n',n}\times {\cal
\tilde{H}}_{n'n}$. A central theme in this work is that since the
primaries $\Phi_{n'n}$ form a closed algebra under the OPE they
can be decomposed into conformal blocks which connect two Hilbert
spaces. Conformal blocks are denoted by $\Phi_{n',n,m'm}^{l',l}$
which describes the restriction of $\Phi_{n',n}$ to a map that
only acts from $\mathcal{H}_{m',m}$ to $\mathcal{H}_{l',l}$. More
details can be found in \cite{felder}.

An illustrative example, which will prove crucial in what follows,
is the conformal block structure of the primary $\Phi_{2,1}$ of weight
$1/10$. General arguments show that the fusion rule of this field
with any other primary $\Phi_{n'n}$
is $\phi_{(2,1)}\quad\times
\quad\phi_{(n',n)}=\phi_{(n'-1,n)}\quad +\quad\phi_{(n'+1,n)}.$
The only non-vanishing conformal blocks in the decomposition of
$\Phi_{2,1}$ are those that connect a primary with the primary
right above it and the primary right below in the Kac table,
namely,
$\phi_{2,1,n',n}^{n'-1,n}$ and $\phi_{2,1,n',n}^{n'+1,n}$. This
can be summarized formally by defining the following
decomposition\footnote{Perhaps the notation with $\downarrow$ and $\uparrow$ is a bit misleading. By $\Phi_{2,1}^\downarrow$, we mean that conformal block of $\Phi_{2,1}$ which maps
\be
{\cal H}_{0} \stackrel{\Phi_{2,1}^\downarrow}{\rightarrow} {\cal H}_{1\over 10} \stackrel{\Phi_{2,1}^\downarrow}{\rightarrow} {\cal H}_{6 \over 10} \stackrel{\Phi_{2,1}^\downarrow}{\rightarrow} {\cal H}_{3 \over 2}
\ee
This is going down only in the first column of the Kac table, but is actually going up in the third column.}
\be\label{updown} \Phi_{2,1}=\Phi_{2,1}^{\downarrow} \oplus
\Phi_{2,1}^{\uparrow}. \ee Similarly, the fusion rule of the
Ramond field $\Phi_{1,2}$ with any primary is
$\phi_{(1,2)}\quad\times
\quad\phi_{(n',n)}=\phi_{(n',n-1)}\quad+\quad\phi_{(n',n+1)}$
showing that it is composed of two blocks, which we denote as
follows \be \label{confblockspin} \Phi_{1,2}=\Phi_{1,2}^{-} \oplus
\Phi_{1,2}^{+} .\ee It is important here to specify on which half
of the Kac table we are acting. We take $\phi_{(n',n)}$ to be
either in the first column or in the top half of the second
column, i.e. in the boldface region of table~1. With this
restriction we denote by $\Phi_{1,2}^{-}$ the conformal block that
takes us to the left in the Kac table and $\Phi_{1,2}^{+}$ the one
that takes us to the right. Conformal blocks transform under
conformal transformations exactly like the primary field they
reside in but are usually not single-valued functions of
$z(\bar{z})$. This splitting into conformal blocks plays a crucial
role in the twisting procedure. The $+$ and $-$ labels will be
clarified further when we consider the Ramond sector of the full
$G_2$ algebra in section \ref{GSO} where we see that these labels
correspond to Ramond sector ground states with different fermion
numbers.

\section{Chiral Primaries, Moduli and a Unitarity Bound}
Having discussed this $c={7\over 10}$ subalgebra we now turn to
the full $G_2$ chiral algebra. We first identify a set of special
states which will turn out to saturate a unitarity bound for the
full $G_2$ algebra. We call these the chiral primary states. This
name seems appropriate since the representations built on chiral
primary states are ``short'' whereas the generic representation is
``long.'' The chiral primary states include the moduli of the
compactification, i.e. the metric and $B$-field moduli that
preserve the $G_2$ holonomy.

\subsection{Chiral Primary States}\label{ssam}

The chiral-algebra associated with manifolds of $G_2$
holonomy\footnote{We loosely refer to it as ``the $G_2$ algebra"
but it should not be confused with the Lie algebra of the group
$G_2$.} allows us to draw several conclusions about the possible
spectrum of such theories. It is useful to decompose the
generators of the chiral algebra in terms of primaries of the
tri-critical Ising model and primaries of the
remainder (\ref{decomp}). The commutation
relations of the $G_2$ algebra imply that  some of the generators
of the chiral algebra decompose as \cite{sv}:
$G(z)=\Phi_{2,1}\otimes\psi_{\frac{14}{10}}\quad,\quad
K(z)=\Phi_{3,1}\otimes\psi_{\frac{14}{10}}$ and
$M(z)=a\Phi_{2,1}\otimes\chi_{\frac{24}{10}}+b[X_{-1},\Phi_{2,1}]\otimes\psi_{\frac{14}{10}},$
with $\psi,\chi$ primaries of the indicated weights in the $T_r$
CFT and $a,b$ constants.

The Ramond sector ground states on a seven dimensional manifold
(so that the corresponding CFT has $c=21/2$) have weight ${7 \over
16}$. This implies that these states, which are labelled by two
quantum numbers (the weights under the tri-critical part and the
remaining CFT), are $|{7\over 16},0\rangle$ and $|{3\over
80},{2\over 5}\rangle$. The existence of the $|{7\over
16},0\rangle$ state living just inside the tri-critical Ising
model is crucial for defining the topological theory.  Coupling
left and right movers, the only possible RR ground states
compatible with the $G_2$ chiral algebra\footnote{Otherwise the
spectrum will contain a 1-form which will enhance the chiral
algebra \cite{sv}. Geometrically this is equivalent to demanding
that $b_1=0$.} are a single $|{7\over 16},0\rangle_L\otimes\
|{7\over 16},0\rangle_R$ ground state and a certain number of
states of the form $|{3\over 80},{2\over 5}\rangle_L\otimes\
|{3\over 80},{2\over 5}\rangle_R$. For a further discussion of the
RR ground states see also section~\ref{GSO} and appendix~C.

By studying
operator product expansions of the RR ground states using the fusion rules
\begin{eqnarray*}
{7 \over 16} \times {7 \over 16} &=& ~0 ~+~ {3 \over 2} \\
{7 \over 16} \times {3 \over 80} & = & {1 \over 10} + {6 \over 10}
\end{eqnarray*}
we get the
following ``special" NSNS states
\be
\label{special}
|0,0\rangle_L\otimes\
|0,0\rangle_R,\ |{1\over 10},{2\over 5}\rangle_L\otimes\ |{1\over
10},{2\over 5}\rangle_R,\ |{6\over 10},{2\over 5}\rangle_L\otimes\
|{6\over 10},{2\over 5}\rangle_R ~{\rm  and}  ~|{3\over
2},0\rangle_L\otimes\ |{3\over 2},0\rangle_R\ee
 corresponding
 to the 4 NS primaries $\Phi_{n',1}$ with $n'=1,2,3,4$ in the tri-critical Ising model.
Note that for these four states there is a linear relation between
the Kac label $n'$ of the tri-critical Ising model part and the
total conformal weight $h_{total}={n'-1\over 2}$. In fact, in section \ref{unitarity}, we show  that
 similar to the BPS bound in the $\mathcal{N}=2$
case, primaries of the $G_2$ chiral algebra satisfy a (non-linear)
bound of the form
\be
h_I + h_r \geq
\frac{1+\sqrt{1+80 h_I}}{8} .
 \ee
which is precisely saturated for the four NS states listed above.
We  will therefore refer to those states as ``chiral primary"
states. Just like in the case of Calabi-Yau, the ${7\over 16}$
field maps Ramond ground states to NS chiral primaries and is thus
an analogue of the ``spectral flow" operators in Calabi-Yau.

\subsection{Moduli}

It was  shown in \cite{sv}\ that the  upper components
\[
\tilde{G}_{-{1\over 2}}|{1\over 10},{2\over 5}\rangle_L\otimes\ G_{-{1\over 2}}|{1\over 10},{2\over 5}\rangle_R
\]
correspond to exactly marginal deformations of the CFT preserving the $G_2$ chiral algebra
\be\label{marg}
\{ G_{-{1\over 2}},\mathcal{O}_{{1\over 10},{2\over 5}} \}=\mathcal{O}_{0,1}.
\ee
and as such, correspond to the moduli of the $G_2$ compactification. As we will see in more detail later, there are $b_2+b_3$ such moduli.

Geometrically, the metric moduli are deformations of the metric
($\delta g_{ij}$) that preserve Ricci flatness (these deformations
also preserve the $G_2$ structure). Such deformations satisfy the
Lichnerowicz equation:
\be \Delta_L \delta g_{ij}\equiv -\nabla^2
\delta g_{ij}+2R_{mijn}\delta g^{mn} +2 R^k_{(i} \delta g_{j)k}=0.
\label{modulieq} \ee
 That there are $b_3$ solutions to this
equation (up to diffeomorphisms) can be seen by relating
(\ref{modulieq}) to an equation for a three-form $\omega$ which is
constructed out of $\delta g$ via $\delta g_{ij}$: $
\omega_{ijk}=\phi_{l[ij} \delta g^l_{k]}. $ Indeed, it can be
shown \cite{joycebook} that for every solution of (\ref{modulieq})
modulo diffeomorphisms there is a corresponding harmonic
three-form: \be \Delta_L \delta g\ =0 \leftrightarrow \Delta
\omega =0. \ee

A natural question is if $\Delta_L$ can be written as the square
of some first order operator. Such a construction exists if the
manifold supports a covariantly constant spinor $\epsilon_0$. We
can construct a spinor valued one-form out of $\delta g_{ij}$ as
$\delta g_{ij}(\Gamma^i \epsilon_0) dx^j$. This is a section of
S(M) $\otimes T^*M$ where $S(M)$ is the spin bundle. There is a
natural $D \hspace{-0.21cm}/$ operator acting on this vector
bundle. It can be shown that $D \hspace{-0.21cm}/^\dagger
D\hspace{-0.21cm}/ = \Delta_L$, which then reduces
(\ref{modulieq}) to \be D \hspace{-0.21cm}/ \Bigl( \delta g_{ij}
\Gamma^i \epsilon_0 dx^j \Bigr) =0 \ee which was shown to imply
\be \nabla_{i} \delta g_{jk}~ \phi^{ij}{}_l =0 \label{mod} \ee in
\cite{math0311253}. This first order condition for the metric
moduli will be beautifully reproduced from our analysis later of
the BRST cohomology of our topologically twisted sigma model.

There is another quick way to see how the condition of being
chiral primary implies the first order condition (\ref{mod}). This
is done using the zero mode of the generator $K(z)$ of the $G_2$
algebra. In the next section we will find that $K_0=0$ for chiral
primaries using some explicit calculations. One can also show this
more generally, since the $K_0$ eigenvalue of highest weight
states  of the $G_2$ algebra can be determined in terms of their
$L_0$ and $X_0$ eigenvalues by using the fact that the null ideal
in (\ref{nullideal}) has to vanish when acting on such states (see
appendix \ref{algebra}). Again this leads to the conclusion that
$K_0=0$ for chiral primaries. Now in the large volume limit the
operator ${\cal O}_{{1\over 10},{2 \over 5}L} \times {\cal O}_{{1
\over 10},{2 \over 5}R} $, correspond to the operator $\delta
g_{ij} \psi^i_L \psi^j_R $. \footnote{The tri-critical Ising model
weight  of this operator can be computed to be ${1 \over 10}$  by
taking the OPE of it with $X$ and then extracting the second order
pole.} The $K_0$ eigenvalue is then easily extracted from the
double pole in the OPE \be K(z) {\cal O}_{{1 \over 10},{2 \over
5}L} (0) \sim \cdots + {{\nabla_{i} \delta g_{jk}~ \phi^{ij}{}_l
}{\cal O}_{{1 \over 10},{2 \over 5}L} (0) \over z^2} + \cdots. \ee
We see that $K_0=0$ implies precisely the first order condition
(\ref{mod}) which is a nice consistency check of the framework.

\subsection{A Unitarity Bound}
\label{unitarity}

The $G_2$ algebra has highest weight representations, made from a
highest weight vector that is annihilated by all positive modes of
all the generators. First, notice that when acting on highest
weight states, the generators $L_0,X_0$ and $K_0$
commute\footnote{ The only subtlety is the $[X_0,K_0]$ commutator.
It does not vanish in general, but it does vanish when acting on
highest weight states.} so a highest weight state can be labelled
by the three eigenvalues $l_0,x_0,k_0$ \footnote{As we mentioned
in the previous subsection, $k_0$ is determined in terms of $l_0$
and $x_0$ by requiring the vanishing of the null ideal
(\ref{nullideal}) when acting on these states. We ignore this in
this subsection, though it does alter the analysis.}. In addition,
$l_0\geq 0$, $x_0\leq 0$, and $k_0$ is purely imaginary. The first
two conditions follow from unitarity (recall that $-5X$ is the
stress tensor of the tri-critical Ising model), the last condition
follows from the hermiticity conditions on $K_0$:
$K_m^{\dagger}=-K_{-m}$.


Next, we want to derive some bounds on $l_0,x_0,k_0$ that come
from unitarity. In particular, we consider the three states $\{
G_{-1/2} |l_0,x_0,k_0\rangle$, $\Phi_{-1/2} |l_0,x_0,k_0\rangle$,
$M_{-1/2} |l_0,x_0,k_0\rangle \}$ and we consider the matrix
${\cal M}$ of inner products of these states with their hermitian
conjugates\footnote{This analysis assumes that $x_0$ is strictly
negative otherwise $\Phi_{-\frac{1}{2}}|l_0,0,k_0\rangle$
vanishes. For $x_0$ we remove this state and consider the matrix
of inner products of the remaining two states, which leads to
exactly the same conclusion.}. This matrix can be worked out using
the commutation relations and we find
\begin{equation} \label{j1}
{\cal M} = \left( \begin{array}{ccc} 2l_0 & k_0 & l_0+2x_0 \\
-k_0 & -6 x_0 & -5k_0/2 \\ 2x_0+l_0 & 5k_0/2 & l_0/2 + 4x_0-8 x_0
l_0 \end{array} \right)
\end{equation}
This matrix is indeed hermitian, and unitarity implies that the
eigenvalues of this matrix should be nonnegative. In particular,
the determinant should be nonnegative
\be
 \label{j2} \det {\cal M}
= (8 l_0 - 6 x_0 - 8 l_0 x_0)k_0^2 + 24 x_0^2 (4 l_0^2 - l_0 +
x_0).
\ee
The piece between parentheses before $k_0^2$ is always
positive, and $k_0^2$ is always negative. Therefore we should in
particular require that (for $x_0\neq 0$)
\be \label{j3}
4l_0^2 -
l_0 + x_0 \geq 0
\ee
which implies
\be \label{j4}
 l_0 \geq \frac{1
+ \sqrt{1-16 x_0}}{8}.
 \ee
Changing basis to eigenvalues of $T_r, T_I$ (see \ref{teerest}) the bound
(\ref{j4}) becomes
\be \label{bound}
h_I + h_r \geq
\frac{1+\sqrt{1+80 h_I}}{8} .
 \ee
This bound will turn out to play an important role. When the bound
is saturated, we will call the corresponding state ``chiral
primary'' in analogy to states saturating the BPS bound in ${\cal
N}=2$. Since in the NS sector of the tri-critical Ising model,
$h_I=0,\frac{1}{10},\frac{6}{10},\frac{3}{2}$ chiral states have
total $h_I+h_r$ scaling dimension $0,\frac{1}{2},1,\frac{3}{2}$
which exactly match the special NSNS states \ref{special}. We will
see that just like for ${\cal N}=2$ theories it is exactly those
chiral states that survive the topological twist. Indeed, in the
Coulomb gas approach they became weight zero after the twist. It
is interesting to see that the definition of chiral primaries
involves a nonlinear identity. This reflects the fact that the
$G_2$ chiral algebra is non-linear. Since $\det {\cal M}=0$ for
chiral primaries, a suitable linear combination of the three
states used in building $\det {\cal M}$ vanishes. In other words,
chiral primaries are annihilated by a combination of fermionic
generators and the representations built from chiral primaries
will be smaller than the general representation, as expected for
BPS states.

When the bound (\ref{bound}) is saturated, $\det {\cal M}$ can
only be nonnegative as long as $k_0=0$. Thus, chiral primaries
necessarily have $k_0=0$, and we will mostly suppress the quantum
number $k_0$ in the remainder.

\section{Topological Twist}
\label{topological}

To construct a topologically twisted CFT, we usually proceed in two steps. First
we define a new stress-energy tensor, which changes the quantum numbers of the fields
and operators of the theory under Lorentz transformations. Secondly, we identify a
nilpotent scalar operator, usually constructed out of the supersymmetry generators of the
original theory, which we declare to be the BRST operator. Often this BRST operator
can be obtained in the usual way by gauge fixing a suitable symmetry. If the new
stress tensor is exact with respect to the BRST operator, observables (which are elements
of the BRST cohomology) are metric independent and the theory is called topological.
In particular, the twisted stress tensor should have a vanishing central charge.

\subsection{Review of twisting the Calabi-Yau ${\mathbf  \sigma}$-model}
\label{cytwist}

In practice \cite{kodira,antoniadis}, for the ${\cal N}=2$ theories,
an n-point correlator on the sphere in the twisted theory can
conveniently be defined\footnote{Up to proper normalization.} as a
correlator in the \emph{untwisted} theory of the same n operators
plus two insertions of a spin-field, related to the space-time
supersymmetry charge, that serves to trivialize the spin bundle. For
a Calabi-Yau 3-fold target space there are two $SU(3)$ invariant
spin-fields which are the two spectral flow operators
$\mathcal{U}_{\pm{1\over 2}}$. This discrete choice in the left and
the right moving sectors is the choice between the $+(-)$ twists
\cite{Witten} which results in the difference between the
topological $A/B$ models.

The action for the $\sigma$-model on a Calabi-Yau is given by
\begin{equation}
S=\int d^2 z {1 \over 2}g_{i\bar{j}}\Bigl(\partial  x^i \bar{\partial}x^{\bar{i}} +\partial  x^{\bar{i}} \bar{\partial}x^{{i}}\Bigr) +
g_{i\bar{j}} \Bigl(i \psi^{\bar{j}}_{-} D \psi_{-}^i + i \psi_+^{\bar{j}}\bar{D}\psi_+^i \Bigr)
+R_{i\bar{j} k \bar{l}}\psi_+^i \psi_+^{\bar{j}}\psi_-^k\psi_-^{\bar{l}}
\end{equation}
Twisting this $\sigma$-model corresponds to adding a background
gauge field for the $U(1)$ which acts on the complex fermions.
Effectively, we change the covariant derivative from $D=\partial +
{\omega \over 2}$ to $D'=\partial+{\omega \over 2}+A$, where we
set the background value of $A={\omega \over 2}$. Similarly,
$\bar{D}$ changes to  $\bar{D}'=\bar{\partial}+{\bar{\omega} \over
2} \pm \bar{A}$, where the $+$ sign refers to the B twist and the
$-$ sign refers to the A twist. This has the effect of changing
the action in the following way:
\begin{equation}
\delta S = \int  g_{i\bar{j}}\psi_+^i
\psi_+^{\bar{j}}{\bar{\omega}\over 2} \pm g_{i\bar{j}}\psi_-^i
\psi_-^{\bar{j}}{\omega\over 2}
\end{equation}
Just considering the left moving sector, and bosonizing the
$\psi_+$'s by defining $g_{i\bar{j}}\psi_+^i \psi_+^{\bar{j}} =i
\sqrt{d} \partial \phi$, where $d$ is the complex dimension of the
Calabi-Yau, we find
\[
\delta S = \int g_{i\bar{j}} \psi_+^i \psi_+^{\bar{j}} {\omega \over
2} =-i{\sqrt{d} \over 2} \int \phi \partial \omega = +i {\sqrt{d}
\over 2} \int \phi R. \] On a genus $g$ Riemann surface, we can
choose $R$ such that it has $\delta$-function support at $2-2g$
points. So for example, on a sphere, we get
\[
e^{-\delta S} =  e^{i {\sqrt{d} \over 2} \phi(0)}e^{i {\sqrt{d}
\over 2} \phi(\infty)}\] which implies that correlation functions
in the twisted theory are related to those in the untwisted theory
by $2-2g$ insertions of the operator (also known as the spectral
flow operator) $e^{i {\sqrt{d} \over 2} \phi}$:
\[
\langle \cdots \rangle_{\rm twisted} = \langle e^{i{\sqrt{d} \over
2} \phi(\infty)} \cdots e^{i{\sqrt{d} \over 2} \phi(0)}
\rangle_{\rm untwisted} \] This effectively adds a background
charge for the field $\phi$ of magnitude $Q=\sqrt{d}$, changing
the central charge of the CFT
\[
c={3\over 2} \times 2d  \rightarrow 1-3Q^2 +3d -1 =0
\]
which is what we expect in a topological theory.

\subsection{The $\mathbf{G_2}$ Twist On The Sphere}\label{g2sphere}

We can apply a similar procedure to the $G_2$ $\sigma$-model. The
role of the operator $e^{i{\sqrt{d} \over 2}\phi}$ will be played
by the conformal block $\Phi_{1,2}^+$  of the primary with
conformal weight ${7\over 16}$ which creates the state $|{7 \over
16}, 0 \rangle $. Notice that this state sits entirely inside the
tri-critical Ising model. Indeed, also in the case of Calabi-Yau
manifolds, the spectral flow operator $e^{i{\sqrt{d} \over 2}
\phi}$, sits purely within the $U(1)={U(d) \over SU(d)}$ part. In
$G_2$ manifolds, the coset ${SO(7)_1 \over (G_2)_1}$ (with central
charge ${7 \over 10}$) plays the same role as the $U(1)$
subalgebra in $\mathcal{N}=2.$ We therefore suggest (refining a
similar suggestion of \cite{sv}) that correlation functions of the
twisted theory are defined in terms of the untwisted theory as

\be\label{twicorsphere}
\begin{split}
\langle &V_1(z_1)\dots V_n(z_n)\rangle_{\mathtt{twisted}}^{\mathtt{plane}}\equiv \\
\prod_{i=1}^{n}z_i^{({h_i-\tilde{h_i}})}
\langle &\Sigma(\infty)V_1(z_1)\dots V_n(z_n)\Sigma(0)\rangle_{\mathtt{untwisted}}^{\mathtt{plane}}
\end{split}
\ee where, $(h) \tilde{h}$ are the weights with respect to the
(un)twisted stress tensor respectively\footnote{The product
$\prod_{i=1}^{n}z_i^{({h_i-\tilde{h_i}})}$ comes about from the
mapping between the flat cylinder and the sphere. Note that this is
not the same as computing the expectation value of $V_1(z_1)\dots
V_n(z_n)$ in the Ramond ground state $\Sigma(0)|0\rangle$ because we
insert the same operator at $0,\infty$ and not an operator and its BPZ
conjugate.} and $\Sigma$ is the conformal block
\begin{equation}
\Sigma=\Phi_{1,2}^+
\end{equation}
defined in (\ref{confblockspin}).

In \cite{sv} further arguments were given, using the Coulomb gas
representation of the minimal model, that there exists a twisted
stress tensor with vanishing central charge. Those arguments, which
are briefly reviewed in appendix A, are problematic because the
Coulomb gas representation really adds additional degrees of freedom
to the minimal model. To properly restrict to the minimal model, one
needs to consider cohomologies of BRST operators defined by Felder
\cite{felder}. The proposed twisted stress tensor of \cite{sv} does
not commute with Felder's BRST operators and therefore it does not
define a bona fide operator in the minimal model. In addition, a
precise definition of a BRST operator for the topological theory was
lacking in \cite{sv}.

We will proceed differently. We formulate our discussion purely in
terms of the tri-critical Ising model itself without ever
referring to the Coulomb gas representation, except by way of
motivation and intuition. We will propose a BRST operator, study
its cohomology, and then use \ref{twicorsphere}\ to compute
correlation functions of BRST invariant observables. The
connection to target space geometry will be made. We will then
comment on the extension to higher genus and on the existence of a
topologically twisted $G_2$ string.

\subsection{The BRST operator}
The basic idea is that the topological theory for $G_2$ sigma
models should be formulated in terms of its
(non-local)\footnote{It should be stressed that this splitting
into conformal blocks is non-local in the sense that conformal
blocks may be multi-valued functions of z ($\bar{z}$). } conformal
blocks and not in terms of local operators. By using the split
(\ref{updown}) into conformal blocks, we can split any field whose
tri-critical Ising model part contains just the conformal family
$\Phi_{2,1}$ into its up and down parts. For example, the ${\cal
N}=1$ supercurrent $G(z)$ can be split as \be\label{gmsplit}
G(z)=G^{\downarrow}(z) + G^{\uparrow}(z). \ee We claim that
$G^{\downarrow}$ is  the BRST current and $G^{\uparrow}$ is a
candidate for the for the anti-ghost\footnote{Incidently, the
Coulomb gas representation indeed assigns the expected conformal
weights after the twist (see appendix A).}. The basic ${\cal N}=1$
relation \be G(z)G(0)=\left(
G^{\downarrow}(z)+G^{\uparrow}(z)\right)\left(
G^{\downarrow}(0)+G^{\uparrow}(0)\right)\sim
\dfrac{2c/3}{z^3}+\dfrac{2T(0)}{z} \ee proves the nilpotency of
this BRST current (and of the candidate anti-ghost) because the
RHS contains descendants of the identity operator only and has
trivial fusion rules with the primary fields of the tri-critical
Ising model and so $(G^{\downarrow})^2=(G^{\uparrow})^2=0$.



An algebraic formulation of the decomposition \ref{gmsplit}\
starts from defining projection operators. Any state in the theory
can be labelled by its eigenvalues under the two commuting
(\ref{teerest}) Virasoro modes of $T_I,T_r$ and perhaps some
additional quantum numbers needed to completely specify the state.
We denote by $P_{n'}\ $ the projection operator on the sub-space
of states whose tri-critical Ising model part lies within the
conformal family of one of the four NS primaries $\Phi_{n',1}$.
The image of $P_{n'}$ is $\mathcal{H}_{n',1}$  which we abbreviate
here to $\mathcal{H}_{n'}$. The  corresponding weights of the
primary fields in the tri-critical Ising model by $\Delta(n')$.
Thus, $\Delta(1)=0$, $\Delta(2)=\frac{1}{10}$,
$\Delta(3)=\frac{6}{10}$ and $\Delta(4)=\frac{3}{2}$. This is
summarized by the equation \be \label{j12}
\Delta(n')=\frac{(2n'-3)(n'-1)}{10}. \ee The 4 projectors add to
the identity \be \label{j11} P_1+P_2+P_3+P_4=1 \ee because this
exhaust the list of possible highest weights in the NS sector of
the tri-critical Ising model\footnote{For simplicity, we will set
$P_{n'}=0$ for $n'\leq 0$ and $n'\geq 5$, so that we can simply
write $\sum_{n'} P_{n'}=1$ instead of (\ref{j11}).}.

We can now define our candidate BRST operator in the NS sector more rigorously
\be\label{BRSTop}
Q = G_{-{1\over 2}}^{\downarrow}\equiv\sum_{n'} P_{n'+1} G_{-{1\over 2}} P_{n'}.
\ee
The nilpotency
 $Q^2=0$ is easily proved:
\be Q^2 = \sum_{n'} P_{n'+2} G_{-{1\over2}}^2 P_{n'} = \sum_{n'}
P_{n'+2} L_{-1} P_{n'}=0 \ee where we could replace the
intermediate $P_{n'+1}$ by the identity because of the property
\ref{gmsplit} and the last equality follows since $L_{-1}$ maps
each ${\cal H}_{n'}$ to itself.

\subsection{BRST Cohomology and Chiral Operators}

Having defined the BRST operator, we can now compute its
cohomology. We first derive the condition on the tri-critical
Ising model weight $h_I$ and its total weight for it to be
annihilated by $Q$. Then we go on to defining the operator
cohomology, which correspond to operators (or conformal blocks of
operators) ${\cal O}$ satisfying $\{Q,{\cal O}\} =0$. We mostly
work in the NS sector. Perhaps it is more appropriate to work in
the Ramond sector since the topological theory computations are
done in the Ramond sector of the untwisted theory (see also
section~\ref{rsec}). We assume here that a version of spectral
flow exists which will map the NS sector to the Ramond sector. We
discuss such a spectral flow in appendix \ref{spectral}.
\subsubsection{State Cohomology}

As a first step in the analysis of the BRST cohomology, we
consider the action of $Q$ on highest weight states
$|h_I,h_r\rangle=|\Delta(k),h_r\rangle$ of the full algebra.
Because $Q$ is a particular conformal block of the supercharge
$G_{-{1 \over 2}}$, to extract the action of $Q$ on a state, we
first act with $G_{-{1 \over 2}}$ on the state and then project on
to the term. As discussed previously, the $\mathcal{N}=1$
supercurrent $G$ can be decomposed as
$\Phi_{2,1}\otimes\psi_{14\over 10}$. The fusion rules of the
tri-critical Ising model then imply that
\begin{eqnarray}
G_{-1/2} | \Delta(k),h_r \rangle &= &c_1 |\Delta(k-1),h_r-\Delta(k-1)+\Delta(k)-\frac{1}{2}\rangle \nonumber\\
&&~~~~ + c_2 |\Delta(k+1),h_r-\Delta(k+1)+\Delta(k)-\frac{1}{2}\rangle \label{j21}
\end{eqnarray}
where the two states on the right are highest weight states of the
$L_m,X_m$ subalgebra (but not necessarily of the full $G_2$
algebra) and which are normalized to have unit norm. Then by
definition \be \label{j24} Q | \Delta(k),h_r \rangle = c_2
|\Delta(k+1),h_r-\Delta(k+1)+\Delta(k)-\frac{1}{2}\rangle . \ee
Using the $G_2$ algebra (appendix \ref{algebra}), we find that \be
\label{j22} \langle \Delta(k),h_r | G_{1/2} G_{-1/2} |
\Delta(k),h_r \rangle = 2 (\Delta(k) + h_r) =|c_1|^2 + |c_2|^2 .
\ee The first answer is obtained using
$\{G_{1/2},G_{-1/2}\}=2L_0$, the second follows from (\ref{j21}).
In a similar way we compute \bea \label{j23} \langle \Delta(k),h_r
| G_{1/2}X_0 G_{-1/2} | \Delta(k),h_r \rangle & =& 9\Delta(k)-h_r-
10\Delta(k) (\Delta(k)+h_r)\nonumber \\ &  = & -5 \Delta(k-1)
|c_1|^2 - 5 \Delta(k+1) |c_2|^2. \eea We can use (\ref{j22}) and
(\ref{j23}) to solve for $c_1$ and $c_2$ up to an irrelevant
phase. In particular, we find that the highest weight state is
annihilated by $Q$, which is equivalent to $c_2=0$, if \be
\label{j25} 9\Delta(k)-h_r- 10\Delta(k) (\Delta(k)+h_r) = -10
\Delta(k-1) (\Delta(k)+h_r). \ee We can rewrite this as \be
\label{j101} \Delta(k)+h_r = \frac{10 \Delta(k)}{10 \Delta(k) + 1
- 10\Delta(k-1)} = \frac{k-1}{2}= \frac{1 + \sqrt{1 +
80\Delta(k)}}{8} \ee where we used (\ref{j12}). This is precisely
the unitarity bound (\ref{bound}). Therefore, the only highest
weight states that are annihilated by $Q$ are the chiral primaries
that saturate the unitarity bound. It is gratifying to see a close
parallel with the other examples of topological strings in four
and six dimensions\footnote{ Strictly speaking the above
derivation is not quite correct for $k=1,4$, since $\Delta(0)$ and
$\Delta(5)$ do not exist. If they would appear, then the
corresponding representations would not be unitary, since they lie
outside the Kac table. This implies that the only representations
with either $k=0$ or $k=3$ that can appear in the theory
necessarily have $h_r=0$, and these are indeed annihilated by the
BRST operator.}. We have shown  so far that all states that are
primary under the $L_m,X_m$ subalgebra and are annihilated by
$G_{1/2}$ are annihilated by $Q$ if they saturate the unitarity
bound. These states, need not be primary with respect to the full
$G_2$ algebra. This is implied by the condition $|c_1|^2 \geq 0$
in (\ref{j22}) and (\ref{j23}).

Of course, to study the full BRST cohomology, much more work is
required, and in particular we would want to prove that BRST
closed descendants are always BRST exact. We don't have such a
proof, but some partial evidence is given in  section
\ref{sl2algebra}. In the RR sector it is much easier to analyze
the BRST cohomology and there one immediately sees that the
cohomology consists of just the RR ground states (see
section~\ref{rsec}).

The geometric meaning of the BRST cohomology will become clear in
the next section. In the remainder of this section, we collect
various other technical aspects of the twisted CFT. Readers more
interested in the more geometrical aspects can jump to section
\ref{geometry}.

\subsubsection{Operator Cohomology}

Let $\mathcal{O}_{n',h,\alpha}$ be the local operator
corresponding to the state
$|\Delta(n'),h,\alpha\rangle$.\footnote{Here $\alpha$ is a formal
label that might be needed to completely specify a state.}
Generically, $Q$ does not commute with the local operators
$\mathcal{O}_{\Delta(1),0},\ \mathcal{O}_{\Delta(2),{2\over 5}},\
\mathcal{O}_{\Delta(3),{2\over 5}}$ and
$\mathcal{O}_{\Delta(4),0}$ corresponding to the chiral states
$|0,0\rangle, |\frac{1}{10},\frac{2}{5}\rangle,
|\frac{6}{10},\frac{2}{5}\rangle, |\frac{3}{2},0\rangle$ (for
brevity we will denote those 4 local operators just by their
tri-critical Ising model Kac index $\mathcal{O}_i,\ i=1,2,3,4).\ $
This is because the topological $G_2$ CFT is formulated not in
terms of local operators of the untwisted theory but in terms of
non-local conformal blocks. It is straightforward to check that
the following blocks, \be\label{cpbloc}
\mathcal{A}_{n'}=\sum_{m}^{}P_{n'+m-1}\mathcal{O}_{n'}P_{m} \ee
which pick out the maximal ``down component" of the corresponding
local operator, do commute with $Q$ and are thus in its operator
cohomology. For example writing explicitly $Q=P_{4}G_{-{1\over
2}}P_{3}+P_{3}G_{-{1\over 2}}P_{2}+P_{2}G_{-{1\over 2}}P_{1}$ it
follows trivially from the definition of the projectors
$P_{I}P_{J}=P_{I}\delta_{I,J}$ that $Q$ commutes with
$\mathcal{A}_4=P_{4}\mathcal{O}_4\ P_{1}.$ To get some familiarity
with the notation we work out another example, \be\label{qandatwo}
\begin{split}
\{ Q,\mathcal{A}_2 \} &=\sum_{n'}P_{n'+1}\left( G_{-{1\over 2}}P_{n'}\mathcal{O}_2+ \mathcal{O}_2P_{n'}G_{-{1\over 2}}\right) P_{n'-1}\\
&=\sum_{n'}P_{n'+1}\left( \{ G_{-{1\over 2}},\mathcal{O}_2 \}\right) P_{n'-1}=\sum_{n'}P_{n'+1}\mathcal{O}_{\Delta(1),1}P_{n'-1}=0
\end{split}
\ee where we repeatedly use the property \ref{gmsplit}\ and the
existence of the marginal operators \ref{marg}. Note that we have
not shown that the blocks \ref{cpbloc} exhaust the $Q$ cohomology
but presumably this is indeed the case.

This algebraic characterization of the conformal blocks
corresponding to chiral primaries fits nicely with the Coulomb gas
approach where the tri-critical Ising model vertex operator (i.e.
block) of the chiral primaries was identified in \ref{neweight}\
to be exactly the unscreened vertex that created the maximal
``down" shift in the Kac table.

\subsection{The Chiral Ring}

In a close parallel to what happens in theories with ${\cal N}=2$
SUSY, the conformal blocks which commute with $Q$ form a ring
under the OPE. Due to the simplicity of the tri-critical Ising
model there are in fact just two non trivial checks which are
$\mathcal{A}_{2}(z)\mathcal{A}_{2}(0)$ and
$\mathcal{A}_{2}(z)\mathcal{A}_{3}(0)$. For example
\be\label{cec}\begin{split}
\mathcal{A}_{2}(z)\mathcal{A}_{3}(0)&=P_{4}\mathcal{O}_2(z)P_3\mathcal{O}_3(0)P_1=P_4\mathcal{O}_2(z)\mathcal{O}_3(0)P_1=\\
&=P_4\mathcal{O}_4(0)P_1=\mathcal{A}_4(0).
\end{split}
\ee The second equality follows because $P_1$ projects on the
identity and the third due to the unitarity bound \ref{bound}\
(which for chiral primaries is just the linear relation
\ref{j101}) implying that in the OPE of two chiral primaries there
can be no poles and the leading regular term is automatically also
a chiral primary.

\subsection{An $\mathbf{sl(2|1)}$ Subalgebra}
\label{sl2algebra} We can construct an interesting $sl(2|1)$
subalgebra of the full algebra, whose commutation relations are
identical to the lowest modes of the $N=2$ algebra. To construct
this subalgebra, we define \be G_r^{\uparrow} = \sum_k P_{k-1} G_r
P_k, \quad G_r^{\downarrow} = \sum_k P_{k+1} G_r P_k, \quad
J_0=L_0-\{G_{-1/2}^{\downarrow}, G_{1/2}^{\uparrow} \} . \ee Using
properties of the $G_2$ algebra, and Jacobi identities, we can
show that the algebra generated by $G^{\downarrow}_{\pm 1/2}$,
$G^{\uparrow}_{\pm 1/2}$, $L_0$, $L_{\pm 1}$ and $J_0$ closes and
forms the algebra $sl(2|1)$. Notice that $Q\equiv
G_{-1/2}^{\downarrow}$ is one of the generators of this algebra.
We know that $sl(2|1)$ has short and long representations, and any
state in the BRST cohomology must necessarily be a highest weight
state of a short representation. This shows that $sl(2|1)$
descendants are never part of the BRST cohomology. This is a hint
that the only elements of the BRST cohomology are the chiral
primaries, but to prove this we would need to extend the above
reasoning to include also elements which are descendants with
respect to the other generators of the $G_2$ algebra, or require
us to determine the precise form of the antighost and twisted
stress tensor.


\subsubsection*{Position Independence of Correlators}
Notice that
the generators of translations on the plane, namely, $L_{-1}$ and $\tilde{L}_{-1}$ are BRST exact:

\be
L_{-1}=\{ Q,G_{-{1\over 2}}^{\uparrow}\}
\ee
It follows that, in the topological $G_2$ theory, genus zero correlation functions of chiral primaries between BRST closed states are position independent.
This is a crucial ingredient of topological theories.

\subsection{A Twisted Virasoro Algebra?}
\label{twistedvirasoro} Above, we constructed an $sl(2|1)$
algebra, and it is natural to ask if it can be extended to a full
$N=2$ algebra. This seems unlikely, but one definitely expects to
find at least all the modes of a twisted stress-tensor, which is
essential for the construction of a topological string theory on
higher genus Riemann surfaces. Since genus zero amplitudes are
independent of the locations of the operators, this suggests that
such a twisted stress tensor should indeed exist.

The construction of the $sl(2|1)$ algebra immediately yields a
candidate for the twisted stress tensor, namely \be
\label{attempt} \tilde{L}_m \equiv \{Q, G_{m+1/2}^{\uparrow} \}
\equiv \{Q, G_{m+1/2} \}. \ee This definition seems to work at
first sight. For example, \be \tilde{L}_{-1} = L_{-1} \ee as
expected for a twisted energy-momentum tensor. In addition, \be
[\tilde{L}_{-1}, \tilde{L}_m ] = (-1-m) \tilde{L}_{m-1}, \ee which
is the correct commutation relation for a Virasoro algebra. In
addition, $[\tilde{L}_m,\tilde{L}_{-m}]$ annihilates chiral
primaries, as expected for a twisted energy-momentum tensor with
zero central charge. However, there is no obvious reason why the
other commutation relations should be valid. Some extremely
tedious calculations reveal that (assuming that we did not make
any mistakes in the lengthy algebra) when acting on primaries of
the full $G_2$ algebra \be \label{jk0} \tilde{L}_0
|\Delta(k+1),h_r\rangle = \frac{4k-2}{4k-1}
((\Delta(k+1)+h_r)-\frac{k}{2}) |\Delta(k+1),h_r\rangle  \ee and \bea & &
[\tilde{L}_2,\tilde{L}_{-2}] |\Delta(k+1),h_r\rangle =c_k
((\Delta(k+1)+h_r)-\frac{k}{2}) \times  \\ & & (-1485+2868 k +2644
k^2 -3392 k^3 - 640 k^4 + 512 k^5 - 72 k (\Delta(k+1)+h_r))
|\Delta(k+1),h_r\rangle \nonumber \eea with \be c_k =
\frac{4k-2}{(k+1)(2k+3)(4k-11)(4k-1)^2(4k+9)} . \ee This clearly
shows that $[\tilde{L}_2,\tilde{L}_{-2}]\neq 4 \tilde{L}_0$. In
addition, we see the shift in $\tilde{L}_0$ would live entirely in
the tri-critical part were it not for the prefactor
$(4k-2)/(4k-1)$ that appears. Having the twist purely in the
tri-critical piece is appealing, as this can easily be implemented
in the Coulomb gas formulation, but further work is required to
prove that such a twisted energy-momentum tensor indeed exists and
is BRST exact. The above proposal is apparently not quite the
correct one.

\subsection{Moduli and Descent Relations}\label{descre}

As mentioned in section \ref{ssam}\ the upper components
$\tilde{G}_{-{1\over 2}}|{1\over 10},{2\over 5}\rangle_L\otimes\
G_{-{1\over 2}}|{1\over 10},{2\over 5}\rangle_R$ where shown in
\cite{sv} \ to be exactly marginal deformations of the CFT
preserving the $G_2$ chiral algebra. We also saw that they are in
one-to-one correspondence with the $b_3$ metric moduli of the
$G_2$ manifold. Once we include the $B$-field the number of such
moduli will turn out to be $b_2+b_3$ as we will see in
section~\ref{brstgeom}. Since both the ordinary and the
topologically-twisted theories should exist on an arbitrary
manifold of $G_2$ holonomy it is important to check that the
moduli space of deformations of the two theories agrees. So far we
have seen that the interesting objects in the twisted theory are
given in terms of non local objects of the original one. We will
now demonstrate that nevertheless the two theories  have the
same moduli space of deformations. In a fashion identical to
\ref{updown}\ we can split the local field $\mathcal{O}_2$ that
creates the chiral primary state $|{1\over 10},{2\over 5}\rangle$
as

\be\label{splitatwo}
\mathcal{O}_2= \mathcal{O}_2^{\downarrow}+\mathcal{O}_2^{\uparrow}=\sum_mP_{m+1}\mathcal{O}_2P_{m}+\sum_mP_{m-1}\mathcal{O}_2P_{m}.
\ee

The first term coincides with $\mathcal{A}_2$ which corresponds to
a chiral operator in the twisted theory so in particular $\{
Q,\mathcal{A}_2\}=0.$ Also, a computation similar to
\ref{qandatwo}\ shows that $\{ G_{-{1\over
2}}^{\uparrow},\mathcal{O}_2^{\uparrow}\}=0.$ Using this we
compute

\be\label{descent}
\begin{split}
[Q,\{ G_{-{1\over 2}},\mathcal{O}_2 \}] &=[Q,\{ G_{-{1\over 2}}^{\downarrow}+G_{-{1\over 2}}^{\uparrow}\ ,\ \mathcal{O}_2^{\downarrow}+\mathcal{O}_2^{\uparrow}\}]\\
&=[Q,\{ Q,\mathcal{O}_2^{\uparrow}\}]+[Q,\{G_{-{1\over 2}}^{\uparrow},\mathcal{A}_2\}]\\
&=[Q,\{G_{-{1\over 2}}^{\uparrow},\mathcal{A}_2 \}]\\
&=[\{ Q,G_{-{1\over 2}}^{\uparrow}\},\mathcal{A}_2]\\
&=[L_{-1},\mathcal{A}_2]=\partial\mathcal{A}_2.
\end{split}
\ee

In other words, we showed that $\partial \mathcal{A}_2= \{Q, {\rm
something } \}$, and the something is the $(1,0)$-form $\{
G_{-1/2} , \mathcal{O}_2\}$. This is a conventional operator that
does not involve any projectors. If we combine this also with the
right-movers, we find that the deformations in the action of the
topological string are exactly the same as the deformations of the
non-topological string.

\subsection{The Ramond Sector}
\label{rsec}

 We have previously given evidence, though no rigorous
proof, that the cohomology in the NS sector of
$G^{\downarrow}_{-1/2}$ is given by the chiral primaries. In the R
sector the situation is somewhat different. There is an obvious
candidate for a BRST operator in the R sector, namely
$Q=G^{\downarrow}_0$. Perhaps this is an even better candidate, as
it is the zero-mode of a field (as it should be in a twisted
theory), and because our twisting essentially boils down to doing
computations in the R sector. It is not immediately clear that
there is an easy map between the action of $G^{\downarrow}_0$ in
the R sector and the action of $G^{\downarrow}_{-1/2}$ in the NS
sector. This would require us to have a suitable isomorphism
between the NS and R sector. Such an isomorphism does exist and is
sometimes referred to as spectral flow (discussed more in appendix
\ref{spectral}), however it is not at all clear that this maps
$G^{\downarrow}_{-1/2}$ to $G^{\downarrow}_0$. It does however map
R ground states to chiral primaries, so this is further evidence
that the BRST cohomology in the NS sector consists of chiral
primaries and nothing else.

As an aside, notice that in the NS sector we found an $sl(2|1)$
subalgebra using some of the modes of $G^{\uparrow}$ and
$G^{\downarrow}$. In the R sector this is no longer the case. In
the R sector the only easy calculation we can readily do is that
\be \label{jk1} \{ G^{\downarrow}_0, G^{\uparrow}_0 \} =
L_0-\frac{7}{16}. \ee This in particular implies that the
$G^{\downarrow}_0$ cohomology is given by the R ground states.
This is an exact statement. Therefore, $G^{\downarrow}_0$ looks
like an excellent candidate BRST operator. It also has the nice
property that the right hand side of (\ref{jk1}) is the most
natural definition of $L^{\rm twisted}_0$ in the R sector in
contrast to the situation in the NS sector.

\subsection{Localization}
It can be shown quite generally \cite{Witten} that the path
integral localizes to fixed points of the BRST symmetry. For the
usual case of the A and B model, this implies that only
holomorphic and constant maps contribute, respectively. To derive
a similar statement for the topological $G_2$ sigma model, we
start by writing the action as

\begin{eqnarray*}
S&=&\int d^2 z {1 \over 2}g_{IJ}\partial  x^I \bar{\partial}x^J +
g_{IJ} \Bigl(i \psi^{\uparrow J}_{L} D \psi_{ L}^{\downarrow I} + i \psi^{\downarrow J}_{L} D \psi_{ L}^{\uparrow I} +
 i \psi^{\uparrow J}_{R} \bar{D} \psi_{ R}^{\downarrow I} +
 i \psi^{\downarrow J}_{R} \bar{D} \psi_{ R}^{\uparrow I}    \Bigr) \\ &&~~~~~~~~
+R_{IJKL}\psi_R^{\uparrow I} \psi_R^{\downarrow J} \psi_L^{\uparrow K}\psi_L^{\downarrow L}
\end{eqnarray*}
This action has the fermionic symmetry
\begin{eqnarray*}
\delta  x^I & =& i \epsilon_L \psi_L^{\downarrow I} +i \epsilon_R \psi_R^{\downarrow I}  \\
\delta \psi_L^{\uparrow I} &= &-\epsilon_L \partial x^J   -\epsilon_R \psi_R^{\downarrow K} \Gamma^I_{KM} \psi_L^{\uparrow M}  \\
\delta \psi_L^{\downarrow I} &= &   - \epsilon_R   \psi_R^{\downarrow K} \Gamma^I_{KM} \psi_L^{\downarrow M}  \\
\delta \psi_R^{\downarrow I} &=& -\epsilon_L \psi_L^{\downarrow K} \Gamma^I_{KM} \psi_R^{\downarrow M} \\
\delta \psi_R^{\uparrow I} &=& -\epsilon_R \bar{\partial} x^J -\epsilon_L \psi_L^{\downarrow K} \Gamma^I_{KM} \psi_R^{\uparrow M}
\end{eqnarray*}
The fixed points of this symmetry satisfy $\partial x^I =
\bar{\partial} x^I =0$, which implies that the path integral
localizes on constant maps. Of course, we should take this
analysis with a grain of salt: the decomposition of the world
sheet fermions $\psi^I$ into conformal blocks $\psi^{\uparrow I}+
\psi^{\downarrow I}$ is inherently quantum mechanical and hence it
is problematic to use this decomposition in path integral
arguments. Nevertheless, we take this argument as at least
suggestive that we are localizing on constant maps.


\section{Relation to Geometry}
\label{geometry}
For a $G_2$ manifold, differential forms of any degree can be decomposed into irreducible representations of $G_2$
\begin{eqnarray*}
\Lambda^0 = \Lambda^0_1 & ~~~~~~~~~& \Lambda^1 = \Lambda^1_7 \\
\Lambda^2=\Lambda^2_7 \oplus \Lambda^2_{14} & ~~~~~& \Lambda^3=\Lambda^3_1 \oplus \Lambda^3_7 \oplus \Lambda^3_{27}
\end{eqnarray*}
This is described in more detail in Appendix \ref{decomposition}.
In a similar spirit as Hodge theory, this decomposes the
cohomology groups as  $H^p=\oplus_{\cal R} H^p_{\cal R}(M)$ where
the sum is over $G_2$ representations ${\cal R}$. The cohomology
turns out to depend solely on the representation ${\cal R}$ and
not on the degree $p$ \cite{joycebook}. For a proper compact $G_2$
manifold, $H^1(M)=0$ and so there is no cohomology in the seven
dimensional representation of $G_2$.  Also, $b_1^3=1$,
corresponding to a unique closed three form $\phi$ which defines
the $G_2$ structure. There are only two independent Betti numbers
left unknown, namely $b^2_{14}$ which is equal to the usual second
Betti number $b_2$ and $b_{27}^3=b_3-1$ with no known restrictions
on these numbers.

\subsection{Dolbeault Complex for $\mathbf{G_2}$-Manifolds}
It is possible to define a refinement of the de Rham complex, in a
spirit somewhat similar to Dolbeault cohomology, as follows:
\begin{equation}
0 \rightarrow \Lambda^0_1 \stackrel{\check{D}}{\rightarrow} \Lambda^1_7
\stackrel{\check{D}}{\rightarrow} \Lambda^2_7\stackrel{\check{D}}{\rightarrow}\Lambda^3_1 {\rightarrow}0
\label{complex}
\end{equation}
where $\check{D} $ is the usual exterior derivative when acting on 0-forms, but is the composition of the exterior derivative and projection to the {\bf 7} and {\bf 1} representations of $G_2$ when acting on 1 and 2 forms respectively:
\begin{eqnarray*}
\check{D}(\alpha) &= &\pi^2_7(d\alpha) ~~~~{\rm for    } ~~\alpha \in \Lambda^1 \\
\check{D}(\beta) &= &\pi^3_1(d\beta) ~~~~{\rm for     }~~ \beta \in \Lambda^2
\end{eqnarray*}
where the projection operators $\pi^p_r$ are defined in appendix
\ref{decomposition}. In local coordinates, these expressions
become
\begin{eqnarray*}
\Bigl(\check{D}(\alpha) \Bigr)_{\mu \nu} dx^\mu \wedge dx^\nu &=& 3 \partial_{[\mu} A_{\nu]} \phi^{\mu \nu}_\rho \phi^\rho_{\eta \chi} dx^{\eta} \wedge dx^\chi ~~~~~~~~~~~~~~~~~\alpha=A_\mu dx^\mu \\
\Bigl(\check{D}(\beta)\Bigr)_{\mu \nu \rho} dx^\mu dx^\nu dx^\rho & = & \partial_{[\xi} B_{\eta \chi]} \phi^{\xi \eta \chi} \phi_{\mu \nu \rho} dx^\mu \wedge dx^\nu \wedge dx^\rho ~~~~~~~ \beta=B_{\mu \nu}dx^\mu \wedge dx^\nu
\end{eqnarray*}

We  will next see that the cohomology of this differential complex
maps to the BRST cohomology in the left (or right) moving sector.
The differential operator $\check{D}$ maps to the BRST operator
$G_{-{1 \over 2}}^{\downarrow}$. This gives a nice and natural
geometric meaning to the BRST operator, and clearly shows we are
on the right track.

\subsection{The BRST Cohomology Geometrically} \label{brstgeom}

In the previous section, we argued that the BRST cohomology
consists of the chiral primary operators of our conformal field
theory. We now proceed to study the sigma model description of
these operators and the geometric meaning of the chiral ring.

To determine whether an operator corresponds to a chiral primary,
we need to find its $L_0$ and $X_0$ quantum numbers. Often in
topological theories, this calculation can be reduced to operators
built out of non-derivative fields only. In our case we also
expect this to be the case, since all elements in the cohomology
are in one-to-one correspondence to R ground states. Also, the argument
that the path integral localizes on constant maps indicates that
 only zero modes appear.

So we proceed by analyzing the action of the BRST
operator at the level of operators that do not contain any
derivatives of fields. In the left-moving sector, such operators
are in one-to-one correspondence with $p$-forms on the target
space:
\be
\omega_{i_1,\ldots,i_p}dx^{i_1} \wedge \ldots \wedge dx^{i_p}
\leftrightarrow \omega(x^{\mu})_{i_1,\ldots,i_p} \psi^{i_1} \ldots
\psi^{i_p} .
\ee
The same is obviously also true in the right-moving sector, but
for simplicity we analyze the left-moving sector first.

The group $G_2$ acts on the tangent space of the manifold, and the
space of $p$-forms at a point can be decomposed in $G_2$
representations as explained above. Since $X_0$ and $L_0$ are $G_2$ singlets, they
take the same value in each of these representations. Some further
explicit calculations \footnote{As an example, we determine the $X_0$ eigenvalue of the operator $A(X)_\mu \psi^\mu$ which corresponds to the one form $A(X)_\mu dx^\mu$. Using the expression for $X(z)$ in (\ref{xdef}), the $X_0$ eigenvalue is given by the coefficient of the second order pole in the OPE
\begin{equation}
X(z) .\Bigl(A(X)_\mu \psi^\mu (0) \Bigr ) \sim \cdots -{1 \over 2} {A(X)_\mu \psi^\mu \over z^2}+ \cdots
\end{equation}
which gives the $X_0$ eigen-value of this operator to be $-{1 \over 2}$ and the tri-critical Ising model weight ${1 \over 10}$.}
involving the precise form of $X_0$ then
reveal that the quantum numbers associated to each representation
are
\begin{equation} \label{quantn}
\begin{array}{|c|c|c|c|c|}
\hline & {\bf 1} & {\bf 7} & {\bf 14} & {\bf 27} \\
\hline p=0 & |0,0\rangle & & & \\
\hline p=1 & & |\frac{1}{10},\frac{2}{5} \rangle & & \\
\hline p=2 & & |\frac{6}{10},\frac{2}{5}\rangle  &
|0,1\rangle & \\
\hline p=3 & |\frac{3}{2},0\rangle & |\frac{11}{10} ,\frac{2}{5}
\rangle &   & |\frac{1}{10},\frac{7}{5} \rangle \\
\hline p=4 & |2,0\rangle & |\frac{16}{10} ,\frac{2}{5}
\rangle &   & |\frac{6}{10},\frac{7}{5} \rangle \\
\hline p=5 & & |\frac{21}{10},\frac{2}{5}\rangle  &
|\frac{3}{2},1\rangle & \\
\hline p=6 & & |\frac{26}{10},\frac{2}{5} \rangle & & \\
\hline p=7 & |\frac{7}{2},0\rangle & & & \\
\hline
\end{array}
\end{equation}

This table also nicely reflects the two maps which take a $p$-form
$\omega$ into a $p+3 $ form given by $\omega \wedge \phi$ and into
a $p+4$ form $\omega \wedge * \phi$ (see appendix
\ref{decomposition}). When restricted to $G_2$ representations,
these operators are either identically zero or act as
isomorphisms. They translate to the action of $\Phi_{-{3 \over
2}}$ and $X_{-2}$ at the level of states. Notice that chiral
primaries appear only in four places in (\ref{quantn}), and
precisely those differential forms enter into (\ref{complex}). Of
course, this is not a coincidence, as we will see below.

In order to construct the precise form of these states, we need to
project the relevant forms on to appropriate $G_2$
representation. All such projectors can be constructed in terms of
the three-form $\phi$ and its Hodge dual, as explained in Appendix
\ref{decomposition}. To find their precise form, various
identities satisfied by $\phi$ are useful, such as \bea
\phi_c{}^{de} \phi_{de}{}^f &  = & \frac{1}{6} \delta_c^f \nonumber \\
\phi_{ab}{}^{cd} \phi_{cd}{}^e & = & \frac{1}{6} \phi_{ab}{}^e \nonumber \\
\phi_{ab}{}^c \phi_c{}^{de} & = & \frac{2}{3} \phi_{ab}{}^{de} + \frac{1}{36} (
\delta^d_a \delta^e_b - \delta^e_a \delta^d_b) \nonumber \\
\phi_{ab}{}^{cd} \phi_{cd}{}^{ef} & = & \frac{1}{12} \phi_{ab}{}^{ef} +
\frac{1}{144} (\delta^e_a \delta^f_b - \delta^f_a \delta^e_b) \nonumber \\
\phi_{abc} \phi^{abc} & = & \frac{7}{6} \nonumber \\
\phi_{[ab}{}^{cd} \phi_{e]cd} & = & \phi_{abe} \nonumber \\
\frac{1}{2} \phi_{[ab}{}^{cd} \phi_{e]cd}{}^f & = & -\frac{1}{4} \phi_{abe}{}^f  .
\label{idlist}
\eea
In these equations, antisymmetrization over $n$ indices does not include a factor of $1/n!$.
They are also useful in order to compute the $X_0$ eigenvalue in each representation. Notice,
however, that the exact quantum $X_0$ eigenstates can not in general be written in terms
of fields without derivatives, typically one needs to add some quantum corrections involving
fewer fermions and a few derivatives as well.

This table allows us to extract the precise action of the BRST
operator on the operators that do not involve derivatives. For
example, \be G_{-1/2} A_{\mu}(X) \psi^{\mu} = \frac{1}{2}
\partial_{[\nu} A_{\mu]} \psi^{\nu} \psi^{\mu} + A_{\mu}(X)
\partial X^{\mu}. \ee In the calculation we get a covariant
derivative, however this is equal to the ordinary derivative when
acting on forms as an exterior derivative. To extract the action
of $G_{-1/2}^{\downarrow}$, we first observe the second term has
$X_0=0$ and therefore only contributes to $G_{-1/2}^{\uparrow}$.
The first term has a part transforming in the ${\bf 7}$ of $G_2$
and a part transforming in the ${\bf 14}$ of $G_2$, and according
to (\ref{quantn}) we need to project on the ${\bf 7}$ to obtain
the action of $G^{\downarrow}_{-1/2}$. The relevant projection
operator is $P_{ab}{}^{de} = 6 \phi_{ab}{}^c \phi_c{}^{de}$, and
we finally get \be \label{aux89} G_{-1/2}^{\downarrow}  A_{\mu}(X)
\psi^{\mu} = 3   \partial_{[\nu} A_{\mu]} \phi^{\nu\mu}{}_{\rho}
\phi^{\rho}{}_{\alpha\beta} \psi^{\alpha} \psi^{\beta} . \ee It is
clear by inspection of table (\ref{quantn}) that chiral primaries,
i.e. non-trivial elements of the BRST cohomology, can either be
singlet 0- or 3-forms, or 1- or 2-forms transforming in the ${\bf
7}$ of $G_2$.

By repeating (\ref{aux89}) for the two form $B_{\mu \nu} \psi^\mu
\psi^\nu$ and the three form $\phi_{\mu \nu \alpha} \psi^\mu
\psi^\nu \psi^\alpha$,  the kernel of $Q_{\rm BRST}$ in the
left-moving sector is then seen to consist of \bea
1 & & \nonumber \\
A_{\mu} \psi^{\mu} & {\rm with} & \phi_{\rho}{}^{\mu\nu} \partial_{[\mu} A_{\nu]} =0
\nonumber \\
B_{\mu\nu} \psi^{\mu} \psi^{\nu} & {\rm with} & \phi^{\rho\mu\nu} \partial_{[\rho}
B_{\mu\nu]} = 0 \nonumber \\
\phi_{\mu\nu\rho} \psi^{\mu} \psi^{\nu} \psi^{\rho}. & & \eea We
should still remove the image of $G_{-1/2}^{\downarrow}$, which
means identifying for example \be A_{\mu} \sim A_{\mu}
+\partial_{\mu} C \ee and \be B_{\alpha\beta} \sim B_{\alpha\beta}
+  3
\partial_{[\nu} D_{\mu]} \phi^{\nu\mu}{}_{\rho}
\phi^{\rho}{}_{\alpha\beta} \ee for arbitrary $C$, $D_{\mu}$.

\begin{figure}
  \begin{center}
\epsfysize=2.2in
\epsfxsize=2.9in
    \mbox{\epsfbox{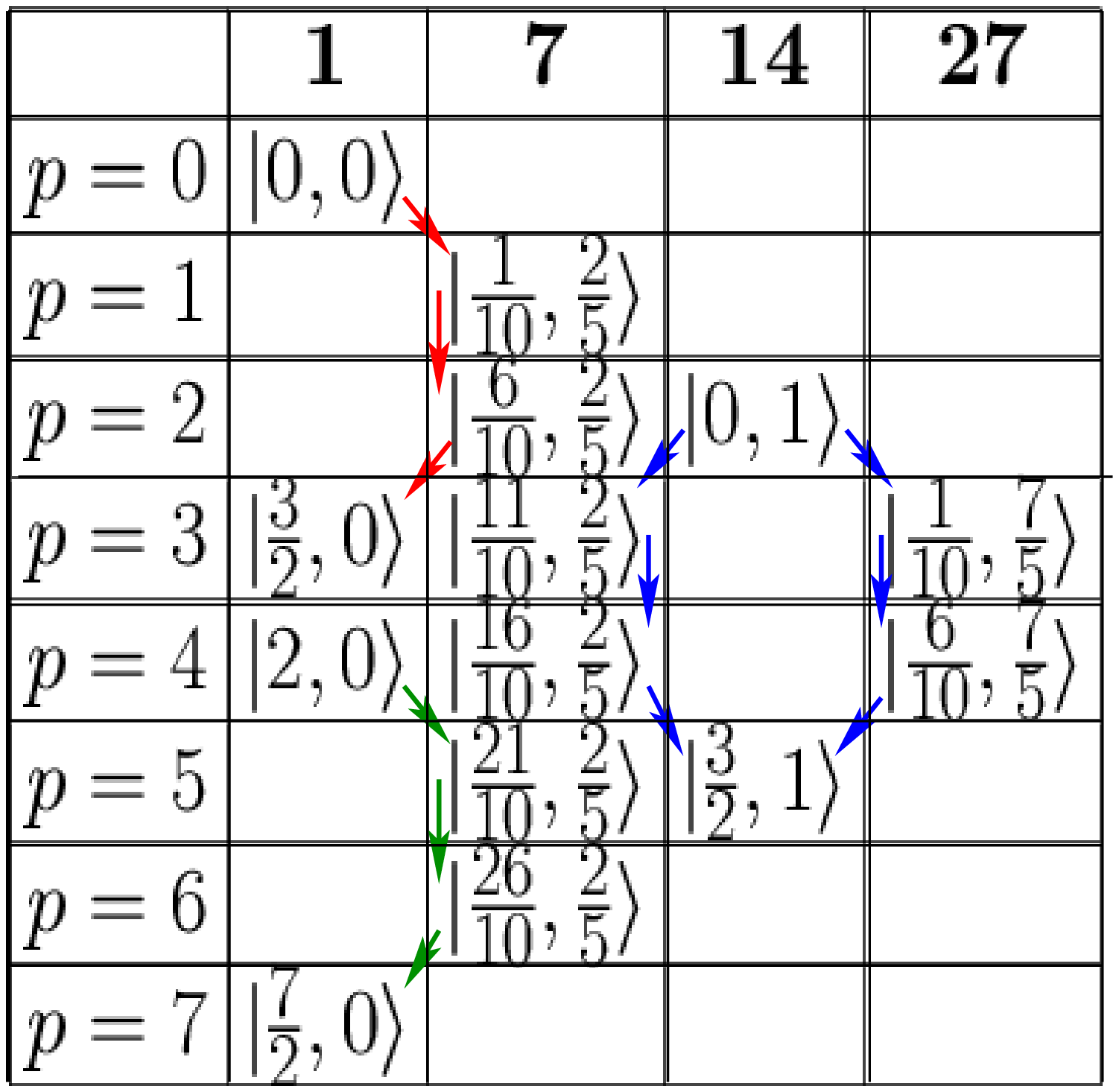}}
  \end{center}
 \caption{Differential complexes and the BRST cohomology}
\label{complexes}
\end{figure}

It is interesting to note that the BRST cohomology in the left
moving sector is just the Dolbeault type cohomology of the
$\check{D}$ operator that we defined in the previous subsection.
The BRST operator $G_{-{1/ 2}}^\downarrow$ naturally maps to the
operator $\check{D}$. In fact, the table \ref{quantn} reveals the
existence of two other differential complexes. One of these is
related to the complex in (\ref{complex}) by the Hodge duality.
The other one is a new complex
\begin{equation}
0 \rightarrow \Lambda^2_{14} \stackrel{\tilde{D}}{ \rightarrow}  \Lambda^3_{7}\oplus \Lambda^3_{27} \stackrel{\tilde{D}}{\rightarrow} \Lambda^4_{7}\oplus \Lambda^4_{27} \stackrel{\tilde{D}}{\rightarrow} \Lambda^5_{14} \rightarrow 0
\end{equation}
where the differential operator $\tilde{D}$ is the composition of
the ordinary exterior derivative with appropriate projection
operators (defined in appendix \ref{decomposition}). This new
complex does not consist of chiral primaries and does not seem to
play any role in the twisted theory we are considering, but it
would still be interesting to know whether it has a distinguished
geometric interpretation.

If we do not combine left and right movers, the cohomology is almost trivial. As we noted earlier,
compact  $G_2$ manifolds have $b_1=0$ and therefore there is no cohomology in
the seven-dimensional representation of $G_2$. As a consequence, only the identity and the three-form
survive if we do not include right-movers.

However,  once we combine left- and right-movers, we obtain a  more
interesting cohomology. The two-form $B$ and one-form $A$ are in one-to-one correspondence
via $ B_{\mu\nu} = \phi_{\mu\nu}{}^{\alpha} A_{\alpha}$ so it is sufficient to consider only the
combination of the left- and right moving one-forms. Each of them transforms in the ${\bf 7}$
of $G_2$, and ${\bf 7}\otimes {\bf 7}= {\bf 1}+ {\bf 7} + {\bf 14} + {\bf 27}$. We get one
non-trivial class from ${\bf 1}$, none from ${\bf 7}$, $b_2$ from ${\bf 14}$ and
$b_3-1$ from ${\bf 27}$. In total, we get $b_2+b_3$, corresponding to the non-trivial
$B$-field and metric deformations of the $G_2$ manifold. This is indeed the set of moduli that we
 expect to find in a topological theory. If we replace the left or right movers by
a two-form, these results do not change. We also get a
contribution to the cohomology from the left-moving zero/three
form times the right-moving zero/three form. The total cohomology
is \bea
0-{\rm form}\,\, \times \,\, 0-{\rm form} & \rightarrow & b_0 \nonumber \\
1-{\rm form}\,\, \times \,\, 1-{\rm form} & \rightarrow & b_2 + b_3 \nonumber \\
2-{\rm form}\,\, \times \,\, 2-{\rm form} & \rightarrow & b_4 + b_5 \nonumber \\
3-{\rm form}\,\, \times \,\, 3-{\rm form} & \rightarrow & b_7
\eea
plus another copy of this if we allow the left and right levels not to match each other.
Either way, we get one or two copies of the full cohomology $H^{\ast}(M)$ of $M$.

We can verify whether we recover known results about the metric moduli of $G_2$ manifolds.
According to the above, metric and $B$-field moduli should be given by operators of the form
\be
(\delta g_{\mu\nu} + \delta B_{\mu\nu}) \psi_R^{\mu} \psi_L^{\nu}
\ee
with
\be  \label{eq101}
\phi_{\alpha}{}^{\lambda\mu} (\nabla_{[\lambda} \delta g_{\mu]\nu} +
\nabla_{[\lambda} \delta B_{\mu]\nu} ) =0.
\ee
Metric moduli are indeed known to satisfy this equation (eq. \ref{mod}) as pointed out in
\cite{math0311253}. To verify that $B$-moduli also satisfy (\ref{eq101}), we first use
the fact that $\phi$ is covariantly constant to rewrite
\be \label{eq102}
\phi_{\alpha}{}^{\lambda\mu}( \nabla_{[\lambda} \delta B_{\mu]\nu})=\nabla_{[\lambda} (\delta B_{\mu]\nu} \phi_{\alpha}{}^{\lambda\mu}).
\ee
Since $B$-moduli transform in the ${\bf 14}$ of $G_2$, they also obey (see appendix \ref{decomposition})
\be
\delta B_{\lambda\mu} \phi_{\alpha}{}^{\lambda\mu}=0.
\ee
We can therefore replace the rhs of (\ref{eq102}) by
\be
\nabla_{[\lambda} (\delta B_{\mu\nu]} \phi_{\alpha}{}^{\lambda\mu})
=\partial_{[\lambda} \delta B_{\mu\nu]}
\phi_{\alpha}{}^{\lambda\mu}=0 \ee since $B$-moduli are closed
two-forms. This shows that the $B$-moduli also satisfy
(\ref{eq101}) and the BRST cohomology consists exactly of the
metric and the $B$-field moduli.

\subsection{Correlation Functions}

In this section we explicitly compute some simple correlation
functions in the $G_2$ sigma model by working in the classical,
large volume approximation.

As we discussed already, the operator cohomology contains only
operators that map ${\cal H}_i$ to ${\cal H}_j$ with $i\leq j$.
Therefore only a finite set of correlation functions will be
nonzero. Let's first consider the left-movers only, and consider a
three-point function of three operators ${\cal O}_k = A^k_{\mu}
\psi^{\mu}$, with $k=1,2,3$, and we assume each to be in the BRST
cohomology. This boils down to the calculation of \be
\label{eq201} \left\langle V_{\frac{7}{16},+} {\cal O}_1 {\cal
O}_2 {\cal O}_3 V_{\frac{7}{16},+} \right\rangle \ee in the
untwisted theory. This object turns out to be a 4-point function
in the R-sector \be \left\langle \Phi_0 {\cal O}_1 {\cal O}_2
{\cal O}_3 \right\rangle_R \ee because
$V_{\frac{7}{16},+}^{\dagger}=V_{\frac{7}{16},-}=\Phi_0
V_{\frac{7}{16},+}$. The operator $\Phi$ is
$\phi_{\alpha\beta\gamma} \psi^{\alpha} \psi^{\beta}
\psi^{\gamma}$, and from the contractions we obtain for the
correlator something proportional to \be \phi_{\alpha\beta\gamma}
g^{\alpha\mu} g^{\beta\nu} g^{\gamma\rho} A^1_{\mu} A^2_{\nu}
A^3_{\rho}. \ee The inverse metrics arise due to the fact that in
this approximation the fermion two-point function is proportional
to the inverse metric.

Combining left and right movers, relabelling everything in terms
of metric and $B$-field moduli, and including an integral over the
seven manifold from the zero mode of $X^{\mu}$, we finally obtain
for the three-point function for metric and $B$-field moduli \be
{\cal F}_{3-{\rm point}} = \int_M d^7 x \sqrt{g}
\phi_{\alpha\beta\gamma} (\delta_1 g^{\alpha\alpha'} + \delta_1
b^{\alpha\alpha`}) (\delta_2 g^{\beta\beta'} + \delta_2
b^{\beta\beta`}) (\delta_3 g^{\gamma\gamma'} + \delta_3
b^{\gamma\gamma`}) \phi_{\alpha'\beta'\gamma'} . \ee

To analyze this expression a bit further, we drop the $B$-field
moduli. In addition, we will take a suitable set of coordinates
$t_i$ on the moduli space of $G_2$ metrics, and denote by $Y_i$
the operator corresponding to sending $t_i\rightarrow t_i+\delta
t_i$. In other words, the three-point function reads
\begin{equation}
\langle Y_i Y_j Y_k \rangle = \int_M d^7 x \sqrt{g} \phi_{\alpha\beta\gamma}
\frac{\partial g^{\alpha\alpha'}}{\partial t_i} \frac{\partial
g^{\beta\beta'}}{\partial t_j}\frac{\partial g^{\gamma\gamma'}}{\partial t_k}
\phi_{\alpha'\beta'\gamma'}.
\end{equation}
One might expect, based on general arguments (see e.g.
\cite{Dijkgraaf}), that this is the third derivative of some
prepotential if suitable `flat' coordinates are used. For example,
consider the manifold $M=T^7$ and choose coordinates such that
$\phi$ is linear in them. We find that
\begin{equation} \label{hitch}
\langle Y_i Y_j Y_k \rangle = -\frac{1}{21}
\frac{\partial^3}{\partial t_i \partial t_j \partial t_k} \int
\phi \wedge \ast \phi.
\end{equation}
This strongly suggests that the same results should also be valid
on general $G_2$ manifolds. In fact, in the next subsection, we
will develop a version of ``special geometry'' for $G_2$ manifolds
and show that with an appropriate
 definition of
flat coordinates for the moduli space of $G_2$ metrics, the three point function can be written as
in (\ref{hitch})


The action \be \label{hitchact} S=\int \phi \wedge \ast \phi \ee
also appears in \cite{hitchin}, where it was shown that the
critical points of this functional, viewed as a functional on the
space of three-forms in a given cohomology class, are precisely
the three-forms of $G_2$ manifolds. It was also the starting point
of topological M-theory in \cite{vafa}, see also \cite{gerasimov}.
It is tempting to speculate that our topological $G_2$ string
provides the framework to quantize topological M-theory, which by
itself is not yet a well-defined quantum theory.

\subsection{$\mathbf{G_2}$ Special Geometry}

To prove in full generality a relation between our topological three point function
and the Hitchin functional we need to develop a version of ``special geometry'' for
$G_2$ manifolds.

First of all we define
\be
{\cal I} = \int \phi \wedge \ast \phi,
\ee
which will be a functional on the space of $G_2$ metrics (or on the space of
the corresponding three-forms).

The most natural choice for flat coordinates, as our torus example
also suggests, is to choose periods, as we do in the case of the
six-dimensional topological string. We thus pick a symplectic
basis of homology three-cycles $C_A$ and dual four cycles $D^A$,
and define coordinates on the moduli space of $G_2$ metrics as \be
t^A = \int_{C_A} \phi. \label{periodphi} \ee For the dual periods
we introduce the notation \be F_A= \int_{D^A} \ast \phi. \ee It is
perhaps tempting to write \be \phi=t^A \chi_A \label{linear} \ee
with $\chi_A$ a basis of three forms Poincare dual to the
four-cycles $D_A$. This is not quite correct as the detailed form
of $\phi$ will in general differ from (\ref{linear}) by an exact
three-form. In most calculations, this exact three-form drops out,
but it is important to keep in mind that $\phi$ cannot simply be
expanded linearly in a given basis of cohomology.

Continuing, we can also write $F_A$ as \be F_A = \int \ast \phi
\wedge
\partial_A \phi \ee Furthermore, by a generalization of the
Riemann bilinear identities we find that \be {\cal I} = t^A F_A.
\ee

Let's now take one derivative of ${\cal I}$. We readily obtain
\be \label{sg1}
\partial_B {\cal I} = F_B + t^A \partial_B F_A.
\ee

We can also perform straightforward explicit computations by using  the
canonical expressions for $\phi$ and $\ast\phi$ in local coordinates:
\bea
\phi&=&dx^{123}+dx^{145}+dx^{167}+dx^{246}-dx^{257}-dx^{347}-dx^{356} \\
\ast \phi & = &
dx^{4567}+dx^{2367}+dx^{2345}+dx^{1357}-dx^{1346}-dx^{1256}-dx^{1247}
\eea Here, $dx^{ijk}=e^i \wedge e^j \wedge e^k$, with
$e^i=e^i_{\mu} \, dx^{\mu}$ a local orthonormal frame, i.e. a set
of vielbeins in which the metric becomes $g_{\mu\nu}=e_{\mu}^i
e_{\nu}^i$. To find the variation of various quantities with
respect to $t^A$, we will need to vary the vielbeins.  We notice
that up to $SO(7)$ rotations rotating the $e^i$ into each other
\be \label{sg9}
\partial_A e^a_{\mu} = \frac{1}{2} \partial_A h_{\mu\nu} h^{\nu\lambda} e^a_{\lambda}
. \ee Then, using the explicit expressions for $\phi$ and $\ast
\phi$ in terms of the vielbeins, we find \be F_A = \int \ast \phi
\wedge \partial_A \phi = {3 \over 2} \partial_A h_{\mu \nu} h^{\mu
\nu} {\cal I} \ee and \be t^B \partial_A F_B = \int \partial_A
\ast \phi \wedge \phi   = {2} \partial_A h_{\mu \nu} h^{\mu \nu}
{\cal I} \ee which implies that \be \label{sg2} t^A \partial_B F_A
= \frac{4}{3} F_B \ee We conclude from (\ref{sg1}) and (\ref{sg2})
that \be \label{sg3} F_B = \frac{3}{7} \partial_B {\cal I}. \ee
Thus we see that ${\cal I}$ is homogeneous of degree $7/3$ in the
coordinates $t^A$, which can also easily be verified explicitly,
but more importantly we have found that the dual periods are the
derivatives of a single function, the prepotential $F$, which is
given by \be \label{sg4} F = \frac{3}{7} {\cal I}. \ee

Next we turn to the second derivative of ${\cal I}$. From the above we readily
obtain
\be \label{sg5}
\int \partial_A \phi \wedge \partial_B \ast \phi =\int \partial_B \phi \wedge \partial_A \ast \phi= 3 \int \phi \wedge \partial_A \partial_B \ast \phi=\frac{3}{7} \partial_A \partial_B {\cal I}.
\ee
We can evaluate the first expression most easily,
 by varying the vielbeins that appear
in the standard expression for $\phi$ and $\ast\phi$, and by
counting the resulting terms. We find
\be \label{sg6}
\int \partial_A \phi \wedge \partial_B \ast \phi =\frac{1}{2} \int
\sqrt{g} ( \partial_A h_{\mu\nu} h^{\mu\nu} \partial_B
h_{\rho\sigma} h^{\rho\sigma} -
\partial_A h_{\mu\nu} h^{\nu\rho} \partial_B
h_{\rho\sigma} h^{\sigma\mu} ) .
\ee

On the other hand, by using the third identity in (\ref{idlist})
we deduce
\be \label{sg7}
\int \sqrt{g} \phi_{abc}
\partial_A h^{aa'} \partial_B h^{bb'} h^{cc'} \phi_{a'b'c'} =\frac{1}{36}\int \sqrt{g} ( \partial_A h_{\mu\nu} h^{\mu\nu}
\partial_B h_{\rho\sigma} h^{\rho\sigma} -
\partial_A h_{\mu\nu} h^{\nu\rho} \partial_B
h_{\rho\sigma} h^{\sigma\mu} ).
\ee

Combining (\ref{sg5}), (\ref{sg6}) and (\ref{sg7}) we finally
obtain
\be \label{sg8}
\int \sqrt{g} \phi_{abc}
\partial_A h^{aa'} \partial_B h^{bb'} h^{cc'}
\phi_{a'b'c'}=\frac{1}{42} \partial_A \partial_B {\cal I}.
\ee

Therefore, the second derivatives of ${\cal I}$ closely resemble
the expression for the three-point function we obtained from the
topological string.

Turning finally to the third derivative, this analysis is a bit
more tedious. In analogy with (\ref{sg5}) we have \be \label{sg9a}
\int \partial_A \phi \wedge \partial_B
\partial_C \ast \phi= -\frac{3}{2} \int \phi \wedge \partial_A
\partial_B \partial_C \ast \phi=\frac{3}{7} \partial_A \partial_B \partial_C {\cal I}.
\ee

The first expression is again the most useful one to manipulate,
and we do this as before in terms of a representation in a local
orthonormal flat frame (i.e. vielbeins). We again use the
variation of the vielbein as given in (\ref{sg9}). We find a new
feature, namely we now also will run into double derivatives of
the metric, due to the double derivative acting on $\ast\phi$ in
the first expression in (\ref{sg9a}). We can get rid of this
double derivative as follows. We write $\partial_{BC}$ for the
double derivative acting on a single vielbeins only. Then it is
easy to  see that \be \label{sg10} \int \partial_A \phi \wedge
\partial_{BC} \ast \phi = \int \partial_{BC} \phi \wedge
\partial_A \ast \phi . \ee Now notice that $\partial_B \partial_C
= \partial_{BC} + \partial_{BC}'$, where $\partial_{BC}'$ is
defined such that the two derivatives never act on the same
vielbein. Thus, for example, \bea
\partial_{BC} e^1 \wedge e^2 & \equiv & \partial_B \partial_C e^1 \wedge e^2 +
e^1 \wedge \partial_B \partial_C e^2 \nonumber \\
\partial_{BC}' e^1 \wedge e^2 & \equiv & \partial_B e^1 \wedge \partial_C e^2 +
\partial_C e^1 \wedge \partial_B e^2
\eea
and clearly these two add up to $\partial_B \partial_C$. Because $\phi$ is linear in
$t^A$ (\ref{linear}), we can replace in (\ref{sg10}) $\partial_{BC} \phi = \partial_B \partial_C \phi -\partial'_{BC} \phi =-\partial'_{BC} \phi $.
%
So we obtain,
\be \label{sg11}
\partial_A \partial_B \partial_C {\cal I} = \frac{7}{3} \left(
\int \partial_A \phi \wedge \partial_{BC}' \ast \phi -
\int \partial_{BC}' \phi \wedge \partial_A \ast \phi \right).
\ee
In this expression no double derivatives of the metric appear anymore.
However it contains a priori all kinds of contractions of the three single derivatives
of the metric. To determine the detailed form of the result, we took (\ref{sg11}),
wrote $\ast\phi$ in terms of $\phi$ using the seven-dimensional completely antisymmetric
$\epsilon$ tensor, and expanded (\ref{sg11}) in terms of all possible contractions that
can appear. After a significant amount of tedious algebra we found, quite surprisingly, that almost
all terms cancel, and that we are left with the simple final result
\be \label{sg12}
\partial_A \partial_B \partial_C {\cal I} = -21 \int \sqrt{g}
\phi_{abc}
\partial_A h^{aa'} \partial_B h^{bb'} \partial_C h^{cc'}
\phi_{a'b'c'}.
\ee

This proves that our topological three-point function is indeed
the third derivative of a single function, which is precisely the
Hitchin functional, viewed as a function on the space of $G_2$
metrics! Notice that (\ref{sg12}) is valid both for the rather
trivial modulus which corresponds to rescaling $\phi$, as well as
for the $b_3-1$ moduli which live in the ${\bf 27}$ of $G_2$. For
the latter moduli an expression similar to (\ref{sg12}) was
written down in \cite{leeleung}, where it was used to describe
fibrations of $G_2$ manifolds by coassociative submanifolds. These
three-point functions were called Yukawa couplings in that paper,
though the relation with the physical Yukawa couplings in M-theory
was not given. Our results shows that the cubic coupling
(\ref{sg12}), which is the topological three-point function, is
indeed closely related to the physical Yukawa couplings that one
obtains in compactifying M-theory on $G_2$ manifolds. This is
because the K\"ahler potential of the resulting four-dimensional
theory is essentially the logarithm of ${\cal I}$, and Yukawa
couplings are given by the third derivative of the K\"ahler
potential. A more detailed discussion can be found in
section~\ref{treelevel}.

\subsection{Inclusion of the $\mathbf{B}$-field}

We next want to see what happens when we include the $B$-field. There is only one relevant
correlator
\be \label{b1}
\int \sqrt{g}
\phi_{abc}
\partial_p B^{aa'} \partial_q B^{bb'} \partial_C h^{cc'}
\phi_{a'b'c'}, \ee since the correlators involving one or three
$B$-field insertions vanish identically due to
symmetry/anti-symmetry properties of the index contractions. We
introduced coordinates $s^p$ on the space $H^2(M)$ of $B$-fields,
but still need to specify how they are defined. To simplify the
above expression, we first observe that since $B^{bb'}$ lives in
the ${\bf 14}$ of $G_2$ (the $B$-field is a closed two form and
the only non-trivial second cohomology transforms as in the ${\bf
14}$ dimensional representation of $G_2$), which means
$\phi_{abb'} B^{bb'}=0$. Therefore, we can antisymmetrize over
$a,a',b,b'$ in the above expression so that it becomes \be
\label{b2} \frac{1}{24} \int \sqrt{g} \phi_{c[ab}\phi_{a'b']c'}
\partial_p B^{aa'} \partial_q B^{bb'} \partial_C h^{cc'}.
\ee
Next, we can use the following identity
\be
\label{b3}
\phi_{a[bc}\phi_{b'c']a'} = -\frac{4}{9} g_{a[b} \phi_{cb'c']a'} -
\frac{4}{9} g_{a'[b} \phi_{cb'c']a} - \frac{2}{9} \delta_{aa'} \phi_{[bcb'c']}
\ee
which we can prove in a local orthonormal frame.
Inserting (\ref{b3}) into (\ref{b2}) leads to
\be
\label{b4}
\int \sqrt{g}
\phi_{abc}
\partial_p B^{aa'} \partial_q B^{bb'} \partial_C h^{cc'}
\phi_{a'b'c'} =
-\frac{1}{9} \frac{\partial^3}{\partial t^C \partial s^p \partial s^q }
\int \sqrt{g} \phi^{abcd} B_{ab} B_{cd}
\ee
where it is crucial that we choose our coordinates $s^a$ such that the periods
of $B\wedge B$ along all four-cycles are purely quadratic expressions in terms
of the $s^p$ that do not depend on the $t^A$. We can rewrite (\ref{b4})
more compactly as
\be
\label{b5}
\int \sqrt{g}
\phi_{abc}
\partial_p B^{aa'} \partial_q B^{bb'} \partial_C h^{cc'}
\phi_{a'b'c'} =
-\frac{1}{216} \frac{\partial^3}{\partial t^C \partial s^p \partial s^q }
\int B\wedge B\wedge \phi
\ee
which is manifestly invariant under $B \rightarrow B+dV$. The expression
on the right hand side of (\ref{b5}) also appeared in \cite{leeleung} as
defining a nice quadratic form on the space of $B$-fields, here we see
that it arises naturally from the topological $G_2$ string. Also notice
that this term is purely cubic in the coordinates, so fourth and higher
derivatives of this terms will vanish identically.

The final generating functional of all correlation functions is an
extension of the Hitchin's functional to include the B-fields: \be
\label{sfinal} {\cal I}_{\rm tot}= \int \phi\wedge \ast \phi +
\frac{7}{72} \int B\wedge B \wedge \phi. \ee

\subsection{What are we quantizing?}
\label{wavefunction}

From the above discussion it seems clear that the prepotential
${\cal I}$ of the  topological string theory that we are studying
can be viewed as a  wave function in the Hilbert space that one
obtains by quantization of the symplectic space $H^2(M,\mathbb
R)\oplus H^3(M,\mathbb R)\oplus H^4(M,\mathbb R)\oplus
H^5(M,\mathbb R)$, with symplectic structure $\omega(\delta
\alpha,\delta\beta)=\int \delta \alpha \wedge \delta \beta$. For
the six-dimensional topological string, this point of view was
taken in \cite{relevantwitten}, see also \cite{gerasimov,erikv},
and it was shown that this is the natural way to understand the
holomorphic anomaly. In our case we do not have a holomorphic
anomaly, so it is not clear how compelling the interpretation of
${\cal I}$ as a wave function is, see also section~\ref{coupcon}.
Still, it is interesting to pursue this idea a little bit and
therefore we will now briefly study the wave function
interpretation restricting to the metric degrees of freedom only,
i.e. we restrict ourselves to $H^3 \oplus H^4$.

In order to be able to define suitable covariant derivatives we
first define a K\"ahler potential \be \label{defk} K=-\frac{3}{7}
\log {\cal I}. \ee This is, up to a numerical factor, precisely
the K\"ahler potential of the 4d theory obtained by compactifying
M-theory on a $G_2$ manifold (see section~\ref{treelevel}). In
fact, the expression in (\ref{b5}) corresponds to the gauge
couplings of the 4d theory\footnote{More precisely
\cite{Gutowski:2001fm}, the gauge couplings are proportional to
$\Bigl( t^A \frac{\partial^3}{\partial t^A
\partial s^p \partial s^q } \int B\wedge B\wedge \phi \Bigr) $ and
the $\theta$ terms are given by $\Bigl( p^A
\frac{\partial^3}{\partial t^A \partial s^p \partial s^q } \int
B\wedge B\wedge \phi \Bigr)$  where $p^A$ are moduli coming from
the $C$ field in M-theory:
\[
C=\sum_{a=1}^{h_2} A^a \wedge \partial_a B + p^A \partial_A \phi
\]
Here $A^a$ are the $h_2$ gauge fields in the four-dimensional theory.  \label{kk} We will come back to this in section \ref{treelevel}. }
 so that at tree level our topological string computes
both the K\"ahler potential and the gauge couplings of the low energy effective field theory.

We can use the K\"ahler potential to define a covariant derivative
\be \label{defdel}
\nabla_A \phi =\partial_A\phi  + \partial_A K \phi
\ee
which has the property that $\nabla_A \phi$ lives purely in the ${\bf 27}$
of $G_2$. In other words, the covariant derivative projects out the $G_2$
singlet contribution. Similarly, we can define a covariant derivative of
$\ast\phi$ via
\be \label{defdel2}
\nabla_A \ast \phi = \partial_A \ast \phi + \frac{4}{3} \partial_A K \ast \phi.
\ee

A useful observation is that \be \label{c1} \nabla_A \ast\phi = -
\ast \nabla_A \phi \ee which can be derived using the calculations
done in the preceding sections, but which also follows from the
identity \cite{hitchin} \be \delta \ast \phi =  \ast(\frac{4}{3}
\pi_1(\delta \phi) + \pi_7(\delta\phi) - \pi_{27}(\delta\phi) )
\ee where $\pi_1,\pi_7$ and $\pi_{27}$ are the appropriate
projections on the corresponding $G_2$ representations, and
$\delta\phi$ is an arbitrary variation.

Turning back to $H^3\oplus H^4$, we wish to consider the
quantization of this space with respect to the symplectic form
\be \label{sefsymp}
\omega = \int_M \delta \alpha_3 \wedge \delta \alpha_4
\ee
for $(\alpha_3,\alpha_4)\in H^3\oplus H^4$.

The simplest quantization, the analogue of the real polarization in the case of the
B-model, is to define
\be
p^A = \int_{C_A} \alpha_3, \qquad q_A = \int_{D^A} \alpha_4
\label{pa}
\ee
for which the symplectic form becomes simply
\be
\omega = \sum_A dp^A \wedge dq_A .
\ee
This structure is manifestly independent of the $G_2$ structure of the manifold, {\em i.e.} it is background independent.

Next, we introduce a different set of coordinates. We pick a fixed reference
$G_2$ structure $\phi$ and choose
\be
(\alpha_3,\alpha_4) = (x^A \partial_A \phi, y^A \ast_{\phi} \partial_A \phi).
\ee
We put the subscript $\phi$ on $\ast$ to indicate that this is defined wrt to the
reference $G_2$ structure. Notice that $\ast_{\phi} \partial_A \phi$ is closed,
this follows from the identity
\be \label{c2}
\ast \partial_A \phi = -\partial_A \ast \phi - \frac{7}{3}  \partial_A K \ast \phi
\ee
and since $d\ast\phi=0$ it is clear that $d \partial_A \ast \phi=0$ as well, so that the
right hand side of (\ref{c2}) is indeed closed.

Combining (\ref{sg5}) and (\ref{c2}) we find that the symplectic form becomes
\be
\omega = e^{-7K/3} \partial_A \partial_B K dx^A \wedge dy^B
\ee
so that after quantization
\be
[x^A,y^B]=-i e^{7K/3} K^{AB}
\ee
with $K^{AB}$ the inverse of $K_{AB}\equiv \partial_A \partial_B K$.

As we vary the background the quantization changes. The coordinate
$x^A$ is independent of the background (in fact, $x^A=p^A$ defined
in (\ref{pa})), since $\phi$ is linear in the background
coordinates $t^A$ (up to possible an exact form). However, $y^A$
changes. Its variation follows by imposing \cite{relevantwitten}
\be \frac{\partial \alpha_4}{\partial t^B} =0. \ee After some
straightforward algebra we obtain \be \label{c4}
\partial_A y^D - \frac{7}{3} \partial_A K y^D = - K_{ABC} K^{CD} y^B,
\ee
where $K_{ABC}\equiv \partial_A \partial_B \partial_C K$. It is interesting
to observe that the answers are naturally expressed in terms
of the K\"ahler potential $K$.

Equation (\ref{c4}) implies that $y$ eigenstates satisfy
\be \label{c5}
\partial_A |y\rangle = \left( - K_{ABC} K^{BD} \frac{\partial}{\partial y^D} Y^C +
\frac{7}{3} K_A \frac{\partial }{\partial Y^B} Y^B \right) |y\rangle .
\ee

The topological string wave function $\psi(y)=\langle \psi_{\rm
top}|y\rangle$ will then satisfy a similar differential equation,
given that $|\psi_{\rm top}\rangle$ does not depend on the choice
of background $G_2$ structure. This is the analogue of the
holomorphic anomaly for the $G_2$ string.

From here on there are many different polarizations one can study.
We can combine $x^A$ and $y^A$ in complex coordinates and work
with the corresponding coherent states, to be closer to what we do
in the case of a Calabi-Yau manifold. We can also separate out the
overall rescalings of the metric and parametrize \be
(\alpha_3,\alpha_4) = (\xi \phi + x^i \nabla_i \phi, \zeta \ast
\phi + y^j \ast_{\phi} \nabla_j \phi) \ee The symplectic form, in
these coordinates, becomes \be \omega=e^{-7K/3}\Bigl(d\xi \wedge d
\zeta + (\partial_i \partial_j K-\partial_i K \partial_j K) dx^i
\wedge dx^j \Bigr) \ee
 The rest of the analysis will be similar to what we did above and we will not work out the details
here. It will be an interesting question to see whether we can use these differential equations to
make an educated guess about the higher genus contributions to the wave function.

To summarize, the topological $G_2$ string can be viewed as a wave
function associated to a certain Lagrangian submanifold of the
symplectic space $H^2\oplus H^3\oplus H^4 \oplus H^5$. The
Lagrangian submanifold consists of the points \be
(B,\phi,\frac{7}{3} \ast_{\phi}\phi+\frac{7}{72} B\wedge
B,\frac{7}{36} B\wedge \phi) \ee where $\phi$ runs over the space
of $G_2$ metrics and $B$ over $H^2(M)$.

\subsection{Topological $\mathbf{G_2}$ strings on $\mathbf{CY\times S^1}$.}

An interesting example to study is the topological $G_2$ string on
$CY\times S^1$. Because of the $S^1$, this seven-manifold is not a
generic $G_2$ manifold. Whereas generic $G_2$ manifolds have no
supersymmetric two-cycles, $CY \times S^1$ does have such
two-cycles and therefore world-sheet instantons will contribute to
the theory. In addition, the analysis of the BRST cohomology will
be modified since $H^1(CY\times S^1,\mathbb R)=\mathbb R$. We will
postpone a detailed discussion of these issues to another
occasion, and here mainly focus on the metric and $B$-field moduli
of $CY\times S^1$.

Any manifold of the form $CY \times S^1$ has a natural $G_2$
structure of the form
\bea \label{g2struc}
\phi & = & {\rm Re}(e^{i\alpha} \Omega) + R \omega \wedge d\theta \nonumber \\
\ast\phi & = & R ~{\rm Im}(e^{i\alpha} \Omega)\wedge d\theta+
\frac{1}{2} \omega \wedge \omega \eea where $\theta$ is a periodic
variable with period $2\pi$, $e^{i\alpha}$ is an arbitrary phase,
$R$ is the radius of the $S^1$, and $\Omega$ and $\omega$ are the
holomorphic three-form and K\"ahler form on the Calabi-Yau
manifold. These are not completely independent, but have to obey
\be \label{cons1} i \int \Omega \wedge \bar{\Omega} = \frac{4}{3}
\int \omega\wedge\omega \wedge \omega. \ee

The $G_2$ BRST complex in say the left-moving sector, acting at
the level of zero modes, involves among other the following
differentials: \be \label{cycomplex} \Omega^0(M,\mathbb R)
\stackrel{d}{\longrightarrow} \Omega^1(M,\mathbb R) \stackrel{\ast
\phi \wedge d}{\longrightarrow} \Omega^6(M,\mathbb R)
\stackrel{d}{\longrightarrow} \Omega^7(M,\mathbb R). \ee where we
used the identification of the ${\bf 7}$ in $\Omega^2(M,\mathbb
R)$ with $\Omega^6(M,\mathbb R)$ and of the ${\bf 1}$ in
$\Omega^3(M,\mathbb R)$ with $\Omega^7(M,\mathbb R)$ (see table
(\ref{quantn})). The complex (\ref{cycomplex}) is equivalent to
(\ref{complex}) for any $G_2$ manifold. Thus, the full BRST
cohomology is obtained by combining two complexes of the form
(\ref{cycomplex}), one for the left-movers and one for the
right-movers. If we specialize to the case of a Calabi-Yau
manifold times a circle using (\ref{g2struc}), (\ref{cycomplex})
reduces to a certain complex involving the differential forms on
the Calabi-Yau manifold. We are not aware of any literature on
Calabi-Yau manifolds where such a complex appears, and this shows
that the topological $G_2$ twist is not in a straightforward way
related to the usual topological twist for Calabi-Yau manifolds.

More generally, complexes of the form (\ref{cycomplex}) can be
constructed for any special holonomy manifold by simply replacing
$\phi$ by a suitable covariantly closed differential form. It is
an interesting question whether such complexes give in general
rise to a new geometric understanding of special holonomy
manifold.

Turning back to the $CY\times S^1$ case, the metric moduli of
$CY\times S^1$ include the $2h^{1,2}$ complex structure moduli and
$h^{1,1}$ K\"ahler moduli of the Calabi-Yau, but also the radius
of the circle $R$. The total number of metric moduli is therefore
$\dim H^3(CY\times S^1,\mathbb R)-1$. The number of three-form
moduli is, however, equal to $\dim H^3(CY\times S^1,\mathbb R)$.
The difference is the parameter $\alpha$ in (\ref{g2struc}).
Strictly speaking $\alpha$ does not correspond to an element of
the BRST cohomology, and we should therefore remove the period of
$\phi$ corresponding to $\alpha$ from our consideration, but since
nothing turns out to depend on $\alpha$ we may as well work with
the full set of $\dim H^3(CY\times S^1)$ periods. The modulus $R$
on the other hand is physical, and this has some interesting
consequences for the relation between the topological $G_2$ string
and the A- and B-model topological string on the Calabi-Yau
manifold.

To study the topological $G_2$ string and its relation to the A- and
B-model, we choose a basis of three-cycles $A^I,B_I$ with intersection number $(A^I,
B_J)=\delta^I_J$ on the Calabi-Yau manifold. Similarly, we choose a basis of
two-cycles $C^a$ and dual four-cycles $D_a$. The cycles on $CY\times S^1$ are then
given by
\bea \label{homology basis}
{\rm two \,\, cycles} & : & C^a \nonumber \\
{\rm three \,\, cycles} & : & C^a \times S^1, \quad
 A^I,\quad B_I \nonumber \\
{\rm four \,\, cycles} & : & D_a, \quad B_I \times  S^1, \quad
-A^I \times S^1 \nonumber \\
{\rm five \,\, cycles} & : & D_a \times S^1 .
\eea
The prepotential of the topological $G_2$ string also depends on the $B$-field.
To take this account we need to improve the
four form to
\be
\ast\phi \rightarrow \phi^{(4)}\equiv - R {\rm Im}(e^{i\alpha}
\Omega)\wedge d\theta - \frac{1}{2} {\rm Re} (\omega+\frac{i}{2}
B) \wedge (\omega+\frac{i}{2} B).
\ee
The various periods, which define coordinates on the moduli space of $G_2$ metrics, are given by
\bea \label{per1}
b^a & = & \int_{C^a} B \nonumber \\
k^a & = & \int_{C^a\times S^1} \phi \nonumber \\
q^I & = & \int_{A^I} \phi \nonumber \\
p_I & = & \int_{B_I} \phi \nonumber \\
\frac{3}{7} \frac{\partial{\cal I}}{\partial k^a} & = &
\int_{D_a} \phi^{(4)} \nonumber \\
\frac{3}{7} \frac{\partial{\cal I}}{\partial q^I} & = &
\int_{B_I\times S^1} \phi^{(4)} \nonumber \\
\frac{3}{7} \frac{\partial{\cal I}}{\partial p_I} & = &
\int_{-A^I \times S^1} \phi^{(4)} \nonumber \\
\frac{1}{2} \frac{\partial{\cal I}}{\partial b^a} & = & \int_{D_a
\times S^1} B\wedge\phi . \eea Now, we want to relate these
variable to the quantities that appear naturally in the A and the
B models on the Calabi-Yau manifold. If we denote by ${\cal F}^A$
and ${\cal F}^B$ the suitably normalized prepotentials of the A-
and B-model, then these obey \bea \label{per2}
X^I & = &  \int_{A^I} \Omega \nonumber \\
\frac{\partial {\cal F}^B}{\partial X^I} & = & \int_{B_I} \Omega \nonumber \\
t^a & = & \int_{C^a} \omega +\frac{i}{2} B \nonumber \\
\frac{\partial {\cal F}^A}{\partial t^a} & = & \int_{D_a} (\omega+\frac{i}{2}B)^2
\eea
with $X^I$ and $t^a$ the complex structure and complexified K\"ahler moduli.
By comparing (\ref{per1}) and (\ref{per2}) we can now determine the relation between
${\cal I}$ and ${\cal F}^A$ and ${\cal F}^B$. This is somewhat subtle due to
the appearance of the parameter $R$ in $\phi$ and $\phi^{(4)}$. $R$ itself is not
an independent period but it appears in (\ref{per1}) in a non-trivial way. We should also
keep in mind that in (\ref{per2}) $\Omega$ and $\omega$ are constrained by (\ref{cons1}),
so that the variables $X^I$ and $t^a$ obey a nontrivial constraint.
To reformulate this constraint we denote
\be \label{aux12}
P(X^I,\bar{X}^I) = 3 i \int \Omega \wedge \bar{\Omega} , \quad
Q(t^a,\bar{t}^a) = 4 \int \omega^3
\ee
so that the constraint is that $P(X^I,\bar{X}^I)=Q(t^a,\bar{t}^a)$. A comparison of the periods yields the
following set of equations (we put $\alpha=0$ here, but it can be
trivially put back into the equations by replacing
$\Omega\rightarrow e^{i\alpha} \Omega$)
\bea \label{sys1}
b^a & = & 2 {\rm Im} (t^a) \nonumber \\
k^a & = & 2\pi R \,{\rm Re}(t^a) \nonumber \\
q^I & = & {\rm Re}(X^I) \nonumber \\
p_I & = & {\rm Re}(\partial_I {\cal F}^B) \nonumber \\
\frac{3}{7} \frac{\partial {\cal I}}{\partial k^a} & = &
-\frac{1}{2} {\rm Re}(\partial_a {\cal F}^A) \nonumber \\
\frac{3}{7} \frac{\partial{\cal I}}{\partial q^I} & = &
-2\pi R \,{\rm Im} (\partial_I {\cal F}^B) \nonumber \\
\frac{3}{7} \frac{\partial{\cal I}}{\partial p_I} & = &
2\pi R \,{\rm Im}(X^I) \nonumber \\
\frac{1}{2} \frac{\partial {\cal I}}{\partial b^a} &  = & 2 \pi
R\, {\rm Im}(\partial_a {\cal F}^A).
\eea

To solve this system of equations, we first express
$P(X^I,\bar{X}^I)$ in terms of $q^I,p_I$. As is well-known, in
terms of $q^I,p_I$ $P$ is equal to the Legendre transform of the
imaginary part of ${\cal F}^B$, \be \label{defpp} P(p_I,q^I) = 3 i
\int \Omega \wedge \bar{\Omega} = 12({\rm Im}({\cal F}^B) - p_I
{\rm Im}(X^I) )_{q^I  =  {\rm Re}(X^I),\,\,\, p_I  =  {\rm
Re}(\partial_I {\cal F}^B) } . \ee We cannot express
$Q(t^a,\bar{t}^a)$ in terms of $k^a$ directly, due to the factor
of $R$ that appears in the relation between $k^a$ and $t^a$.
However, the following is a function of just the $k^a$: \be
\label{defss} S(k^a) = 4 \int (2\pi R\omega)^3. \ee The constraint
$P=Q$ now implies that $R$ is a nontrivial function of
$q^I,p_I,k^a$, given by \be \label{defrr} 2\pi R(p_I,q^I,k^a) =
\left( \frac{S(k^a)}{P(p_I,q^I)} \right)^{1/3} . \ee We also
define \be \label{deftt} T(p_I,q^I,k^a,b^a) = 12 {\rm Re}({\cal
F}^A) _{t^a =\frac{k^a}{2\pi R(p_I,q^I,k^a)} + \frac{ib^a}{2} }
\ee so that \be \label{aux31} S(k^a) = (2\pi R(p_I,q^I,k^a))^3
T(p_I,q^I,k^a,b^a)_{b^a=0}. \ee

We now claim that \bea \label{finali} {\cal I} & = & 2\pi
R(p_I,q^I,k^a) \left( -\frac{7}{36} P(p_I,q^I) -
\frac{7}{72} T(p_I,q^I,k^a,b^a) \right) \nonumber \\
& = & -\frac{7}{3} (2\pi R) \left(
 ({\rm Im}({\cal F}^B) - p_I {\rm Im}(X^I) ) + \frac{1}{2} {\rm
 Re}({\cal F}^A) \right) .
\eea This shows that the prepotential of the topological $G_2$
string is indeed a combination of the $A$- and $B$-model
topological string, but the complex and K\"ahler moduli of the
Calabi-Yau manifold get mixed in a rather intricate way due to the
presence of the radius $R$. $R$ is closely related to the volume
of the Calabi-Yau manifold, and it would be interesting to see if
this is related to and/or can resolve the gravitational anomaly
found in the one-loop calculation in the six-dimensional Hitchin
system in \cite{recentwitten}. The non-trivial role that $R$ plays
in the above also manifests itself in the analysis of
four-dimensional supergravity, see e.g. \cite{recentlouis}.

To show that (\ref{finali}) solves (\ref{sys1}) is somewhat
complicated due to the dependence of $R$ on $p_I,q^I,k^a$.
However, one may check that \be \label{aux21} \frac{\partial {\cal
I}}{\partial (2\pi R)} = -\frac{7}{36} ( P(p_I,q^I)
-T(p_I,q^I,k^a,b^a)_{b^a=0}) \ee where it is important to
differentiate not just the explicit $R$ that appears in
(\ref{finali}), but also the $R$ that appears in the definition of
$T$ in (\ref{deftt}). The right hand side of (\ref{aux21}) is
precisely the original constraint (\ref{cons1}) and therefore
vanishes identically. In other words, the radius seems to play the
role of a Lagrange multiplier that imposes the volume constraint
(\ref{cons1}). Because of this, we can treat $R$ as a constant when
verifying (\ref{sys1}), and with this simplification it is
straightforward to verify that (\ref{finali}) solves (\ref{sys1}).

From (\ref{finali}) we also find, using (\ref{defrr}) and (\ref{aux31}), that
\be \label{prodform}
{\cal I}_{b^a=0} = -\frac{7}{12} S(k^a)^{1/3} P(p_I,q^I)^{2/3}.
\ee
Thus, the topological $G_2$ string is not just the sum of $A$- and $B$-model,
but it can also be written as the product of fractional powers of the $A$- and $B$-model.
It would be interesting to know whether either the combinations (\ref{finali})
and (\ref{prodform}) have any distinguished meaning for six-dimensional topological strings.



\section{The Topological $\mathbf{G_2}$
String}\label{topologicalstrings} We have so far been considering
a topologically twisted $\sigma$ model of maps from a sphere into
a $G_2$ manifold. However, on higher genus Riemann surfaces, there
is nothing interesting to compute in the $\sigma$-model. To get
interesting amplitudes, we need to couple the $\sigma$ model to
two dimensional gravity, and integrate over the moduli space of
Riemann surfaces. This will define the topological $G_2$ string.
In the following, we first give a preliminary discussion the
topological $\sigma$-model at higher genus and then construct a
measure on the moduli space of Riemann surfaces to define the
topological string amplitudes.

\subsection{Twisting the $\mathbf{\sigma}$ Model At Higher Genus.}\label{higherg}

Generalizing the sphere computation to higher genera
\cite{kodira,antoniadis}, n-point correlators on a genus-g Riemann
surface in the twisted theory are defined as a correlator in the
untwisted theory of the same n operators plus $(2-2g)$ insertions
of the spin-field that is related to the space-time supersymmetry
charge. For a Calabi-Yau 3-fold target space on a Riemann surface
with $g>1$ the meaning of the above prescription is to insert
$2g-2$ of the conjugate spectral flow operator ($e^{-i{\sqrt{d}
\over 2} \phi}$ in the notation of section \ref{cytwist}). To
generalize this to the $G_2$ situation, we will do something
similar.  However,
 there is only a single $G_2$ invariant spinfield. This is where the decomposition in conformal
blocks in section \ref{tim} is useful: the spin-field $\Phi_{1,2}$
(which corresponds to the particular Ramond sector ground state
$|{ 7 \over 16},0 \rangle $) could be decomposed  in a block
$\Phi_{1,2}^+$ and in a block $\Phi_{1,2}^-$ (see eq
\ref{confblockspin} and \ref{threepnine}, and also
section~\ref{GSO}). At genus zero we needed two insertions of
$\Phi_{1,2}^+$, so the natural guess is that at genus $g$ we need
$2g-2$ insertions of $\Phi_{1,2}^-$. We will demonstrate shortly
that with this guess the topological $G_2$ strings are indeed
``critical" in 7 dimensions.

\subsection{Topological Strings}
To go from a topological $\sigma$ model  to topological strings,
we need to integrate over the moduli space of Riemann surfaces,
${\cal M}_g$. To construct a measure on the moduli space of
Riemann surfaces, we need an anti-ghost $G_*^\uparrow$, such that
$\{Q,G_*^\uparrow\}=T$ where $T$ is the twisted stress tensor and
$Q$ is the BRST operator. We use the notation $G_*^\uparrow$ for
the anti-ghost because the conformal block $G^\uparrow$ defined
previously almost does the job, as discussed in section
\ref{twistedvirasoro}. In the following, we assume that a suitable
modification $G_*^\uparrow$ of $G^\uparrow$ exists which we can
use to define the topological string amplitudes. With this
important assumption
 we can define the genus-g free energy $F_g$ of the $G_2$ topological string by integrating over the $3g-3$ dimensional moduli space of genus-g Riemann surfaces $\mathcal{M}_g$ along with $3g-3$ insertions of the anti-ghost folded against Beltrami differentials giving the appropriate measure of integration
\be\label{effgee}
F_g=\int_{\mathcal{M}_g}\langle\prod_{i=1}^{3g-3} |(\mu_i, G_*^{\uparrow})|^2\rangle_{g}
\ee
where the folded anti-ghosts are defined by integrating them over the genus-g worldsheet against the Beltrami differentials\ $(\mu_i, G_*^{\uparrow})=\int d^2z~\mu_i(z) G_*^{\uparrow}(z)$.

\paragraph{Critical Dimension}
The usual topological strings on Calabi-Yau manifolds have a
``critical dimension" $d=6$ (complex dimension 3). This is because
essentially all the higher genus free energies $F_g$  vanish when
the target space is a complex manifold of (complex) dimension
other than 3. The $G_2$ string is critical in 7 dimensions.
Indeed, we can use the fusion rules of the tri-critical Ising
model to show that there is a non-vanishing contribution to
correlation functions of $2g-2$ $\Phi_{1,2}$'s and $3g-3$
$G^\uparrow$. We can also show that their correlation functions
are non-zero by considering the Coulomb gas representation of the
tri-critical Ising model (which is useful to compute correlation
functions). From that perspective the $2g-2$ insertions of
$\Phi_{1,2}^-$ and $3g-3$ insertions of $G^\uparrow$ yield a total
$\phi$ charge of \be\label{critG2} (2-2g){5\over
2\sqrt{10}}+(3g-3){2\over \sqrt{10}}=(g-1){1\over \sqrt{10}}\ee
which is exactly the correct amount needed to cancel the existing
background charge (${1 \over \sqrt{10}}$)  of the tri-critical
Ising model on a genus-g Riemann surface. Here we used that the
anti-ghost $G^{\uparrow}$ has weight two in the Coulomb gas
representation (see appendix A).

The $G_2$ topological string partition function is defined as an
asymptotic series in a coupling constant $\lambda$ \be
\mathcal{Z}=e^{\mathcal{F}},\quad \mathtt{where}\quad
\mathcal{F}=\sum_{g=0}^{\infty}\lambda^{2-2g}F_g. \ee The descent
relations introduced in section \ref{descre}\ enable us to now
define correlation functions of chiral primaries just like in the
${\cal N}=2$ topological string.

\section{Physics in Three Dimensions}

Since we are discussing type II string theory compactified on a
manifold of $G_2$ holonomy, we expect the topological $G_2$ string
to be of relevance for the resulting three-dimensional effective
field theory. In this section we will explore some properties of
this effective field theory and how they are related to
topological $G_2$ strings. Since $G_2$ compactifications preserve
four supercharges, the resulting three-dimensional theory will
have ${{\cal N}=2}$ supersymmetry.

\subsection{Massless fields and the GSO projection}
\label{GSO} We are dealing with an odd dimensional
compactification of string theory. Therefore, the GSO projection
is particularly subtle. In order to define it, we need a notion of
fermion number. We will first define this in the NS sector of the
internal CFT corresponding to the sigma model on the $G_2$
manifold. As discussed in some detail in \cite{sv}, we can assign
a fermion number to a state by assigning a fermion number to the
tri-critical Ising part of the state. In the NS sector, there is a
tri-critical Ising model notion of fermion number in which we
associate fermion number $(-1)^{n+1}$ for states in Hilbert space
${\cal H}_n$. with  $n=1,\ldots,4$ ($n=1$ corresponding to the
identity, $n=2$ to the primary ${1 \over 10}$ etc) . The fermion
number in the 3d spacetime part of the compactification in the NS
sector is the usual one.

In the R sector, things are less straightforward. In three
dimensions, the representations of the Clifford algebra are
two-dimensional, and there are no chiral spinors. The same holds
true in seven dimensions. Therefore, in order to have a
well-defined fermion number, we need to take a reducible
representation of the Clifford algebra in three dimensions which
consists of two spinors which we will call $|3,+\rangle$ and
$|3,-\rangle$ where the sign indicates fermion number. Similarly,
we need two spinors coming from the seven-dimensional part, which
we will call $|7,+\rangle$ and $|7,-\rangle$. The zero modes of
the three-dimensional fermions map $|3,+\rangle$ to $|3,-\rangle$
and vice versa. With this doubling we have a well defined action
of $(-1)^F$ given by $(-1)^F |3,\pm\rangle = \pm |3,\pm\rangle$. A
similar remark applies to the seven-dimensional part. When we
combine the three and seven-dimensional part, we find that if we
take all possible combinations, we obtain a reducible
representation. The smallest irreducible representation, which
still allows for a proper action of $(-1)^F$, is obtained by
taking e.g. the combinations \bea
|\chi,+\rangle & = & |3,+\rangle \otimes |7,+\rangle + |3,-\rangle \otimes |7,-\rangle \nonumber \\
|\chi, -\rangle & = &  |3,+\rangle \otimes |7,-\rangle +
|3,-\rangle \otimes |7,+\rangle \eea where fermion number acts as
$(-1)^F |\chi,\pm\rangle =\pm |\chi,\pm \rangle$. The GSO
projection projects on one of the two chiralities and results in a
single two component spinor in three dimensions. From the right
movers we get another two-component spinor and this is how we
arise at $N=2$ supersymmetry in three dimensions. \footnote{
Notice that this also resolves the peculiar feature that
representations in the R sector (discussed in appendix
\ref{ramond}) of the $G_2$ algebra can be one-dimensional, but
once we combine left and right movers they should be
two-dimensional. As the above shows, the R sector really involves
two-dimensional representations, and the left-right sector
four-dimensional ones. No strange enhancement is necessary once we
combine left and right movers.}

If we just quantize the seven-dimensional sigma model, the above suggests that we get two
copies of each R representation, together with a label $\pm$.
The natural interpretation from the point
of view of the tri-critical Ising model, is that $\pm$ corresponds to the decomposition of $R$ ground states
in two conformal blocks. In this way, the fusion rules of the tri-critical Ising model can be made to
agree with the fermion number assignment, up to an extra minus sign for the product of
two fields in the RR sector. For example,
\bea
\left[ \frac{7}{16},\pm \right] \otimes \left[ \frac{7}{16},\mp \right]  &= & \left[ 0,+ \right] \nonumber \\
\left[ \frac{7}{16},\pm \right] \otimes \left[ \frac{7}{16},\pm \right]  &= & \left[ \frac{3}{2},- \right] \nonumber \\
\left[ \frac{7}{16},\pm \right] \otimes \left[ \frac{3}{80},\mp \right]  &= & \left[\frac{6}{10},+  \right] \nonumber \\
\left[ \frac{7}{16},\pm \right] \otimes \left[ \frac{3}{80},\pm
\right]  &= & \left[ \frac{1}{10},- \right] ,
\eea
etcetera.

Using these fusion rules, it is easy to see that tree level
correlation functions only vanish if the total $(-1)^F$ of the
operators in the correlation function is equal to $(-1)^p$, where
$p=n_R/2$ is half the number $n_R$ of R fields. This applies to
both the left and right movers separately. At higher genus
correlation functions also involve a choice of spin structure.

We can now also properly define operators like $G^{\downarrow}$
and $G^{\uparrow}$ in the R sector. We decompose the R Hilbert
space as
\be {\cal H}_R \equiv
{\cal H}_{R,1} \oplus {\cal H}_{R,2} \oplus {\cal H}_{R,3} \oplus
{\cal H}_{R,4} = {\cal H}_{\frac{7}{16},+} \oplus {\cal
H}_{\frac{3}{80},-} \oplus {\cal H}_{\frac{3}{80},+} \oplus {\cal
H}_{\frac{7}{16},-}
\ee
and define the up and down projections exactly as in the case of
the NS sector in terms of the action on ${\cal H}_i$. For example,
$G^{\downarrow}$ will only map ${\cal H}_i \rightarrow {\cal
H}_{i+1}$.

\subsection{Relation of the Topological $\mathbf{G_2}$ String to Physical Amplitudes}

An important application of topological strings stems from the
realization \cite{Witten,kodira,antoniadis} that its amplitudes
agree with certain amplitudes of the physical superstring. The
usual topological strings on Calabi-Yau manifolds  compute F-terms
in four dimensional compactification of the physical superstrings.
A natural question is: What physical amplitudes does the
topological $G_2$ string compute  in three dimensional ${\cal
N}=2$ compactifications of superstring theories. As we will see,
at genus zero, the topological string indeed computes certain
Yukawa couplings. However, at higher genus, unlike the usual
topological string theories, the topological $G_2$ string does not
compute F-terms in three dimensions. As we will see, this failure
to compute such terms can be traced to the absence of chiral
spinors in three dimensions.

Comactification of type II superstrings on $G_2$ holonomy
manifolds leads to ${\cal N}=2$ supergravity in three dimensions,
where a single supercharge arises from each world sheet chirality.
The (e.g. left moving) supersymmetry generator is constructed
according to the standard FMS ansatz \cite{fms}\
\be\label{spinfield1} Q^{\alpha}=\oint e^{-{\varphi\over
2}}\Bigl(S_{3+}^{\alpha}\Sigma_+ + S_{3-}^{\alpha}\Sigma_- \Bigr)
\ee where $S_{3{\pm}}$ is a spin-field in $R^{1,2}$ (corresponding
to the states $|3, \pm \rangle$ in section \ref{GSO}) and
$\Sigma_{\pm}$ are operators corresponding to the states $|{7},
\pm \rangle$  in section \ref{GSO}.  Also,  $\varphi$ is the
bosonized super-ghost arising in the standard BRST quantization of
type II superstrings.
%

Which physical amplitudes can we possibly relate to the
topological string? These should be amplitudes  involving Ramond
sector vertex operators which, in their $G_2$ factor have the
field $\Sigma_-$ inserted an appropriate number of times to give a
topological amplitude.\footnote{In the case of Calabi-Yau 3-folds,
analogous amplitudes which are related to the topological string
consist of $2g-2$ gravi-photons, which suggests a F-term in the
four dimensional effective action of the form $W^{2g}$, where $W$
is the Weyl super-multiplet of ${\cal N}=2$ supergravity. Here,
$W$ is the chiral superfield of ${\cal N}=2$ supergravity
multiplet whose first component is the graviphoton field strength
$T_{\mu \nu}$. In components, the $W^{2g}$ term gives a coupling
between two gravitons and $2g-2$ graviphotons: $R^2 T^{2g-2}$, and
it can be shown that the coefficient of this term is the
topological string partition function $F_g(t,\bar{t})$.} In
addition, in order to have some non-trivial dynamics in three
dimensions, we need a field which sits in $(\bf{3},\bf! {1})$ of
$SO(3) \times G_2 \subset SO(10)$.  A singlet under the $SO(3)$
factor would imply a non-dynamical degree of freedom in three
dimensions.

\paragraph{The RR sector}
The RR vertex operators have spinor bilinears. We are looking for
singlets under $G_2$. These will come from the spinor bilinears
made out of the covariantly constant spinor on the $G_2$ manifold.
As discussed before, this can only generate a three form or a four
form. All other combinations vanish. Then, there remains a unique
field which sits in the $(\bf{3},\bf{1})$ of $SO(3) \times G_2$.
For type IIA and type IIB, this corresponds to a scalar field
$\rho$ such that \be \label{j301}\begin{split}
{\rm type\ IIA}\quad\quad\partial_{\mu}  \rho &= \int_{M_7} F^{(4)}_{RR} \wedge\ast \phi,\\
{\rm type\ IIB}\quad\quad\partial_{\mu} \rho &= \int_{M_7} F^{(5)}_{RR} \wedge \phi .\end{split}
\ee
where $\phi$ is the 3-form that defines the $G_2$ structure.
The vertex operator (in type IIB)  corresponding to these spacetime fields in the $-1/2$ picture is
\be\label{gravphot}
V^i= e^{-{\varphi+\tilde{\varphi}\over 2}}\Bigl( S_{3+}^{\alpha} (\tau^i_{\alpha \beta}) \tilde{S}_{3+}^{\beta} \Sigma_+\tilde{\Sigma}_+ + S_{3-}^{\alpha} (\tau^i_{\alpha \beta}) \tilde{S}_{3-}^{\beta} \Sigma_-\tilde{\Sigma}_- \Bigr)  \ee
where (non)tilde denotes (left) right-movers and $\tau^i$ are the Pauli matrices.\footnote{For type IIA, we need to change $\tilde{\Sigma}_{\pm}$ to ${\tilde{\Sigma}_{\mp}}$.}

At first sight, it might seem that $2g-2$ insertions of this
operator would twist the $G_2$ part of the CFT by appropriate
insertions of the spin field $\Sigma_-$. However, this is of
course incorrect, because the vertex operator in (\ref{gravphot})
is a sum of two terms. Therefore,  in addition to getting terms
with $\Sigma_-^{2g-2}\tilde{\Sigma}_-^{2g-2}$  which can be simply
related to the topological amplitudes, we get terms with
$\Sigma_+^{2g-2}\tilde{\Sigma}_+^{2g-2}$ insertions and also all
possible cross terms which are non-topological in nature. At a
generic genus, generally these non-topological terms are
non-vanishing, with the result that the total amplitude is
non-topological in nature. For type II strings on Calabi-Yau
manifolds, there is a natural way to restrict to one of the two
terms in such a vertex operator \ref{gravphot}, and that is by
looking at self-dual (or anti self-dual) graviphoton field
strengths. In three dimensions, there is no natural way to
restrict to one of the two terms in the vertex operator.
Therefore, we conclude that generically, the topological string
does not seem to compute F-terms in the three dimensional
effective action. There is an exception, though, at genus 0.

\subsection{Tree level effective action and the topological $\mathbf{G_2}$ string}
\label{treelevel}

In order to describe the three-dimensional effective action it is
convenient to first work with 11d supergravity compactification on
$G_2$ manifolds down to four dimensions. The three dimensional
action can then be obtained by a dimensional reduction. The four
dimensional theory has $b_3$ chiral multiplets and $b_2$ vector
multiplet. The scalars in the chiral multiplets are complex
combinations of the metric moduli and the three form 11
dimensional $C$-field moduli: $S^A=t^A+i p^A$, where $p^A$ is
defined in footnote \ref{kk}. The K\"ahler potential for the
scalars is a function of the real part of $S^A$ and is given by
\cite{Gutowski:2001fm} \be K(S+\bar{S})= -3 \log ({1 \over 7} \int
\phi \wedge \ast \phi) \ee The kinetic terms for the $b_2$ gauge
fields are given by \be {\mathrm{Im}}~ \int d^4 x d^2 \theta
~\tau_{ab} ~W_\alpha^a W^{\alpha b} \ee which can be dimensionally
reduced to three dimensions \be S  = {\mathrm{Im}} \int d^3 x d^2
\theta \tau_{ab} W^a_\alpha W^{b~\alpha} \label{vec3d} \ee where
$W_\alpha^a$ is the field strength superfield, the gauge coupling
is $\tau_{ab}= S^A{\partial_A\partial_a
\partial_b \Bigl(\frac{36}{7}~ {\cal I_{\rm tot}} \Bigr)}$, where ${\cal
I}_{\rm tot}$ is defined in eq (\ref{sfinal}) and
$\partial_a={\partial \over {\partial s^a}}$.


This action is written in terms of dimensionally reduced 4d vector
multiplet as an integral over a chiral half of superspace. In 3
dimensions, vectors multiplets are dual to the chiral multiplet
and it is interesting to determine the K\"ahler potential for
these chiral multiplets. To this end, we need to perform the
duality transformation and it is convenient to do this directly in
superspace. Four-dimensional vector multiplets are not the most
convenient way to define gauge theories in three dimensions. Gauge
theories in three dimensions are usually formulated in terms of
linear multiplets. We therefore first rewrite (\ref{vec3d}) in
terms of linear multiplets $G^a$ in terms of which the action
becomes \be S= \int d^3x ~d^4 \theta (\tau_{ab}(S) +
\bar{\tau}_{ab}(\bar{S})) G^a G^b \label{ourcase}. \ee We can
write the B-field as $G^a \omega_a$ and $\phi=  (S^A +
\bar{S}^A)\chi_A$, where $\omega_a$ and $\chi_A$ are  bases of
$H^2$ and $H^3$ respectively,  of the $G_2$ manifold.  Then, the
superspace action can be formally written as \be S=\int d^3x
~d^4\theta  ~\int B \wedge B \wedge \phi \label{baction} \ee which
is exactly the second term which appears in ${\cal I}_{\rm
total}$.

To perform the duality transformation explicitly between the
linear and the chiral multiplets (see e.g. \cite{deboeroz}), we
can even start from a more general action \be S=\int d^3 x ~d^4
\theta f(G^{a},S, \bar{S}) \label{generalaction} \ee This action
can be rewritten as \be S=\int d^3x ~d^4\theta~ f(\tilde{G}^a, S,
\bar{S}) -\tilde{G}^a (Y_a + \bar{Y}_a) \label{dualityaction} \ee
where the superfields $\tilde{G}^a$ are unconstrained real
superfields, and the $Y_a$ are chiral superfields. Extremizing the
action with respect to $Y_a$ constrains $\tilde{G}^a$ to be linear
superfields from which we obtain (\ref{generalaction}) back. We
can also vary this action with respect to $\tilde{G}^a$ which
yields the equation \be Y_a +\bar{Y}_a={\partial f(\tilde{G}^a, S,
\bar{S}) \over \partial \tilde{G}^a} \ee By solving for
$\tilde{G}^a$ in terms of $S$ and $\bar{S}$ and substituting in
(\ref{dualityaction}) gives the dual description in terms of a
K\"ahler potential $K(Y_a+\bar{Y}_a, S, \bar{S})$ for the chiral
multiplets $Y_a$: \be S= \int d^3 x ~d^4 \theta~K(Y_a +\bar{Y}_a,
S, \bar{S}). \ee Here, $K$ is the Legendre transform of $f$. For
our case (\ref{ourcase}), $f= \bigl(\tau_{ab}(S)+\bar{\tau}_{ab}
(\bar{S})\bigr) \tilde{G}^a \tilde{G}^b$, so \be K(Y_a
+\bar{Y}_a,S, \bar{S})=  (Y_{a}+\bar{Y}_a)
\bigl(\Re\tau(S)^{-1}\bigr)^{ab} (Y_{b}+\bar{Y}_b) \ee This is
simply the Legendre transform of (\ref{baction}) with respect to
the $B$ field moduli.

\section{Discussion, open questions and future directions}

In this concluding section, we list and discuss several interesting issues and future
directions.

\subsection{The coupling constant}
\label{coupcon}

The partition function for the ordinary topological string on
Calabi-Yau manifolds is better thought of as a wave function. This
picture   emerges from the holomorphic anomaly, where the
holomorphic anomaly equation is interpreted as describing the
change in basis (an infinitesimal fourier transform) in the
quantum mechanics whose phase space is given by $H^3(M)$
\cite{relevantwitten}. It remains an interesting question whether
the partition function of our topological string should naturally
have a wave function interpretation. In our case, there is no
corresponding holomorphic anomaly equation. Also, when we consider
our topological string on CY $\times ~ S^1$, it naturally contains
both the holomorphic and anti-holomorphic A and B models. These
facts  suggests an interpretation as a partition function as
opposed to a wave function.

However, we also argued in section \ref{wavefunction} that we
could view the topological $G_2$ string as a wavefunction
corresponding to a lagrangian submanifold of $H^2+H^3+H^4+H^5$.
From this perspective, it is interesting to note that we can
naturally incorporate the string coupling in the framework.
Consider again our function \be {\cal I} = \frac{1}{g_s^2} \int
\phi\wedge \ast\phi + \frac{7}{24 g_s^2} \int B\wedge B \wedge
\phi \ee where we have now included the string coupling constant.
We can associate to it a Lagrangian submanifold of $H^{\ast}(M)$
which now also includes $H^0$ and $H^7$, namely \be
\Bigl(\frac{1}{g_s},B,\phi,\frac{\partial {\cal I}}{\partial\phi},
\frac{\partial {\cal I}}{\partial B}, \frac{\partial {\cal
I}}{\partial \frac{1}{g_s}}\Bigr) \ee In this way the string
coupling gets naturally associated to $H^0(M)$. This is similar to
what is done in the A model.  In the B model, the string coupling
is related to one particular component of $H^3$, namely the one
proportional to the holomorphic three form. At first sight, it
does not seem to be the case here. However, as discussed in
Appendix \ref{decomposition}, there is an isomorphism between
$H^0$ and $H^3_1$, {\em i.e.} those elements of the third
cohomology which transform as the singlet under the group $G_2$.
The moduli space has a projective structure. We can view the $t^A$
defined in (\ref{periodphi}) as providing real projective
coordinates on the $b^3_{27}=b_3-1$ dimensional moduli space of
$G_2$ metrics which correspond to deformations of the $G_2$
structure which are not rescalings of the metric. The partition
function of the topological $G_2$ string is then a section of a
real line bundle of degree ${7 \over 3}$. Though this is not the
structure that we find in the topological string, it may naturally
emerge when we try to lift it to M-theory.



\subsection{Strong coupling limit}
The construction of the topological string theory that we have given
is a perturbative one. The strong coupling limit and a non-perturbative
completion remains an interesting question. A strong coupling limit, if
well defined, could naturally be topological M-theory
\cite{gerasimov,vafa,nekrasov}.
 An obvious strong
coupling limit is one where we scale $\phi$ with $\lambda^{3/7}$
and $g_s$ with $\lambda$, after which we send $\lambda\rightarrow
\infty$. This does not change the form of ${\cal I}$. It is not
clear whether the result should be viewed as a string theory. In
fact, it is perhaps more appropriate to think of this topological
theory as describing certain sector of M-theory compactification
on $G_2$ manifolds down to 4 dimensions. The number of variables
that remain will be one-less compared to the number of variables
in three dimensions -- we lose the degree of freedom corresponding
to the rescaling of $\phi$, the three-form which defines the $G_2$
structure; or equivalently, the string coupling.

Another limit we can study is the theory on $CY \times S^1$. In
this case we can try to decompactify the $S^1$, which is related
via a 9-11 flip to the strong coupling limit above. Since $R$
depends non-trivially on all moduli, it is not immediately clear
what is a natural set of variables that survives. Perhaps we
should keep all $H^3$ except the class proportional to $\phi$, as
we do for the complex structure in the B-model?

\subsection{Relation to black holes and Hitchin flows}

Notice that our function $P(q^I,p_I)$ (eq. \ref{defpp}) is the Legendre
transform of the free energy of the B-model, which is  exactly
the expression that appears in the recent discussions of the
relation between topological strings and black hole entropy
\cite{osv}. This is perhaps not that surprising given that
$P(q^I,p_I)$ is the volume of the
CY at the horizon of the black hole through the attractor
mechanism.
 Yet, one may wonder
whether the circle in the 7d theory on $CY \times S^1$ can be
interpreted as a Euclidean time direction so that the theory can
be directly viewed as a thermal system with nonzero entropy, giving
a microscopic description of the black hole entropy. Perhaps our topological
twist can be interpreted as counting BPS states in a black hole background.

In \cite{mohaupt}, domain wall solutions of ${\cal N}=2$ gauged four-dimensional supergravity
were constructed, where the supergravity theory was obtained by the dimensional
reduction of type IIA on ``half-flat'' six manifolds. These are manifolds which
have a particular type of $SU(3)$ structure. The domain walls are determined
by flow equations which govern the dependence of scalars (corresponding to the
moduli of the internal manifold) in the direction transverse to the domain wall. These
flow equations were shown to be equivalent to Hitchin's flow equations, which implies
that the transverse direction to the domain wall combines with the internal manifold
to give a $G_2$ manifold. A natural question is  whether the black
hole attractor flows have a similar interpretation in terms of Hitchin flows which
may then admit a re-interpretation of these in terms of a manifold with $G_2$ structure.
We leave this interesting point for a future investigation.

Notice that in M-theory on $G_2$ manifolds there are no
supersymmetric black holes, so we do not expect the existing relation between
topological strings and BPS black holes to generalize to this setup.

\subsection{An analogue of KS theory?}
The topological A and B model are defined perturbatively in an on
shell formalism which studies maps from the world sheet to a
target space. Perturbative computations can be done using
world-sheet methods. However, for the B-model, there is a target
space ``string field theory'' (though for the B-model, this
reduces to a field theory),  namely the Kodaira Spencer theory
which presumably yields exactly the same results as the
world-sheet calculations. This is a theory of complex structure
deformations of the Calabi-Yau manifold. The fundamental variable
of Kodaira Spencer theory corresponds to an infinitesimal change
of the complex structure of the Calabi-Yau manifold. The equation
of motion of this theory is equivalent to the complex structure
being integrable. The action, which can be  written down by
following the standard rules of string field theory
\cite{wittencsasst}, consists of a  quadratic kinetic term and a
cubic interaction term. There are no higher point interaction
terms since four and higher point correlation functions in the
world sheet theory vanish.

One may hope that the target space theory of the topological $G_2$
string is a seven dimensional  theory of deformations of $G_2$
structures, a version of the Kodaira Spencer theory that lives in
seven dimensions. The fundamental variable should be an
infinitesimal metric deformation, i.e. a symmetric two-tensor
$A_{\mu\nu}$. If we again follow the standard  string field theory
logic, the action would take the form \be S=S_2(A)+S_3(A) \ee with
$S_2(A)\sim \int A \frac{G_0^-}{b_0^-} A= \int A
\frac{G^\downarrow_0}{G^\uparrow_{*0}}A$  and with \be S_3(A)=\int
d^7x \, \sqrt{g} \phi^{\alpha\beta\gamma} A_{\alpha\alpha'}
A_{\beta\beta'} A_{\gamma\gamma'} \phi^{\alpha'\beta'\gamma'}. \ee
The equation of motion of this theory, if correct, should
correspond to the equation for integrability of $A$ to a $G_2$
metric. Such a quadratic equation is unknown to us so it would be
interesting to study further. Notice that for the A-model such a
simple cubic theory does not exist.

There is yet another theory in the case of the B-model which has
been proposed as a possible equivalent space-time theory, which is
a six-dimensional Hitchin functional. This is proposed in
\cite{vafa} and studied and refined in \cite{recentwitten}. In the
latter paper it is also pointed out that the six-dimensional
Hitchin theory has a one-loop gravitational anomaly which again
suggests that complex and K\"ahler moduli cannot be treated
independently. This agrees nicely with the analysis of our model
on $CY\times S^1$ and clearly it is worth trying to understand
whether our theory on $CY\times S^1$ is free of any such one-loop
anomalies. What is confusing and begs for clarification is the
fact that the six-dimensional theory has a Kodaira Spencer
formulation and a Hitchin formulation and both are supposed to
reproduce the prepotential (see also \cite{gerasimov}), whereas in
seven dimensions, we only have the prepotential itself and that is
the Hitchin functional. It would be quite interesting if the 7d
Hitchin functional would also be the effective spacetime theory,
since that would mean that prepotential obtained from Hitchin's
functional would again be Hitchin's functional. We clearly need to
sort all this out if we want to make progress in ``topological M
theory" (see also \cite{gerasimov,vafa,nekrasov}).

\subsection{Branes}

Though our theory does not have world-sheet instantons (since
there are no supersymmetric 2-cycles), it does have supersymmetric
branes, namely $0,3,4$ and $7$-branes, that will give rise to
non-perturbative corrections. Presumably, the formulation of
topological M-theory is in terms of topological membranes.
However, strings and membranes are dual in seven dimensions. It is
for these reasons that the 3 brane is specially interesting. Its
world-volume theory is a candidate topological membrane theory
that might give rise to an alternative definition of a 7d theory (
see also \cite{harveymoore,beaslywitten} for further discussions
of membranes in $G_2$ manifolds). In some examples one can see
that membranes should play an important role. For example, if one
considers topological strings on orientifolds of CY
compactifications, one finds a version of Gromow-Witten invariants
coming from oriented and unoriented string world-sheets. As the
theory is equivalent to M-theory on $(CY\times S^1)/\mathbb Z_2$,
from the M-theory point of view we are counting membranes wrapping
the $S^1$ \cite{marinoetc}. We leave a detailed discussion of the
branes in the theory to a future publication.

\subsection{Open problems and future directions}

There are several further open problems. Perhaps the most
important one is to find a twisted stress tensor which is crucial
for the definition of the topological string beyond genus zero. It
is also interesting to understand the geometric meaning of the
higher genus amplitudes. In the case of the A-model, the higher
genus amplitudes roughly compute the number of holomorphic maps
from a genus $g$ Riemann surface into the Calabi-Yau. Such an
interpretation is less clear for the B-model for $g>1$ (the genus
$0$ result reproduce the special geometry relations and the genus
1 result is related to the holomorphic  Ray-Singer torsion). For
example, are there interesting indices (like the elliptic genus)
that we can define and study in this context? Perhaps related to
this, we would like to understand better the localization
arguments.


Mirror symmetry for $G_2$ manifolds will be interesting to investigate
in the context of our topological twist. A version of mirror symmetry for
$G_2$ manifolds was studied in \cite{roiban,gaberdiel,9707186,Aganagic}. In \cite{roiban}, an
analogue of Witten index was introduced that counts the total number of ground states
and not just ground states weighted with $(-1)^F$, where $F$ counts the fermion number.
This was defined by using a $Z_2$ automorphism $L$ of the $G_2$ algebra under which the currents
$K$ and $\Phi$ change signs, and the index was defined as ${\mathrm {Tr}}(L (-1)^F)$.
This index will count the total number of chiral primary states in our topological theory.
In fact, in \cite{gaberdiel}, it was argued that acting with $L$ in the left sector and the
identity in the right sector corresponds to the mirror automorphism of the $G_2$ algebra, which
can then be geometrically interpreted as mirror symmetry for $G_2$ manifolds.

We list several other related questions that still remain open.
For example, are there other relations to the low energy effective
action? Is there a Berkovits formulation in three dimensions? Is
the Dolbeault-like complex for $G_2$ manifolds that corresponds to
the BRST cohomology in the left or the right sector  useful in
other contexts? It is also perhaps worthwhile to investigate more
concrete world-sheet models of theories based on the $G_2$
algebra, for example using minimal models and discrete torsion,
see e.g.
\cite{0108091,0110302,0111012,0111048,Aganagic,roiban,0204213,0301164,gaberdiel}.
It is also interesting to extend this construction to more general
setting which involve turning on the NS-NS background fields. As
discussed in \cite{witt}, this setup involves a study of $G_2
\times G_2$ structures, and it would be interesting to understand
how our topological twist is modified in this context.

A natural extension of this work is to study topological strings on spin(7) manifolds. This
may reveal interesting extensions of Hitchin's functionals to such manifolds. We will report
these results elsewhere \cite{spin7}.

\subsection*{Acknowledgments}

It is a pleasure to thank Nathan Berkovits, David Berman, Volker Braun, Robbert
Dijkgraaf, Anton Gerasimov, Thomas Grimm, Sergei Gukov, Hirosi
Ooguri, Samson Shatashvili, Annamaria Sinkovics and Erik Verlinde
for useful discussions. We also thank Sheer El-Showk for finding typographical errors in the earlier version of this paper.  This research is partially supported by
the stiching FOM.

\appendix

\section{The Coulomb Gas Representation}

A useful (though subtle) representation of minimal models is the
``Coulomb gas''  representation. Much of the evidence pointing at
a possible topological twisting for $G_2$ manifolds was
constructed in \cite{sv}\ using this approach. For reasons that
will become apparent defining the topological theory in this
representation is very difficult. Although we proceeded in the
main text to define the topological construction in an independent
way which avoids many of the complications of the Coulomb Gas
Representation, we summarize it here for completeness as well as
for a useful source of intuition for the results we obtained in
the main text.

In the Coulomb gas representation minimal model primaries are represented as
vertex operators in a theory of a scalar coupled to a background
charge. The holomorphic energy momentum tensor in such theories is
given by
\begin{equation}\label{actionwithbgdcharge}
T(z)=-\frac{1}{2}\left(
\partial\phi(z)\partial\phi(z)+iQ\partial^2\phi(z)\right)
\end{equation}
with central charge
\begin{equation}\label{centralchargewithbgd}
c=1-3Q^2.
\end{equation}
Primaries are the ``vertex operators" \be\label{veenen}
V_{n'n}(z)\equiv e^{i\alpha_{n'n}\phi(z)} \ee where
\begin{equation}\label{alphanprimen}
\alpha_{n'n}=\frac{1}{\sqrt{2}}[(n'-1)\alpha_-+(n-1)\alpha_+].
\end{equation}
The conformal dimension of these operators
\begin{equation}\label{weightvertex}
h(V_{n'n})=\frac{1}{2}\alpha_{n'n}(\alpha_{n'n}+Q).
\end{equation}
In the Tri-critical Ising model we choose $Q={1\over\sqrt{10}}$
which sets $\alpha_+ = {4 \over \sqrt{10}}$ and $\alpha_-=-{5 \over
\sqrt{10}}$ and one can easily verify that \ref{weightvertex}\
correctly reproduce the conformal weights inside the tri-critical Ising model.

An important subtlety arises because one can construct two weight
$1$ vertex operators $V_{\pm}\equiv V_{\pm 1,\mp
1}=e^{-i\sqrt{2}\alpha_{\pm}}$ called screening operators.
Integrating $V_{\pm}$ against the vertex operators \ref{veenen}\
gives screened vertex operators which have the same conformal
weight as \ref{veenen}\ but a different ``charge'' under
$\phi\rightarrow\phi +const$. More precisely, these operators are
defined as
\begin{equation}
V_{n'n}^{r'r}(z) = \int \prod_{i=1}^{r'}du_i  \prod_{j=1}^{r} dv_j
V_{n'n}(z) V_{+}(u_1)\cdots V_{+}(u_{r'})V_{-}(v_1)\cdots V_{-}(v_r)
\end{equation}
where the contours of the $u$ and $v$ integrations have been defined
carefully in \cite{felder}. Each screened vertex operator
$V_{n'n}^{r'r}$ correspond to a  different conformal block of the
operator $V_{n'n}$. So, for example, in (\ref{updown}), the two
conformal blocks, in the Coulomb gas picture are given by
\begin{equation}
\Phi_{2,1}^{\uparrow}=P_\vdash V_{21}^{10} P_\vdash + P_\dashv
V_{21}^{00} P_\dashv~,~~~~~~\Phi_{2,1}^\downarrow=P_\vdash
V_{21}^{00} P_\vdash + P_\dashv V_{21}^{10} P_\dashv
\end{equation}
where we have been careful to put in projectors $P_\vdash$ and
$P_\dashv$. $P_\vdash$ projects to the states corresponding to the
first column of the Kac table and the first two entries of the
second column, whereas $P_\dashv$ projects to the last two entries
of the middle column and the third column of the Kac table. In
this way we unambiguously embed the minimal model Hilbert space in
the Hilbert space of the scalar field. Similarly, for the
conformal blocks of $\Phi_{1,2}$ we have the following Coulomb gas
representations:
\begin{equation}
\Phi_{1,2}^+ =P_\vdash V_{12}^{00} P_\vdash + P_\dashv V_{12}^{01}
P_\dashv~,~~~~~~\Phi_{1,2}^-=P_\vdash  V_{12}^{01} P_\vdash +
P_\dashv V_{12}^{00} P_\dashv \label{threepnine}
\end{equation}
 In the Coulomb gas representation of the Tri-critical Ising model, the field
$\phi$ has a background charge $Q={1 \over \sqrt{10}}$. If we just
consider the subspace of the Hilbert space corresponding to the
projection $P_\vdash$, we can write $ P_\vdash V_{12}^{00} P_\vdash
= e^{i { 5 \over 2 \sqrt 10} \phi}$ and then in this sector,
 insertions of two $\Sigma$ fields on a sphere effectively changes the background charge from
\begin{equation}
Q={1 \over \sqrt{10}} \rightarrow {6 \over \sqrt{10}}
\end{equation}
The central charge of the total CFT changes from $c={21 \over 2}$ to
zero:
\begin{equation}
c={3 \over 2} \times 7 ={7 \over 10}+{98 \over 10} \rightarrow
1-3({6 \over \sqrt{10}})^2 +{98 \over 10} =0
\end{equation}
which hints strongly at the existence of a topological theory.

Changing the background charge changes the weights of various
fields. The change in weight depends on the charge of the field.
In fact, since different conformal blocks of the same field carry
different charges, their weights shift by different amounts after
the twist. {\em The twisting acts differently on the conformal
blocks of the same operator}. For example, the new weights of some
of the blocks after the twist are
\be\label{neweightblock}\begin{split} G^{\downarrow}\quad
&\rightarrow\quad 1\quad,\quad
G^{\uparrow}\ \quad \rightarrow\quad 2\\
M^{\downarrow}\quad &\rightarrow\quad 2\quad,\quad
M^{\uparrow}\quad \rightarrow\quad 3\\
\end{split}
\ee

Using  \ref{weightvertex} one finds the conformal weights of Coulomb gas
vertex operators in the twisted theory shifted
\be\label{neweight}\begin{split}
V_{21}^{00}=e^{-2i\over\sqrt{10}},V_{31}^{00}=\ e^{-4i\over\sqrt{10}}\quad &\rightarrow\quad -{2\over 5}\\
V_{31}^{00}=e^{-6i\over\sqrt{10}},\quad\quad \mathbf{1}\quad &\rightarrow\quad\quad 0\\
V_{21}^{10} \sim e^{2i\over\sqrt{10}}\quad\quad\quad\quad
&\rightarrow \quad\quad {3 \over 5}
\end{split}
\ee

Notice that the blocks corresponding to the unscreened vertex
operators in the Coulomb gas representation, dressed with the appropriate
weight in the remainder CFT of the ``chiral" states \ref{special}\
become weight $0$ after the twist. Similar arguments were used in
\cite{sv}.

A few words about the Coulomb gas approach are however in order.
The Hilbert space of the free theory with a background charge is
larger than that of the minimal model. To go from the free theory
to the minimal model, we need to consider cohomologies of approach
BRST operators defined by Felder \cite{felder}. So while the
Coulomb gas representation is useful in doing computations, it
cannot be used to construct new operators unless they commute with
Felder's BRST operators. We thus emphasize that these arguments
should be taken as inspirational rather than rigorous.

\section{The $G_2$ Algebra}
\label{algebra}
The $G_2$ algebra is given by \cite{sv}
\begin{equation}\label{mgg}{\{G_n,G_m\} = \frac{7}{2}(n^2-\frac{1}{4})\delta_{n+m,0}+2L_{n+m}}\end{equation}
\begin{equation}\label{mTT}{[L_n,L_m]=\frac{21}{24}(n^3-n)\delta_{n+m,0}+(n-m)L_{n+m}}\end{equation}
\begin{equation}\label{mTg}{[L_n,G_m]=(\frac{1}{2}n-m)G_{n+m}}\end{equation}
\begin{equation}\label{phiph}{\{\Phi_n,\Phi_m\}=-\frac{7}{2}(n^2-\frac{1}{4})\delta_{n+m,0}+
6X_{n+m}}\end{equation}
\begin{equation}\label{Xphi}{[X_n,\Phi_m]=-5(\frac{1}{2}n-m)\Phi_{n+m}}\end{equation}
\begin{equation}\label{XX}{[X_n,X_m]=\frac{35}{24}(n^3-n)\delta_{n+m,0}-5(n-m)X_{n+m}}\end{equation}
\begin{equation}\label{TX}{[L_n,X_m]=-\frac{7}{24}(n^3-n)\delta_{n+m,0}+(n-m)X_{n+m}}\end{equation}
\begin{equation}\label{gphi}{\{G_n,\Phi_m\}=K_{n+m}}\end{equation}
\begin{equation}\label{gt}{[G_n,K_m]=(2n-m)\Phi_{n+m}}\end{equation}
\begin{equation}\label{gX}{[G_n,X_m]=-\frac{1}{2}(n+\frac{1}{2})G_{n+m}+M_{n+m}}\end{equation}
\begin{equation}\label{gM}{\{G_n,M_m\}=-\frac{7}{12}(n^2-\frac{1}{4})(n-\frac{3}{2})
\delta_{n+m,0}+
(n+\frac{1}{2})L_{n+m}+(3n-m)X_{n+m}} \end{equation}
\begin{equation}\label{phit}{[\Phi_n,K_m]=\frac{3}{2}(m-n+\frac{1}{2})G_{n+m}-3M_{n+m}}\end{equation}
\begin{equation}\label{phiM}{\{\Phi_n,M_m\}=(2n-\frac{5}{2}m-\frac{11}{4})K_{n+m}-3:G\Phi:_{n+m}}\end{equation}
\begin{equation}\label{xt}{[X_n,K_m]=3(m+1)K_{n+m}+3:G\Phi:_{n+m}}\end{equation}
\begin{eqnarray}
\label{XM}
[X_n,M_m]&=&[\frac{9}{4}(n+1)
(m+\frac{3}{2})-\frac{3}{4}(n+m+\frac{3}{2})
(n+m+\frac{5}{2})]G_{n+m}\\&&~~~~~~~-[5(n+1)-
\frac{7}{2}(n+m+\frac{5}{2})]M_{n+m}+4:GX:_{n+m} \nonumber \end{eqnarray}
\begin{equation}\label{tt}{[K_n,K_m]=-\frac{21}{6}(n^3-n)\delta_{n+m,0}+3(n-m)(X_{n+m}-L_{n+m})}\end{equation}
\begin{equation}\label{tM}{[K_n,M_m]=[\frac{11}{2}(n+1)(n+m+\frac{3}{2})-\frac{15}{2}(n+1)n]
\Phi_{n+m}+3:GK:_{n+m}-6:L\Phi:_{n+m}} \end{equation}
\begin{eqnarray}\label{MM}
\{M_n,M_m\}&=&-\frac{35}{24}(n^2-\frac{1}{4})(n^2-\frac{9}{4})
\delta_{n+m,0}+[\frac{3}{2}(n+m+2)(n+m+3)\\
&& ~~~~~-10(n+\frac{3}{2})(m+\frac{3}{2})]X_{n+m}+[\frac{9}{2}(n+
\frac{3}{2})(m+\frac{3}{2})\nonumber \\&
&~~~~~-\frac{3}{2}(n+m+2)(n+m+3)]L_{n+m} -4:GM:_{n+m}+8:LX:_{n+m} \nonumber \end{eqnarray}

An important property of the algebra is the fact that it contains a null ideal,
generated by \cite{blumenhagen,figueroa}
\be
{\cal N}= 4(GX)-2(\Phi K)-4 \partial M - \partial^2 G.
\label{nullideal}
\ee
This null ideal has various consequences. For example, it allows us to determine
the eigenvalue of $K_0$ on highest weight states in terms of their $L_0$ and $X_0$
eigenvalues. Thus, $K_0$ is not an independent quantum number in the theory.

In \cite{blumenhagen} a two-parameter family of chiral algebras was found, with the
same generators as the $G_2$ algebra. However, the $G_2$ algebra is the only one among this
family which has the right central charge $c=21/2$ and contains the tri-critical Ising model
as a subalgebra. The latter is needed for space-time supersymmetry, and therefore the
$G_2$ algebra appears to be uniquely fixed by these physical requirements.

The representation theory of the $G_2$ algebra was studied in some
detail in \cite{noyvert}. Both in the NS and R sector there are
short and long representations. We will discuss the
representations of the latter in the next section~\ref{ramond}. In
the NS sector the short representations correspond to what we
called chiral primaries, whereas in the R sector the short
representations correspond to R ground states.

Character formulae for the $G_2$ algebra are unknown. In
\cite{0108091} the partition functions for string theory on
particular non-compact $G_2$ manifolds were found, and from these
one can extract candidate character formulas for some of the
representations of the $G_2$ algebra. It would be nice to have
general explicit expressions for the characters. One may try to
obtain these by using the fact that the $G_2$ algebra can be
obtained by quantum Hamiltonian reduction (see e.g.
\cite{deboertjin}) from the affine super Lie algebra based on
$D(2,1,\alpha)$, as suggested in \cite{mallwitz}. Following the
strategy in \cite{frenkel} one expects that the characters can be
expressed in terms of highest weight characters of the
$D(2,1,\alpha)$ affine super Lie algebra, but we have not explored
this in this paper.

\section{R sector}
\label{ramond}

In this section we will be completely pedantic. In the R sector we
have the following commutation relations of the zero modes ($L_0$
commutes with everything) \bea
\{ G_0,G_0\} & = & 2(L_0-\frac{7}{16}) \nonumber \\
\{ G_0,\phi_0\} & = & K_0 \nonumber \\
{}[ G_0,X_0] & = & -\frac{1}{4} G_0 + M_0 \nonumber \\
\{ G_0,M_0\} & = & \frac{1}{2}(L_0-\frac{7}{16})  \\
{}[ G_0,K_0] & = & K_0 \nonumber \\
{}[X_0,K_0] & = & \frac{3}{2} K_0 - 3 \phi_0 G_0 \nonumber \\
{}[X_0,\phi_0] & = & 0 \nonumber \\
{}[X_0,M_0] & = & \frac{21}{16} G_0 - \frac{9}{4} M_0 + 4 G_0 X_0 \\
{}[K_0,\phi_0] & = & -\frac{3}{4} G_0 + 3 M_0 \nonumber \\
{}[K_0,M_0] & = & 3 G_0 K_0 - 6 \phi_0 (L_0- \frac{7}{16}) \nonumber \\
\{ \phi_0,\phi_0 \} & = & \frac{7}{8} + 6 X_0 \nonumber \\
\{ \phi_0,M_0 \} & = & \frac{7}{4} K_0 - 3 G_0 \phi_0 \nonumber \\
 \{M_0,M_0\} & = & \frac{21}{8}(L_0-\frac{7}{16}) + 8 (L_0-\frac{7}{16})X_0 - 4 G_0 M_0 .
\eea In addition, there is the operator \be {\cal N}= \frac{3}{2}
M_0 - 3 K_0 \phi_0 + 6 G_0 X_0 \ee which should be null when
acting on highest weight states. To extract this algebra from the
operator product expansion one needs to use a suitable normal
ordering prescription. One may check that this algebra is
consistent with hermiticity, associativity, and yields the right
spectrum for $X_0$.

To build representations, we first consider a highest weight
vector of the form $|7/16,h_r\rangle$. One may check that
$(\frac{7}{4} G_0 + M_0)|7/16,h_r\rangle$ has $X_0$ eigenvalue
equal to $-99/16$. This is outside the Kac table for the
tri-critical Ising model. Therefore, this vector has to be null.
Given this null vector, we find that the representation a priori
has four states remaining. Notice that, as we will discuss
momentarily, these representations may still be reducible.

We introduce the basis
\be
\left( \begin{array}{c} |7/16,h_r\rangle
\\ (-\frac{17}{4} G_0 + M_0)|7/16,h_r\rangle \\ \phi_0 |7/16,h_r\rangle \\
(-\frac{17}{4} G_0 + M_0) \phi_0 |7/16,h_r\rangle
\end{array} \right)
\ee
In this basis the various generators look like (with
$\hat{l}=L_0-\frac{7}{16}$)
\bea
G_0 & = & \left( \begin{array}{cccc} 0 & -6 \hat{l} & 0 & 0 \\
-\frac{1}{6} & 0 & 0 & 0 \\
0 & 0 & 0 & -6\hat{l} \\
0 & 0 & -\frac{1}{6} & 0 \end{array}\right) \nonumber \\
M_0 & = & \left( \begin{array}{cccc} 0 & -\frac{27}{2} \hat{l} & 0 & 0 \\
\frac{7}{24} & 0 & 0 & 0 \\
0 & 0 & 0 & -\frac{27}{2} \hat{l} \\
0 & 0 & \frac{7}{24} & 0 \end{array}\right) \nonumber \\
\phi_0 & = & \left( \begin{array}{cccc}
0 & 0 &  -\frac{49}{8} & 0 \\
0 & 0 & 0 & \frac{7}{8} \\
 1 & 0 & 0 & 0 \\
0 & -\frac{1}{7} & 0 & 0  \end{array}\right) \nonumber \\
X_0 & = & \left( \begin{array}{cccc}
-\frac{35}{16} & 0 & 0 & 0 \\
0 & -\frac{3}{16} & 0 & 0 \\
0 & 0 & -\frac{35}{16} & 0 \\
0 & 0 & 0 & -\frac{3}{16}  \end{array}\right) \nonumber \\
K_0 & = & \left( \begin{array}{cccc}
0 & 0 & 0 & \frac{63}{2} \hat{l} \\
0 & 0 & \frac{7}{8} & 0 \\
0 & -\frac{36}{7} \hat{l}  & 0 & 0 \\
-\frac{1}{7} & 0 & 0 & 0   \end{array}\right) .
\eea
There is a two-parameter family of possible metrics compatible
with unitarity, namely
\be
g=\left( \begin{array}{cccc} \frac{8a}{49} & 0 & -ib & 0 \\
0 & \frac{288 a \hat{l}}{49} & 0 & -36 i \hat{l} b \\
ib & 0 & a & 0 \\
0 & 36 i \hat{l} b & 0 & 36 \hat{l} a \end{array} \right) .
\ee
These representations are not irreducible. Indeed, we can go to an
eigenbasis of $\phi_0$. To do this we define a new basis as
\be
\left( \begin{array}{cccc} \frac{7i}{\sqrt{8}} & 0 & 1 & 0 \\
-\frac{7i}{\sqrt{8}} & 0 & 1 & 0 \\
0 & -\frac{7i}{\sqrt{8}} & 0 & 1 \\
0 & \frac{7i}{\sqrt{8}} & 0 & 1
 \end{array} \right)\left( \begin{array}{c} |7/16,h_r\rangle
\\ (-\frac{17}{4} G_0 + M_0)|7/16,h_r\rangle \\ \phi_0 |7/16,h_r\rangle \\
(-\frac{17}{4} G_0 + M_0) \phi_0 |7/16,h_r\rangle
\end{array} \right).
\ee
Then the generators become
\bea
G_0 & = & \left( \begin{array}{cccc} 0 & 0 & 0 & -6 \hat{l} \\
0 & 0 & -\hat{l} & 0 \\
0 & -\frac{1}{6} & 0 & 0 \\
-\frac{1}{6} & 0 & 0 & 0 \end{array}\right) \nonumber \\
M_0 & = & \left( \begin{array}{cccc} 0 & 0 & 0 & -\frac{27}{2} \hat{l} \\
0 & 0 & -\frac{27}{2} \hat{l} & 0 \\
0 & \frac{7}{24} & 0 & 0 \\
\frac{7}{24} & 0 & 0 & 0 \end{array}\right) \nonumber \\
\phi_0 & = & \left( \begin{array}{cccc} \frac{7i}{\sqrt{8}} & 0 &
0 & 0 \\ 0 & -\frac{7i}{\sqrt{8}} & 0 & 0 \\ 0 & 0 &
\frac{i}{\sqrt{8}} & 0 \\ 0 & 0 & 0 & -\frac{i}{\sqrt{8}}
 \end{array}\right) \nonumber \\
X_0 & = & \left( \begin{array}{cccc}
-\frac{35}{16} & 0 & 0 & 0 \\
0 & -\frac{35}{16} & 0 & 0 \\
0 & 0 & -\frac{3}{16} & 0 \\
0 & 0 & 0 & -\frac{3}{16}  \end{array}\right) \nonumber \\
K_0 & = & \left( \begin{array}{cccc}
0 & 0 & 0 & -\frac{18i}{\sqrt{2}} \hat{l} \\
0 & 0 & \frac{18i}{\sqrt{2}} \hat{l} & 0 \\
0 & \frac{i}{\sqrt{8}}  & 0 & 0 \\
-\frac{i}{\sqrt{8}}  & 0 & 0 & 0   \end{array}\right) .
\eea
The metric becomes
\be
g=\left( \begin{array}{cccc} c_1 & 0 & 0 & 0 \\
0 & c2 & 0 & 0 \\
0 & 0 & 36 c_2 \hat{l}  & 0 \\
0 & 0 & 0 & 36 c_1 \hat{l} \end{array} \right)
\ee
where $c_1,c_2$ are arbitrary constants related to $a,b$ in some
way which is not terribly important. We therefore see that the
representation splits into two complex conjugate ones which are
each two dimensional. For $\hat{l} \neq 0$ this is the complete
story, i.e. the zero modes are represented as two complex
conjugate two-dimensional representations. One is spanned by the
first and fourth vector, the other one by the second and the
third.

In the case we have R ground states, i.e. $\hat{l}=0$, we see that
the system degenerates further. We can consistently decouple the
third and fourth vector and find two complex conjugate
one-dimensional representations of the algebra. These correspond
to the $h_I=\frac{7}{16}$ R ground state that is purely internal.
In this representation, $G_0=M_0=K_0=0$.

The null module generated by the third and fourth vector also
provides two one-dimensional complex conjugate representations.
Taking $c_1$ and $c_2$ to scale as $1/\hat{l}$, we see that this
gives rise to one-dimensional representations of the form
$|\frac{3}{80},\frac{2}{5}\rangle$. In these representation also
$G_0=M_0=K_0=0$.

In short, in the R sector we have massless and massive
representations. If we combine the left and right movers, things
change a little bit. We cannot use eigenvectors of $\phi_0$ and
$\bar{\phi}_0$ with nonzero eigenvalue simultaneously, since that
is inconsistent with $\{\phi_0,\bar{\phi}_0\}=0$. The smallest
unitary representation of this algebra is two-dimensional.
Therefore, combining left and right massless representations leads
to a two-dimensional representation. Combining massless and
massive to a four-dimensional representation, and combining two
massive representations to a eight-dimensional representation.

\section{Decomposition of differential forms into irreps of $G_2$}
\label{decomposition}
In this appendix, we review the decomposition of differential forms into irreducible representations of the group $G_2$. Our discussion follows the one in \cite{Karigiannis}

For a $G_2$ manifold, differential forms of any degree can be decomposed into irreducible representations of $G_2$
\begin{eqnarray*}
\Lambda^0 = \Lambda^0_1 & ~~~~~~~~~& \Lambda^1 = \Lambda^1_7 \\
\Lambda^2=\Lambda^2_7 \oplus \Lambda^2_{14} & ~~~~~& \Lambda^3=\Lambda^3_1 \oplus \Lambda^3_7 \oplus \Lambda^3_{27}
\end{eqnarray*}

This decomposition is compatible with the Hodge star operation, so
$ *\Lambda^n_m = \Lambda^{7-n}_m$. It is useful to define 
this decomposition into irreducible representations explicitly.

\paragraph {2-forms and 5-forms}
The 2-forms decompose into a {\bf  7} and {\bf 14} of $G_2$. These spaces can be characterized as follows:
\begin{eqnarray*}
\Lambda^2_7 &=& \{\omega \in \Lambda^2;~*(\phi \wedge \omega)=2 \omega \} \\
\Lambda^2_{14}
& = & \{\omega \in \Lambda^2; ~*(\phi \wedge \omega) =-\omega \}
\end{eqnarray*}
It is useful to write expressions for projector operators $\pi_7$ and $\pi_{14}$. These  project onto the appropriate subspaces:
\begin{eqnarray*}
\pi^2_7(\omega)&=&{\omega + *(\phi \wedge \omega) \over 3} \\
\pi^2_{14}(\omega) & = & {2 \omega -*(\phi \wedge \omega) \over 3}
\end{eqnarray*}
where the superscript $2$ on $\pi^2_k$ indicates that this is the projector when acting on 2-forms.
In local coordinates, these can be written as
\begin{eqnarray*}
(\pi^2_7)_{ab}^{de} &=& 6 \phi_{ab}^c \phi_{c}^{de} = 4 \phi_{ab}^{de} +{1 \over 6} (\delta^d_a \delta^e_b-\delta^e_a\delta^d_b)  \\
(\pi^2_{14})_{ab}^{ef}&=&- 4 \phi_{ab}^{ef} +{1 \over 3}(\delta^e_a \delta^f_b-\delta^d_a\delta^e_b)
\end{eqnarray*}
Similarly, for 5 forms, we have the decomposition:
\begin{eqnarray*}
\Lambda^5_7&=&\{\omega \in \Lambda^5; ~ \phi \wedge * \omega = 2 \omega \} \\
\Lambda^5_{14} & = & \{ \omega \in \Lambda^5, ~ \phi \wedge *\omega = - \omega \}
\end{eqnarray*}
which implies the projectors
\begin{eqnarray*}
\pi^5_7(\omega) &= &{\omega + \phi \wedge *\omega \over 3} \\
\pi^5_{14}(\omega) & = & {2 \omega -\phi \wedge *\omega \over 3}
\end{eqnarray*}
\paragraph{3-forms and 4-forms}
The three forms decompose into {\bf 1}, {\bf 7} and {\bf 27} dimensional representations of $G_2$. Explicitly, these spaces are given by
\begin{eqnarray*}
\Lambda_1^3 &=& \{\omega \in \Lambda^3: ~\phi \wedge(*(*\phi \wedge \omega)) =7 \omega \} \\
\Lambda^3_7 & = & \{ \omega \in \Lambda^3;~*(\phi \wedge *(\phi \wedge \omega))=-4 \omega \}\\
\Lambda^3_{27} & = & \{ \omega \in \Lambda^3;~ \phi \wedge \omega = * \phi \wedge \omega =0 \}
\end{eqnarray*}
We also define projection operators:
\begin{eqnarray*}
\pi^3_1(\omega) & = & {1 \over 7} \phi \wedge (*(*\phi \wedge \omega)) \\
\pi_7^3(\omega) & = & -{1 \over 4}(*(\phi \wedge *(\phi \wedge \omega))) \\
\pi_{27}^3 (\omega)& = & \omega -\pi_1^3(\omega) -\pi_7^3(\omega)
\end{eqnarray*}
For four forms, we have the decomposition
\begin{eqnarray*}
\Lambda_1^4 &=& \{\omega \in \Lambda^4: ~*\phi \wedge(*(\phi \wedge \omega)) =7 \omega \} \\
\Lambda^4_7 & = & \{ \omega \in \Lambda^4;~\phi \wedge *(\phi \wedge *\omega)=-4 \omega \}\\
\Lambda^4_{27} & = & \{ \omega \in \Lambda^4;~ \phi \wedge \omega = * \phi \wedge \omega =0 \}
\end{eqnarray*}
and the projectors
\begin{eqnarray*}
\pi^4_1(\omega) & = & {1 \over 7} *\phi \wedge (*(\phi \wedge \omega)) \\
\pi_7^4(\omega) & = & -{1 \over 4}(\phi \wedge *(\phi \wedge *\omega)) \\
\pi_{27}^4 (\omega)& = & \omega -\pi_1^4(\omega) -\pi_7^4(\omega)
\end{eqnarray*}


There are natural $G_2$-equivariant isomorphisms between these
spaces. For example, the map $\omega \rightarrow \phi \wedge
\omega$ is an isomorphism between $\Lambda^p_r \cong
\Lambda^{p+3}_r$ if $\phi \wedge \omega_p$ is non-zero when
$\omega \in \Lambda^p_r$:
\begin{eqnarray*}
\Lambda^0_1 &\cong& \Lambda_1^3 ~~~~~~~~~~\Lambda^1_7 \cong \Lambda^4_7 \\
\Lambda^2_7 &\cong& \Lambda_7^5 ~~~~~~~~~~\Lambda^2_{14} \cong \Lambda^5_{14} \\
\Lambda^3_7 &\cong& \Lambda_7^6 ~~~~~~~~~~\Lambda^4_1 \cong \Lambda^7_1
\end{eqnarray*}
Also, the map $\omega \rightarrow *\phi \wedge \omega$ is an
isomorphism between $\Lambda^p_r\cong \Lambda^{p+4}_r$ when  $*
\phi \wedge \omega_p$ is non-zero when $\omega \in \Lambda^p_r$:
\begin{eqnarray*}
\Lambda^0_1 &\cong& \Lambda^4_1 ~~~~~~~~~~ \Lambda^1_7 \cong \Lambda^5_7 \\
\Lambda^2_7 &\cong& \Lambda^6_7 ~~~~~~~~~~ \Lambda^3_1 \cong \Lambda^7_1
\end{eqnarray*}

\section{Some correlation functions}

We can use the expression (\ref{twicorsphere}) to compute some
correlation functions in the twisted theory in terms of
correlation functions of the untwisted theory. For example, the
two point function of operators
\[
{\cal O}_2 = \Phi_{2,1} \otimes \psi_{h},~~~~~~~{\cal O}_3 =
\Phi_{3,1} \otimes \psi_{h}
\]
can be written in terms of a four-point function of the tri-critical Ising model
\begin{eqnarray*}
\langle {\cal O}_2(z_1) {\cal O}_3(z_2) \rangle &=& z_1^{-{1 \over
2}} z_2^{-1} (z_1-z_2)^{-2h} \times
\langle \Phi_{1,2}(\infty) \Phi_{2,1}(z_1) \Phi_{3,1}(z_2) \Phi_{1,2}(0) \rangle_{\rm tri-critical} \\
& = & {c \over (z_1-z_2)^{2h-{4 \over 5}}}
\end{eqnarray*}
where $c$ is a constant. This is independent of position if $h={2
\over 5}$, which is what we need for the operators ${\cal O}_2$
and ${\cal O}_3$ to be chiral in the topological theory. This
correlation functions gets contributions from only one conformal
block, precisely the one that is kept in the topological theory.
On the other hand, consider the two point function of operators
whose tri-critical Ising model weight is ${1 \over 10}$:
\[
{\cal O}={\Phi_{2,1}} \otimes \psi_{h}
\]
The two point function of this operator with itself can be written
in terms of a four-point function of the tri-critical Ising model:
\begin{eqnarray*}
\langle {\cal O}(z_1) {\cal O}(z_2) \rangle &=& z_1^{-{1 \over 2}} z_2^{-{1\over 2}} (z_1-z_2)^{-2h} \times
\langle \Phi_{1,2}(\infty) \Phi_{2,1}(z_1) \Phi_{2,1}(z_2) \Phi_{1,2}(0) \rangle_{\rm tri-critical} \\
& = & {c \over (z_1-z_2)^{2h+{1 \over 5}}} \times {z_1+z_2 \over {z_1 z_2}}
\end{eqnarray*}
This is not even translationally invariant! However, it is easy to
see that the conformal block that contributes to this correlation
function is
\[  \langle \Phi_{1,2} {\cal O}^\uparrow {\cal O}^\downarrow \Phi_{1,2} \rangle \]
but ${\cal O}^\uparrow$ is not a chiral operator. Correlation
functions of chiral operators obey all the properties of a usual
CFT. However, correlation functions of non-chiral operators in the
twisted theory are not that of a CFT. This is qualitatively
different from what happens in the usual ${\cal N}=2$ twisting. In
that case, the twisted theory makes sense as a CFT, even before we
restrict ourselves to chiral operators. This intermediate CFT does
not seem to exist for us.

\section{Spectral flow and the twist}
\label{spectral}
Whether or not the twisted stress tensor exists, and if so what
its precise form is remains for now an open problem. In the case
of Calabi-Yau manifolds, the existence of spectral flow was useful
in order to construct the twisted stress tensor, so it is worth
considering what precisely the analogue of spectral flow is in our
case.

Spectral flow, a word used rather loosely, refers to a particular
isomorphism between the R and NS sector of an $N=2$ conformal
field theory. What it does is easily illustrated in case of a free
scalar field $\varphi$. Denote by $\hat{p}=i \oint \partial
\varphi$ the zero mode of the momentum operator, and  by $\hat{x}$
the conjugate coordinate. Then spectral flow by the amount $\eta$
is simply implemented by the operator \be S_1=e^{i\eta \hat{x}}.
\ee Spectral flow maps representations with momentum eigenvalue
$p$ to representations with momentum eigenvalue $p+\eta$. If we
bosonize the $U(1)$ current in $N=2$ theories then this $S_1$
precisely implements what is usually referred to as spectral flow.

This is not quite the same as the statement that some particular R
operator generates spectral flow. In that case, we are talking
about an operator in the theory, and not a simple object
constructed out of zero modes only such as $S_1$. It is this full
operator, and not $S_1$, that appears in the generator of
space-time supersymmetry. It is again easy to illustrate this in
the case of a free scalar field. Instead of $S_1$ we consider the
operator \be S_2=\oint \frac{dz}{z^{\eta q+1}} e^{i\eta\phi} :
{\cal H}_p \rightarrow {\cal H}_{\eta+p} \ee acting on
representations with momentum eigenvalue $p$ and mapping them to
representations of eigenvalue $p+\eta$. On highest weight states,
$S_1$ and $S_2$ are identical, but on descendants they are not.
The new stress tensors obtained by spectral flow are obtained
using $S_1$. One can also define new stress tensors using the
action of $S_2$, simply as $L_n'=S_2^{-1} L_n S_2$, but this is
not usually done. One can explicitly work out the difference
between the two prescriptions, but that is not very insightful.
The modes of the twisted stress tensors of the A and B-model are
linear combinations of the modes of the initial stress tensor and
its spectrally flown version. This is spectral flow with respect
to $S_1$. Whether the twisted stress tensor have any relation to
the new stress tensor obtained through $S_2$ is not known.

In the case of $G_2$ manifolds, the situation is different. We no
longer have a version of $S_1$, but we do have a version of $S_2$,
where the exponential of the field is now replaced by the R vertex
operator $V_{7/16,+}$. It maps chiral primaries to R ground states
and vice versa. It should induce an isomorphism between the NS and
R sector of the theory, otherwise the theory would not be
space-time supersymmetric. In particular, this implies that we can
define a new stress tensor in say the NS sector via $L_n'=S_2^{-1}
L_n S_2$. Clearly, $L_0'$ annihilates all chiral primaries and is
a good candidate for a the zero mode of a twisted stress tensor.
Whether the highest modes of $L_n'$ can also be used to construct
the modes of a twisted stress tensor still remain to be worked
out, even in the case of Calabi-Yau manifolds. We leave this as an
interesting direction to explore.


%

%



\newpage

\begin{thebibliography}{100}

\bibitem{Witten}
E.~Witten, ``Mirror manifolds and topological field theory,''
arXiv:hep-th/9112056.

\bibitem{sdual}
A.~Neitzke and C.~Vafa,
``N = 2 strings and the twistorial Calabi-Yau,''
arXiv:hep-th/0402128;
N.~Nekrasov, H.~Ooguri and C.~Vafa,
``S-duality and topological strings,''
JHEP {\bf 0410}, 009 (2004)
[arXiv:hep-th/0403167].


\bibitem{gerasimov}
A.~A.~Gerasimov and S.~L.~Shatashvili,
``Towards integrability of topological strings. I: Three-forms on Calabi-Yau
manifolds,''
JHEP {\bf 0411}, 074 (2004)
[arXiv:hep-th/0409238].

\bibitem{vafa}
R.~Dijkgraaf, S.~Gukov, A.~Neitzke and C.~Vafa,
``Topological M-theory as unification of form theories of gravity,''
arXiv:hep-th/0411073.

\bibitem{nekrasov}
N.~Nekrasov,
``A la recherche de la m-theorie perdue. Z theory: Chasing m/f theory,''
arXiv:hep-th/0412021.

\bibitem{grassi}
P.~A.~Grassi and P.~Vanhove,
``Topological M theory from pure spinor formalism,''
arXiv:hep-th/0411167.


\bibitem{sinkovics}
L.~Anguelova, P.~de Medeiros and A.~Sinkovics,
``On topological F-theory,''
arXiv:hep-th/0412120.

\bibitem{sv}
S.~L.~Shatashvili and C.~Vafa, ``Superstrings and manifold of
exceptional holonomy,'' arXiv:hep-th/9407025.


\bibitem{dolbeault}
S. Salamon, ``Riemannian Geometry and Holonomy Groups'', Pitman Research Notes in Math. Series 201, Longman, Harlow, 1989; R. Reyes, ``Some special geometries defined by Lie Groups'', Phd Thesis, Oxford Univ., 1993; M.  Fernandez, L.  Ugarte,`` Dolbeault Cohomology for $G_2$-Manifolds'', Geometriae Dedicata, Volume 70, Issue 1, 57 - 86, 1998.

\bibitem{blumenhagen}
R.~Blumenhagen,
``Covariant construction of N=1 superW algebras,''
Nucl.\ Phys.\ B {\bf 381}, 641 (1992).


\bibitem{figueroa}
J.~M.~Figueroa-O'Farrill,
``A note on the extended superconformal algebras associated with  manifolds of
exceptional holonomy,''
Phys.\ Lett.\ B {\bf 392}, 77 (1997)
[arXiv:hep-th/9609113].


\bibitem{noyvert}
D.~Gepner and B.~Noyvert,
``Unitary representations of SW(3/2,2) superconformal algebra,''
Nucl.\ Phys.\ B {\bf 610}, 545 (2001)
[arXiv:hep-th/0101116];
B.~Noyvert,
``Unitary minimal models of SW(3/2,3/2,2) superconformal algebra and
manifolds of G(2) holonomy,''
JHEP {\bf 0203}, 030 (2002)
[arXiv:hep-th/0201198];


\bibitem{9604133}
B.~S.~Acharya,
``N=1 M-theory-Heterotic Duality in Three Dimensions and Joyce Manifolds,''
arXiv:hep-th/9604133.

\bibitem{9707186}
B.~S.~Acharya,
``On mirror symmetry for manifolds of exceptional holonomy,''
Nucl.\ Phys.\ B {\bf 524}, 269 (1998)
[arXiv:hep-th/9707186].


\bibitem{0108091}
T.~Eguchi and Y.~Sugawara,
``CFT description of string theory compactified on non-compact manifolds  with
G(2) holonomy,''
Phys.\ Lett.\ B {\bf 519}, 149 (2001)
[arXiv:hep-th/0108091].

\bibitem{0110302}
R.~Roiban and J.~Walcher,
``Rational conformal field theories with G(2) holonomy,''
JHEP {\bf 0112}, 008 (2001)
[arXiv:hep-th/0110302].

\bibitem{0111012}
T.~Eguchi and Y.~Sugawara,
``String theory on G(2) manifolds based on Gepner construction,''
Nucl.\ Phys.\ B {\bf 630}, 132 (2002)
[arXiv:hep-th/0111012].


\bibitem{0111048}
R.~Blumenhagen and V.~Braun,
``Superconformal field theories for compact G(2) manifolds,''
JHEP {\bf 0112}, 006 (2001)
[arXiv:hep-th/0110232];
``Superconformal field theories for compact manifolds with Spin(7)
holonomy,''
JHEP {\bf 0112}, 013 (2001)
[arXiv:hep-th/0111048].


\bibitem{Aganagic}
  M.~Aganagic and C.~Vafa,
  ``G(2) manifolds, mirror symmetry and geometric engineering,''
  arXiv:hep-th/0110171.


\bibitem{roiban}
R.~Roiban, C.~Romelsberger and J.~Walcher,
``Discrete torsion in singular G(2)-manifolds and real LG,''
Adv.\ Theor.\ Math.\ Phys.\  {\bf 6}, 207 (2003)
[arXiv:hep-th/0203272].

\bibitem{0204213}
K.~Sugiyama and S.~Yamaguchi,
``Coset construction of noncompact Spin(7) and G(2) CFTs,''
Phys.\ Lett.\ B {\bf 538}, 173 (2002)
[arXiv:hep-th/0204213].

\bibitem{0301164}
T.~Eguchi, Y.~Sugawara and S.~Yamaguchi,
``Supercoset CFT's for string theories on non-compact special holonomy
manifolds,''
Nucl.\ Phys.\ B {\bf 657}, 3 (2003)
[arXiv:hep-th/0301164].

\bibitem{gaberdiel}
M.~R.~Gaberdiel and P.~Kaste,
``Generalised discrete torsion and mirror symmetry for G(2) manifolds,''
JHEP {\bf 0408}, 001 (2004)
[arXiv:hep-th/0401125].

\bibitem{0409191}
B.~S.~Acharya and S.~Gukov,
``M theory and Singularities of Exceptional Holonomy Manifolds,''
Phys.\ Rept.\  {\bf 392}, 121 (2004)
[arXiv:hep-th/0409191].



\bibitem{Howe}
  P.~S.~Howe and G.~Papadopoulos,
  ``Holonomy groups and W symmetries,''
  Commun.\ Math.\ Phys.\  {\bf 151}, 467 (1993)
  [arXiv:hep-th/9202036].

\bibitem{Odake}
  S.~Odake,
  ``Extension Of N=2 Superconformal Algebra And Calabi-Yau Compactification,''
  Mod.\ Phys.\ Lett.\ A {\bf 4}, 557 (1989).

\bibitem{felder}
G.~Felder, ``Brst Approach To Minimal Methods,'' Nucl.\ Phys.\ B
{\bf 317}, 215 (1989) [Erratum-ibid.\ B {\bf 324}, 548 (1989)].

\bibitem{joycebook}
D.~Joyce,
``Compact manifolds with special holonomy'',  Oxford University Press, 2000.

\bibitem{math0311253}
X.~ Dai, X.~Wang, G.~Wei,
``On the Stability of Riemannian Manifold with Parallel Spinors,'' arXiv:math.dg/0311253.

\bibitem{kodira}
M.~Bershadsky, S.~Cecotti, H.~Ooguri and C.~Vafa,
``Kodaira-Spencer theory of gravity and exact results for quantum string
amplitudes,''
Commun.\ Math.\ Phys.\  {\bf 165}, 311 (1994)
[arXiv:hep-th/9309140].



\bibitem{antoniadis}
I.~Antoniadis, E.~Gava, K.~S.~Narain and T.~R.~Taylor,
``Topological amplitudes in string theory,''
Nucl.\ Phys.\ B {\bf 413}, 162 (1994)
[arXiv:hep-th/9307158].

\bibitem{Dijkgraaf}
  R.~Dijkgraaf, H.~Verlinde and E.~Verlinde,
  ``Topological Strings In D < 1,''
  Nucl.\ Phys.\ B {\bf 352}, 59 (1991).

\bibitem{hitchin}
N.~Hitchin,
``The geometry of three-forms in six and seven dimensions,''
arXiv:math.dg/0010054;
N.~Hitchin,
``Stable forms and special metrics,''
arXiv:math.dg/0107101.

\bibitem{leeleung}
  J.~H.~Lee and N.~C.~Leung,
  ``Geometric structures on G(2) and Spin(7)-manifolds,''
  arXiv:math.dg/0202045.

\bibitem{relevantwitten}
  E.~Witten,
  ``Quantum background independence in string theory,''
  arXiv:hep-th/9306122.


\bibitem{erikv}
E.~Verlinde,
  ``Attractors and the holomorphic anomaly,''
  arXiv:hep-th/0412139.



\bibitem{Gutowski:2001fm}
  J.~Gutowski and G.~Papadopoulos,
  ``Moduli spaces and brane solitons for M theory compactifications on
  holonomy G(2) manifolds,''
  Nucl.\ Phys.\ B {\bf 615}, 237 (2001)
  [arXiv:hep-th/0104105].






\bibitem{recentwitten}
  V.~Pestun and E.~Witten,
  ``The Hitchin functionals and the topological B-model at one loop,''
  arXiv:hep-th/0503083.

\bibitem{recentlouis}
  T.~W.~Grimm and J.~Louis,
  ``The effective action of type IIA Calabi-Yau orientifolds,''
  arXiv:hep-th/0412277.


\bibitem{fms}
  D.~Friedan, E.~J.~Martinec and S.~H.~Shenker,
  ``Conformal Invariance, Supersymmetry And String Theory,''
  Nucl.\ Phys.\ B {\bf 271}, 93 (1986).



\bibitem{deboeroz}
  N.~J.~Hitchin, A.~Karlhede, U.~Lindstrom and M.~Rocek,
  ``Hyperkahler Metrics And Supersymmetry,''
  Commun.\ Math.\ Phys.\  {\bf 108}, 535 (1987).



\bibitem{osv}
  H.~Ooguri, A.~Strominger and C.~Vafa,
  ``Black hole attractors and the topological string,''
  Phys.\ Rev.\ D {\bf 70}, 106007 (2004)
  [arXiv:hep-th/0405146].

\bibitem{mohaupt}
  C.~Mayer and T.~Mohaupt,
  ``Domain walls, Hitchin's flow equations and G(2)-manifolds,''
  Class.\ Quant.\ Grav.\  {\bf 22}, 379 (2005)
  [arXiv:hep-th/0407198].

\bibitem{wittencsasst}
  E.~Witten,
  ``Chern-Simons gauge theory as a string theory,''
  Prog.\ Math.\  {\bf 133}, 637 (1995)
  [arXiv:hep-th/9207094].


\bibitem{harveymoore}
  J.~A.~Harvey and G.~W.~Moore,
  ``Superpotentials and membrane instantons,''
  arXiv:hep-th/9907026.

\bibitem{beaslywitten}
  C.~Beasley and E.~Witten,
  ``A note on fluxes and superpotentials in M-theory compactifications on
  manifolds of G(2) holonomy,''
  JHEP {\bf 0207}, 046 (2002)
  [arXiv:hep-th/0203061].


\bibitem{marinoetc}
  V.~Bouchard, B.~Florea and M.~Marino,
  ``Counting higher genus curves with crosscaps in Calabi-Yau orientifolds,''
  JHEP {\bf 0412}, 035 (2004)
  [arXiv:hep-th/0405083];
  V.~Bouchard, B.~Florea and M.~Marino,
  ``Topological open string amplitudes on orientifolds,''
  JHEP {\bf 0502}, 002 (2005)
  [arXiv:hep-th/0411227].

\bibitem{witt}
  F.~Witt,
  ``Generalised $G_2$-manifolds,''
  arXiv:math.dg/0411642;
  F.~Witt,
  ``Special metric structures and closed forms,''
  arXiv:math.dg/0502443.


\bibitem{spin7}
J.~de Boer, A.~Naqvi and A.~Shomer, ``Topological Strings on Exceptional Holonomy Manifolds'', to appear.


\bibitem{deboertjin}
  J.~de Boer and T.~Tjin,
  ``Quantization and representation theory of finite W algebras,''
  Commun.\ Math.\ Phys.\  {\bf 158}, 485 (1993)
  [arXiv:hep-th/9211109];
  J.~de Boer and T.~Tjin,
  ``The Relation between quantum W algebras and Lie algebras,''
  Commun.\ Math.\ Phys.\  {\bf 160}, 317 (1994)
  [arXiv:hep-th/9302006].

\bibitem{mallwitz}
  S.~Mallwitz,
  ``On SW minimal models and N=1 supersymmetric quantum Toda field theories,''
  Int.\ J.\ Mod.\ Phys.\ A {\bf 10}, 977 (1995)
  [arXiv:hep-th/9405025].


\bibitem{frenkel}
  E.~Frenkel, V.~Kac and M.~Wakimoto,
  ``Characters and fusion rules for W algebras via quantized Drinfeld-Sokolov
  reductions,''
  Commun.\ Math.\ Phys.\  {\bf 147}, 295 (1992).







\bibitem{Karigiannis}
S.~Karigiannis,
``Deformations of $G_2$and Spin(7) Structures on Manifolds,''
arXiv: math.DG/0301218.






\end{thebibliography}
\end{document}